\documentclass[fleqn,usenatbib]{mnras}

\usepackage[T1]{fontenc}
\usepackage{ae,aecompl}

\usepackage{graphicx}	
\usepackage{amsmath}	
\usepackage{amssymb}	
\usepackage{gensymb}
\usepackage{rotating}
\usepackage{float}
\restylefloat{table}
\restylefloat{figure}
\usepackage{enumerate}
\usepackage[utf8]{inputenc}
\usepackage{caption}
\usepackage{verbatim}
\DeclareUnicodeCharacter{2212}{-}

\title[High Mass X-ray Binaries with Fermi-LAT]{The Galactic high mass X-ray binary population with \textit{Fermi}-LAT}

\author[M. Harvey et al.]{
Max Harvey,$^{1}$\thanks{E-mail: max.harvey@durham.ac.uk}
Cameron B. Rulten,$^{1}$
and Paula M. Chadwick$^{1}$
\\
$^{1}$Centre for Advanced Instrumentation, Department of Physics, University of Durham, South Road, Durham DH1 3LE, United Kingdom
}

\date{Accepted XXX. Received YYY; in original form ZZZ}

\pubyear{2021}

\begin{document}
\label{firstpage}
\pagerange{\pageref{firstpage}--\pageref{lastpage}}
\maketitle

\begin{abstract}
We search for $\gamma$-ray emission from 114 Galactic high mass X-ray binaries, including 4 well studied catalogued sources, in 12.5 years of \textit{Fermi}-LAT data in conjunction with the 10-year point source catalogue. Where a $\gamma$-ray excess appears to be spatially coincident with an X-ray binary, further investigation is performed to ascertain whether this excess is the product of physical processes within the binary system itself. We identify $\gamma$-ray excesses coincident with 20 high mass X-ray binaries where there is little or no prior evidence for $\gamma$-ray emission. However, we find that many of these are false positives caused by source confusion or the $\gamma$-ray background. Nonetheless, tentative but promising indicators of $\gamma$-ray emission are identified for several new systems, notably including 1A\,0535+262, RX\,J2030.5+4751 and SAX\,J1324.4-6200. 

\end{abstract}



\section{Introduction}

X-ray binaries (e.g. \citealt{verbunt_origin_1993} \& \citealt{casares_x-ray_2016}) are systems where a compact object (the accretor), either a stellar mass black hole or neutron star, and a companion star (the donor) are in orbit around a common gravitational barycentre. They can be divided into two broad sub-populations based on the mass of the donor star. The high mass X-ray binaries (HMXBs) (e.g. \citealt{reig_bex-ray_2011}, \citealt{walter_high-mass_2015} \& \citealt{kretschmar_advances_2019}) have massive companion stars which continually lose mass through stellar winds; these winds are then accreted onto the compact object and are heated, causing X-ray emission. The low mass X-ray binaries (LMXBs) \citep{van_paradijs_low-mass_2001} have lower mass companion stars which fill, and exceed, their Roche lobe as part of the latter stages of stellar evolution. As a result, matter from these companion stars overflows to the compact object through the first Lagrangian point, and thence is accreted (\citealt{paczynski_evolutionary_1971} \& \citealt{tout_wind_1991}).

Gamma-rays have been detected from a variety of binary star systems, including X-ray binaries. In the most recent catalogue from the \textit{Fermi} Large Area Telescope (\textit{Fermi}-LAT) \citep{atwood_large_2009}, the 4FGL-DR2 (\citealt{abdollahi_fermi_2020}, \citealt{ballet_fermi_2020}), a total of 22\footnote{We consider the sources with either the `HMB', `LMB', `NOV' or `BIN' source class in the 4FGL-DR2 to be \textit{Fermi}-LAT detected binary systems. Two further 4FGL sources are known: 4FGL J1405.1-6119 \citep{corbet_discovery_2019} and HESS\,J1832−093/4FGL\,J1832.9-0913 \citep{marti-devesa_hints_2020}. We do not discuss these in depth as they do not have the HMB source class.} binary systems are listed as $\gamma$-ray sources, 8 of which are HMXBs. These 8 objects can be divided broadly into two classes: $\gamma$-ray emitting microquasars and $\gamma$-ray binaries (\citealt{mirabel_gamma-ray_2012}, \citealt{dubus_gamma-ray_2015} \& \citealt{paredes_phenomenology_2019}). These classes are not mutually exclusive with one another, with microquasars being distinguished by their physical properties and $\gamma$-ray binaries being an exclusively phenomenological label. Additionally, $\gamma$-rays are seen from 13 nearby $\gamma$-ray novae\footnote{Only one $\gamma$-ray nova is included in the 4FGL-DR2, V5668\,Sgr.} (e.g. \citealt{morris_gamma-ray_2017} \& \citealt{franckowiak_search_2018}) and the colliding wind binaries, $\eta$-Carinae \citep{abdo_fermi_2010-1} and $\gamma ^{2}$-Velorum (WR11)\footnote{Of these two binaries, only $\eta$-Carinae is included in the 4FGL-DR2.} (\citealt{pshirkov_fermi-lat_2016} \& \citealt{marti-devesa_hints_2020}). We do not consider these here, nor do we deal with the LMXB population \citep{liu_catalogue_2007}, which will be discussed in a later paper.

\subsection{Microquasars}
In both low and high mass X-ray binaries, matter falling towards the accretor releases large amounts of gravitational potential energy, primarily in the form of X-ray emission, the distinguishing feature of these systems. This emission is not constant; variability is a common feature of X-ray binary systems as X-ray emission is fundamentally linked to accretion rate, which is itself inherently variable. The behaviour of X-ray binary systems is described by the Hardness-Intensity model \citep{fender_towards_2004}, where accretion discs build over time and then drain onto the central object, producing a relativistic jet similar to those seen in active galactic nuclei (AGN). This jet gives rise to a class of X-ray binaries known as the microquasars (e.g \citealt{mirabel_superluminal_1994} \& \citealt{corbel_microquasars_2010}). Typical radio-loud AGN are distinguished by their radio jets (for example in the well-studied radio galaxy M\,87 (\citealt{turland_observations_1975}, \citealt{biretta_hubble_1999} \& \citealt{walker_observations_2016})). The microquasar binary systems also have strong radio emission from their jets in addition to luminous X-ray emission from the central binary system, making these systems stellar mass analogues to the AGN population. However, the microquasar population is small and diverse; while there are hundreds of known X-ray binary systems identified in the Milky Way and in the Large and Small Magellanic Clouds (\citealt{liu_catalogue_2006} and \citealt{liu_catalogue_2006}), only ~10s of these have been identified as microquasars. 

The first microquasar discovered, and one of the best studied, is SS433, a HMXB system with resolvable jets \citep{fabian_ss_1979}. SS433 is thought to be a unique object, with persistent jets due to a persistently super-Eddington accretion disc \citep{fabrika_jets_2004}. Several studies have identified evidence for $\gamma$-ray emission from the jets of SS433 (e.g. \citealt{bordas_detection_2015}, \citealt{rasul_gamma-rays_2019}, \citealt{li_gamma-ray_2020}). However, neither the binary nor any components of the jets are catalogued $\gamma$-ray sources in the 4FGL-DR2 due to the fact that this emission does not reach the typical $z = 5 \sigma$ level required for a conventional claim of detection at the catalogue position of SS\,433 over the energy range used in the production of the 4FGL. 

Two\footnote{Although they may also be microquasars, we discuss LS\,5039 and LS\,I+61\,303 as $\gamma$-ray binaries in Section \ref{sec:grbs}; here we discuss the microquasars Cyg\,X-1 and Cyg-X-3 which show emission characteristics different from the $\gamma$-ray binaries \citep{chernyakova_gamma-ray_2020}.} microquasars are listed in the 4FGL-DR2: Cyg X-3 \citep{abdo_fermi_2009} and Cyg X-1 \citep{bodaghee_gamma-ray_2013}. In these microquasars, particle acceleration occurs down the jet, resulting in a non-thermal electromagnetic (EM) emission component which is seen from radio through to $\gamma$-ray wavelengths (\citealt{orellana_leptonic_2007}, \citealt{araudo_high-energy_2009}). Unlike SS433, neither Cyg~X-1 or Cyg~X-3 has persistent jets, and $\gamma$-ray emission is seen only when a jet is present or ejections take place from the system. \cite{bodaghee_gamma-ray_2013} report that for Cyg X-3 (the more significant $\gamma$-ray emitter of the two) these $\gamma$-ray emitting episodes last for \~10s of days, with intervals of \~100s of days between them, and that there is strong evidence for multi-wavelength correlation between the soft X-ray and radio emission \citep{corbel_giant_2012}, with $\gamma$-ray emission occurring at least when there is recurring radio emission \citep{abdo_fermi_2009}. In the case of Cyg X-1, transient emission is also detected with \textit{Fermi}-LAT on daily timescales, but at a lower statistical significance than Cyg X-3, and occurs during the low-hard X-ray state in the Cyg~X-1 system \citep{zanin_gamma_2016}.

\subsection{$\gamma$-ray binaries}
\label{sec:grbs}
The remaining 6 HMXBs in the 4FGL-DR2 fall into a broad category known as the $\gamma$-ray binaries. These have the peak of their emission in the $\gamma$-ray waveband, compared to the microquasars which generally have the peak of their emission in X-rays \citep{dubus_gamma-ray_2013}, although there is some degeneracy between the two categories. Of these 6 systems, PSR\,B1259-63 (\citealt{aharonian_discovery_2005}, \citealt{abdo_fermi_2010-1}), 1FGL J1018.6-5856 \citep{corbet_1fgl_2011}, LS\,5039 \citep{paredes_discovery_2000}, HESS\,J0632+057 (\citealt{aharonian_discovery_2007}, \citealt{hinton_hess_2009}), and LS\,I\,+61\,303 \citep{lamb_point_1997} are in the Milky Way and LMC P3 is in the Large Magellanic Cloud \citep{corbet_luminous_2016}. All of these systems are also detected by the current generation of ground-based TeV observatories.

Whilst $\gamma$-ray production in microquasars is most likely due to accretion onto the compact object and subsequent particle acceleration in a jet, $\gamma$-ray binaries are distinct in that their $\gamma$-ray emission comes from shocks between the wind of the accretor and the stellar wind of the companion star \citep{dubus_gamma-ray_2015}, or possibly through an accretion-ejection regime if the source is also a microquasar.

Young pulsars, such as PSR\,B1259-63, continually lose kinetic energy in the form of a pulsar wind. For an isolated pulsar, this results in the formation of a pulsar wind nebula (PWN), a cloud of relativistic particles accelerated by the central pulsar. PWN are luminous across the EM spectrum, and produce non-thermal emission through shocks with the interstellar medium (\citealt{amato_theory_2014}, \citealt{amato_theory_2020}). 17 PWN are seen with \textit{Fermi}-LAT and recorded in the 4FGL-DR2, with 14 of these being extended sources. When a rotation-powered pulsar forms a wind in a binary system around a high mass star, this interacts with the dense wind of the companion star, producing shocks between the two winds within the binary system rather than at the extended scales observed in PWN. This leads to orbitally-modulated $\gamma$-ray emission in some systems; for example, the light-curve of PSR\,B1259-63 shows increased $\gamma$-ray emission at periastron (\citealt{aharonian_discovery_2005}, \citealt{chang_gev_2018}). 

Whilst this scenario requires the accreting compact object to be a neutron star rather than a black hole, the existence of a neutron star is confirmed in only PSR\,B1259-63 and possibly LSI\,+61\,303 \citep{torres_magnetar-like_2011}, although there is an ongoing debate regarding the phenomenology of this source and whether an accreting compact object is present (e.g. \citealt{massi_evidence_2020}). This is thought to be because the pulsars in the remaining systems are so deeply embedded within their systems' circumstellar winds that coherent radio pulsations cannot be detected \citep{dubus_gamma-ray_2006}. Nevertheless, this model is favoured for the $\gamma$-ray HMXB systems other than the microquasars. Better knowledge of these systems and an expanded catalogue of $\gamma$-ray binaries are needed to build an improved picture \citep{dubus_gamma-ray_2015}.

\subsection{Surveying the HMXB population}
In this paper we present an independent survey of Galactic high mass X-ray binary systems using the \textit{Fermi} Large Area Telescope and the data from \cite{liu_catalogue_2006}, with the intention of increasing the catalogue of potential $\gamma$-ray emitting X-ray binaries. The catalogue of \cite{liu_catalogue_2006} contains 114 HMXBs, including 4 sources already identified as $\gamma$-ray emitters in the 4FGL (Cyg X-1, Cyg\,X-3, LS\,5039 and LSI\,+61\,303). This leaves 110 HMXBs to be surveyed which are not previously detected with \textit{Fermi}-LAT. We build individual models of the region of interest around each binary system, and use maximum likelihood estimation to fit this model to the data. We use our model to test the hypothesis that there is a $\gamma$-ray point source coincident with the position of a HMXB. If we find it likely that this is the case, then we investigate the source's spectral properties and temporal variability. We also consider the possibility that some weak $\gamma$-ray emitting objects may only be seen sporadically and would not reach the statistical threshold for detection ($5 \sigma$) when integrated over the full mission duration of Fermi-LAT (12.5 years).

\section{\textit{Fermi}-LAT observations and data analysis}

\subsection{Data Reduction and Modelling}
The vast majority of X-ray binary systems are located close to the Galactic plane, which itself is an extremely luminous background source of $\gamma$-rays when observed with \textit{Fermi}-LAT. Modelling the Galactic plane accurately is non-trivial; many extended sources (such as supernova remnants and pulsar wind nebulae) are present on the Galactic plane, in addition to a densely-packed field of point sources. In addition to these, the Galactic plane diffuse background is still poorly understood and is distinctly non-uniform, although the most recent Galactic diffuse model provides a better representation of the background when compared to previous versions \citep{acero_development_2016}. 

We follow the maximum likelihood modelling method of \cite{mattox_likelihood_1996} in order to model the \textit{Fermi}-LAT data in a region of interest (ROI) centered around the position of each HMXB. Although there is considerable overlap between ROIs, for simplicity and clarity we treat each system independently rather than considering multiple HMXBs in the same ROI simultaneously. We use the  \texttt{Fermitools v1.2.23} in conjunction with the Python module \texttt{Fermipy v0.19.0} \citep{wood_fermipy:_2017}. We then follow a standard data reduction chain consisting of photon selection followed by computing instrument exposure and livetime. The selected photons are binned into spatial bins of $0.1 \degree$ width and into 8 energy bins per decade. We then set up our model using the parameters described in Table \ref{tbl:params}, including the most recent point source catalogue, the 10 year 4FGL-DR2. The \texttt{gta.optimize} routine is used to push the parameters of the model closer to their global maxima iteratively, and the \texttt{gta.find\_sources} routine is then used to populate the model with any additional, uncatalogued sources detected that are more than $0.5 \degree$ away from the nearest neighbour. We free the normalisation of all sources within $1 \degree$ of the centre of the ROI, including the isotropic and Galactic background components, and execute a full likelihood fit, using the \texttt{MINUIT} optimiser, until an optimal fit quality of 3 is obtained.

In order to test the accuracy of our model, we generate residual maps of each ROI, which reflect the difference between the model and the data. Certain regions of the Galactic plane are prone to over- or under-fitting, particularly at higher photon energies where statistics are poor. However, in the vast majority of cases the model reflects the data accurately enough for our purposes. We also generate test statistic maps, which indicate the positions and significance of any excess $\gamma$-rays which are not accounted for in our model. 

On completion of this procedure, we have a fully-fitted model centered around the position of each HMXB, with statistical maps to test how accurately the model reflects the LAT data.

\begin{table}
\centering
\begin{tabular}{cc}
\hline \hline
Observation Period (Dates) & 04/08/2008 - 05/02/2021 \\
Observation Period (MET) & 239557417 - 600307205 \\
Observation Period (MJD) & 54682 - 58423 \\
Energy Range (GeV) & 0.1 - 500 \\
Data ROI width & $10\degree$ \\
Model ROI Width & $15\degree$ \\
Zenith Angle & $< 90\degree$ \\
GTI Filter & \texttt{DATA\_QUAL>0 \&\& LAT\_CONFIG==1} \\
Instrument Response  & \texttt{P8R3\_SOURCE\_V2} \\
Isotropic Diffuse Model & \texttt{iso\_P8R3\_SOURCE\_V2\_v1} \\
Galactic Diffuse Model & \texttt{gll\_iem\_v07} \\
Point Source Catalogue & 4FGL-DR2 \\
Extended Source Templates & 8 Year Templates \\
\hline
\end{tabular}
\caption{The parameters used in the likelihood analysis of the regions of interest around the X-ray binary systems in the Liu et al. catalogue.}
\label{tbl:params}
\end{table}

\subsection{Testing for persistent $\gamma$-ray emission}
\label{sec:test_pers}

To assess whether $\gamma$-ray emission is detected from the position of a HMXB, we perform a statistical test of significance. For a likelihood model such as ours we can use a hypothesis test, which provides us with a test statistic (TS) measuring the goodness-of-fit of an alternate hypothesis ($\Theta_{1}$) against a null hypothesis ($\Theta_{2}$). In this case, our alternate hypothesis is that there is a $\gamma$-ray point source present at a particular position in our model, and the null hypothesis is that there is not. The TS is given by Equation \ref{eqn:TS}:

\begin{equation}
    \label{eqn:TS}
    \mathrm{TS} = 2 \ln \frac{L(\Theta_{1})}{L(\Theta_{2})}
\end{equation}

Here $L(\Theta_{1})$ and $L(\Theta_{2})$ are the likelihoods of the two hypotheses. The TS itself is distributed as a $\chi^{2}$ statistic for $k$ statistical degrees of freedom between the two hypotheses (see Wilks' Theorem; \citealt{wilks_large-sample_1938}). As a result, the TS directly translates to a $z$-statistic, a more universally understood measure of statistical significance. 

The \texttt{gta.find\_sources} algorithm from \texttt{Fermipy} generates a TS map of our ROI and then iteratively fits point sources to the most significant peaks of $\gamma$-ray emission which cannot be accounted for by existing model components. The algorithm will fit point sources to the 4 highest TS peaks, and then repeat either 5 times, or until there are no more peaks above a user-defined minimum TS value, which we set to $\mathrm{TS} = 9$, being equivalent to $z = 3 \sigma$. As a measure to avoid source confusion (where $\gamma$-ray emission between two close sources becomes indistinguishable), we define a minimum separation for our algorithm, whereby a source cannot be fitted within $0.5 \degree$ of a higher TS peak. 

Each $\gamma$-ray source added to our model by the \texttt{gta.find\_sources} algorithm has a positional uncertainty, which reflects the systematic uncertainty of the instrument. We define a $\gamma$-ray source as being spatially coincident with the position of a HMXB if the angular offset from the position of the HMXB is less than the $95 \%$ angular positional uncertainty\footnote{The PSF of \textit{Fermi}-LAT is large and energy-dependent and therefore the `position' of a source is simply the point where the origin of the $\gamma$-rays is most likely to be, given the PSF. The angular positional uncertainty is the statistical uncertainty on this likely position.} of the point source, i.e. the HMXB lies within the positional uncertainty of the $\gamma$-ray source. 

In most cases, we do not expect to see a $\gamma$-ray source coincident with the position of a binary through the use of the \texttt{gta.find\_sources} algorithm alone. Whilst many of the HMXBs simply will not produce any detectable $\gamma$-ray emission, 
the use of the \texttt{gta.find\_sources} algorithm has some limitations. In each ROI we are limited to fitting sources to only the 20 highest TS peaks. This excludes any sources that are not necessarily one of these 20, but still have a $\mathrm{TS} >9$ and would otherwise be considered $\gamma$-ray point sources within our survey. Additionally, with a minimum separation of $0.5 \degree$, it is entirely possible for a HMXB to be emitting detectable and distinguishable $\gamma$-ray emission, but to be within this minimum angular separation. This is a particular issue on the Galactic plane, where catalogued point sources are packed closely together, and would exclude many of our HMXBs from the discovery of $\gamma$-ray emission.

To mitigate this, following our full likelihood fit (and after \texttt{gta.find\_sources} has been run), we add a point source at the position of the central HMXB manually in each ROI if there is no coincident source identified by the \texttt{gta.find\_sources} algorithm. This added source has an initial soft power law spectrum with spectral index $\Gamma = 2.0$. We then free all parameters of this added source and the normalisation of the sources within $1 \degree$ of it, and both components of the $\gamma$-ray background. We then execute another maximum likelihood fit which gives a TS for this added source. 

As is conventional in $\gamma$-ray astronomy, we use a TS threshold of 25 ($z=5\sigma$) to claim full detection of a $\gamma$-ray source coincident with the position of a HMXB. We also report $\gamma$-ray fluxes for $\gamma$-ray excesses in the $9 \leq \mathrm{TS} < 25$ ($3 \sigma \leq z < 5 \sigma$) range as, while these do not meet the conventional significance for detection and often lack the photon statistics for further analysis, they may be worthy of further study. For those sources (either detected using the \texttt{gta.find\_sources} algorithm, or added later) which exceed the $3 \sigma$ threshold, we carry out further investigation detailed in Section \ref{sec:asa}, in order to explore whether this $\gamma$-ray emission is likely to originate from the binary.

\subsection{Testing for transient/variable $\gamma$-ray emission}

 Three of the eight HMXB systems catalogued in the 4FGL-DR2 have a variability index $> 18.48$, indicating a less than $1\%$ chance of their being steady $\gamma$-ray sources \citep{abdollahi_fermi_2020}, on monthly timescales. These 3 variable sources are: PSR B1259-63 (4FGL\,J1302.9-6349), Cygnus X-3 (4FGL\,J2032.6+4053) and LSI +61 303 (4FGL\,J0240.5+6113) \citep{ballet_fermi_2020}. In all cases, their $\gamma$-ray emission correlates with multi-wavelength emission, firmly identifying these 4FGL sources as the $\gamma$-ray counterparts of the HMXBs. Additionally, possible orbital modulation in SS\,433 (\citealt{li_gamma-ray_2020} and \citealt{rasul_gamma-rays_2019}), the intermittent nature of the $\gamma$-ray emission from Cyg\,X-1 correlating with the low-hard X-ray state \citep{bodaghee_gamma-ray_2013}, and periodic variability from LS 5039 (\citealt{abdo_fermilat_2009}, \citealt{hadasch_long-term_2012} and \citealt{yoneda_sign_2020}) suggest that the HMXB population is generally variable at $\gamma$-ray wavelengths irrespective of emission mechanisms, although short-term variability may not be detectable with \textit{Fermi}-LAT due to sensitivity limitations. 
                                   
In order to check for transient or variable $\gamma$-ray emission from the positions of the HMXBs, we construct a light-curve at their positions using either the coincident source found by the \texttt{gta.find\_sources} algorithm or the source we add afterwards. We use approximately 6 month time bins (25 bins over our observation period) in all cases. To produce the light-curves we use the \texttt{gta.lightcurve} algorithm which carries out a full likelihood fit of the ROI in each time bin to calculate an integrated flux and TS value for the source on a bin-by-bin basis. For each bin we report an energy flux value if $\mathrm{TS} \geq 4$, or a $95\%$ confidence limit otherwise.

In order to see whether a light-curve shows evidence for emission in any bin(s), we employ a mathematical `light-curve condition' which, when satisfied, indicates that there are bins in a light-curve with significant $\gamma$-ray flux. If the condition is satisfied, we can then examine these bins to see whether $\gamma$-ray emission is constant (non-variable), the flux values vary (variable), or a $\gamma$-ray flux appears only during a for a certain time interval (transient). The likelihood fitting used in \texttt{gta.lightcurve} results in a TS value for each bin giving us a measure of how significant the $\gamma$-ray flux is in this bin. Using Wilks' Theorem, we are then able to calculate a p-value for each bin, which gives the probability that each bin arises by chance. We then take all of the p-values for each bin with $\mathrm{TS} > 4$, and use Equation \ref{eqn:lcp} below to calculate a p-value for all the significant ($>2 \sigma$) bins: 

\begin{equation}
    \label{eqn:lcp}
    p_{lc} = 1 \times \prod_{i = 1}^{n} p_{i},
\end{equation}

where $n$ is the number of bins with $\mathrm{TS} > 4$, and $p_{i}$ is the p-value of each bin. We exclude bins with $\mathrm{TS} < 4$, because these have insufficient statistics to provide a reliable flux value and are therefore generally less useful for trying to understand the properties of a source. This is particularly true on the Galactic plane (where most of the HMXB population lies) where the luminous diffuse $\gamma$-ray emission and crowded field mean that bins with $\mathrm{TS} < 4$ are likely to be noise-dominated.

We consider a source to have satisfied the light-curve condition and to have evidence for significant emission in its light curve if $p_{lc} < 5 \times 10^{-7}$, the p-value for $z = 5 \sigma$. This is consistent with our threshold for testing for persistent $\gamma$-ray emission, as described in Section \ref{sec:test_pers}. It is important to note that the light-curve condition does not provide information on the nature of any transient emission or variability, but only that significant $\gamma$-ray emission is present in the light-curve itself.

\onecolumn
\begin{table}
\centering
\begin{tabular}{ccccccc}
\hline \hline
Binary name & TS & $z$-Score & Energy Flux ($\mathrm{MeV}\,\mathrm{cm}^{-2} \, \mathrm{s}^-1$) & LC Condition & Section & Radio\\
\hline
SAX J1324.4-6200 & 12.8 & $3.6 \sigma$ & $2.98 \times 10^{-6}$ & Yes & \ref{bin:SAX13} & No\\
1H 0749-600 & 14.4 & $3.8 \sigma$ & $7.23 \times 10^{-7} $ & No & \ref{bin:1H07} & No \\
1H 1238-599 & 10.6 & $3.3 \sigma$ &$1.70 \times 10^{-6}$ & Yes & \ref{bin:1H12} & No \\
GRO J1008-57 & 24.3 & $4.9 \sigma$ & $3.24 \times 10^{-6}$ & Yes & \ref{bin:GRO10} & Yes  \\
IGR J16320-4751 & 31.1 & $5.6 \sigma$ & $8.31 \times 10^{-6}$ & Yes & \ref{bin:IGR1632} & No \\
IGR J16358-4726 & 9.5 & $3.1 \sigma$ & $4.86 \times 10^{-6}$ & Yes & \ref{bin:IGR1635} & No \\
IGR J16465-4507 & 50.8 & $7.1 \sigma$ & $9.71 \times 10^{-6}$ & Yes & \ref{bin:IGR1646} & No\\
1WGA J0648.0-4419 & 18.5 & $4.3 \sigma$ & $8.64 \times 10^{-7}$ & No & \ref{bin:1WGA06} & No\\
AX J1740.1-2847 & 7.2 & $2.7 \sigma$ & N.A. & Yes & \ref{bin:AX17} & No \\
IGR J17544-2619 & 19.7 & $4.4 \sigma$ & $4.33 \times 10^{-6}$ & No & \ref{bin:IGR17} & No\\
H 1833-076 & 29.2 & $5.4 \sigma$ & $6.74 \times 10^{-6}$ & Yes & \ref{bin:H18} & No\\
GS 1839-04 & 17.8 & $4.2 \sigma$ & $4.76 \times 10^{-6}$ & Yes & \ref{bin:GS18} & No\\
IGR J19140+0951 & N.A. & N.A & N.A. & Yes & \ref{bin:IGR19} & No\\
1A 0535+262 & 12.4 & $3.5 \sigma$ & $1.45 \times 10^{-6}$ & Yes & \ref{bin:1A05} & Yes\\
GRO J2058+42 & 16.4 & $4.0 \sigma$ & $2.44 \times 10^{-6}$ & No & \ref{bin:GRO20} & No\\
W63 X-1 & 13.1 & $3.6 \sigma$ & $1.51 \times 10^{-6}$ & Yes & \ref{bin:W63} & No\\
SAX J2103.5+4545 & N.A. & N.A. & N.A. & Yes & \ref{bin:SAX21} & No\\
RX J2030.5+4751 & 30.8 & $5.5 \sigma$ & $2.10 \times 10^{-6}$ & Yes & \ref{bin:RX20} & No\\
4U 2206+543 & 30.5 & $5.5 \sigma$ & $1.71 \times 10^{-6}$ & Yes & \ref{bin:4U22} & No\\
IGR J00370+6122 & 7.3 & $2.7 \sigma$ & N.A. & Yes & \ref{bin:IGR00} & No\\

\hline
\end{tabular}
\caption{All high mass X-ray binaries with a spatially-coincident $\gamma$-ray source (of $\mathrm{TS} \geq 9$), and/or the light-curve (LC) condition is fulfilled, along with the calculated TS, $z$-score and energy flux value (units of $\mathrm{MeV}\,\mathrm{cm}^{-2} \, \mathrm{s}^-1$) integrated across the analysis energy range (100 MeV - 500 GeV). Those sources which show evidence for \textit{only} intermittent $\gamma$-ray emission (i.e. the LC condition is fulfilled) and have TS < 9, or where no source is fitted over the full observation period, do not have values of energy flux listed here. The Radio column indicates whether there is a measured radio flux for the associated HMXB in the SIMBAD astronomical database. In the case of these HMXBs we calculate 95\% confidence upper limits on energy flux, which are given in Appendix A. The `section' column gives the subsection of the paper in which more detailed discussion of each ROI is given. As in the Liu et al. catalogue, HMXBs are ordered by declination, from south to north. This convention is used throughout this paper.}
\label{tbl:pers_res}
\end{table}

\begin{table}
\centering
\begin{tabular}{ccccc}
\hline \hline
Binary name & TS & $z$-Score & 4FGL $z$-Score & Energy Flux ($\mathrm{MeV}\,\mathrm{cm}^{-2} \, \mathrm{s}^-1$)\\
\hline
LS 5039 & 18000 & $130 \sigma$ & $62 \sigma$ & $2.14 \times 10^{-4}$\\
Cyg X-1 & 88 & $9.4 \sigma$ & $8.6 \sigma$ & $3.49 \times 10^{-6}$\\
Cyg X-3 & 860 & $29 \sigma$ & $11 \sigma$ & $2.42 \times 10^{-5}$\\
LS I +61 303 & 170000 & $410 \sigma$ & $250 \sigma$ & $4.67 \times 10^{-4}$\\

\hline
\end{tabular}
\caption{The four 4FGL sources included in the HMXB catalogue, together with the TS value and corresponding $z$-score from our analysis over the full 12.5 years of LAT data. For comparison, we include the detection $z$-score for each source provided in the 4FGL-DR2. It should be noted when comparing the $z$-score values that the photon selection for the computation of the 4FGL parameters is different (100~MeV - 1~TeV compared with 100~MeV - 500~GeV for our analysis), and the observation time is lower (10 years vs. 12.5). Our analysis methodology is also different.}
\label{tbl:known_binaries}
\end{table}

\twocolumn

\section{Survey Results}
We consider any ROI from our modelling which has a source coincident with the position of the binary and $\mathrm{TS} \geq 25$ to show significant evidence for $\gamma$-ray emission from the binary's position. 
These sources are significant enough (and have enough photon statistics) for spectral analysis and $\gamma$-ray localisation using the \texttt{gta.localize} algorithm. We describe the results on a source-by-source basis in Section \ref{sec:asa}, and discuss whether the $\gamma$-ray emission detected in each ROI is likely to be from the spatially-coincident HMXB. In total, we detect 5 significant new $\gamma$-ray sources coincident with the positions of HMXBs with $\mathrm{TS} \geq 25$ in the 13-year dataset. These are listed with their TS values and integrated energy fluxes in Table \ref{tbl:pers_res}.

For ROIs where a $\gamma$-ray excess coincident with the position of the binary lies in the $9 \leq \mathrm{TS} < 25$ range it is often impossible to carry out meaningful spectral analysis or localisation. Additionally, these sources lie below the threshold for a conventional detection, and therefore ascertaining the presence of $\gamma$-ray emission from such systems is challenging. Nonetheless, temporal qualities such as flares or phased emission can still be used to associate a sub-threshold $\gamma$-ray excess with an X-ray binary.
We find a total of 11 $\gamma$-ray sources in the $9 \leq \mathrm{TS} < 25$ range. These are also listed with their TS and integrated energy flux in Table \ref{tbl:pers_res}.

For ROIs from which no significant $\gamma$-ray flux ($\mathrm{TS} < 9$) is detected, we report a $95\%$ confidence upper limit on flux in Table \ref{tbl:sample}. 

Several X-ray binaries with known $\gamma$-ray emission are also included in the \cite{liu_catalogue_2006} HMXB catalogue. These are the microquasars Cyg~X-1, Cyg~X-3 and SS433, and the $\gamma$-ray binaries LS~5039 and LS~I~+61 303. With the exception of SS433, we detect all of these sources with a $z$-score within a factor of 3 of their 4FGL-DR2 values. These values are shown in Table \ref{tbl:known_binaries}. That we do not detect SS433 in our analysis in contrast to previous studies is unsurprising as \cite{li_gamma-ray_2020}, the most recent study of the system, used a phased analysis in order to resolve the extended emission of SS 433 from the highly luminous, nearby pulsar PSR J1907+0602, which we do not use here. Previous studies, including \cite{rasul_gamma-rays_2019} use different background models, and a different catalogue without the inclusion of maximum likelihood weighting, which makes a direct comparison difficult. Finally, the position of the $\gamma$-ray emission from SS\,433 appears to correspond to the jet termination lobe, itself offset from the central position of the binary which we analyse \citep{rasul_gamma-rays_2019}.

We find 16 HMXB systems where the light-curve condition is met. The majority of these (12 systems) also have persistent $\gamma$-ray emission with $\mathrm{TS} \geq 9$, including all of the HMXBs with persistent $\mathrm{TS} \geq 25$ emission. We also find significant ($z>2 \sigma$) bins in the light-curves of two additional sources with $\mathrm{TS} < 9$, and two where no fit is found for a point source at the position of the HMXB. Despite the lack of a persistent $\gamma$-ray excess coincident with these binaries, multi-wavelength data can be used to identify features (such as flares) across different wavebands coincident with the apparent $\gamma$-ray emission. To this end, we use light-curve data (where available) for these 16 systems in the X-ray waveband from the \textit{Swift} Burst Alert Telescope (BAT) which operates in the 15 - 50 keV range and the Monitor of All Sky X-ray Image (MAXI), which operates in the 2 - 20 keV energy range. We also use any available V-band optical photometry data from the American Association of Variable Star Observers (AAVSO). These sources, along with those which show persistent $\gamma$-ray emission, are considered in Section \ref{sec:asa} and Appendix \ref{appendix:false_positive}.

\section{Discussion}
\label{sec:asa}

\subsection{SAX\,J1324.4-6200}
\label{bin:SAX13}
\begin{figure}
    \centering
    \includegraphics[width=240pt]{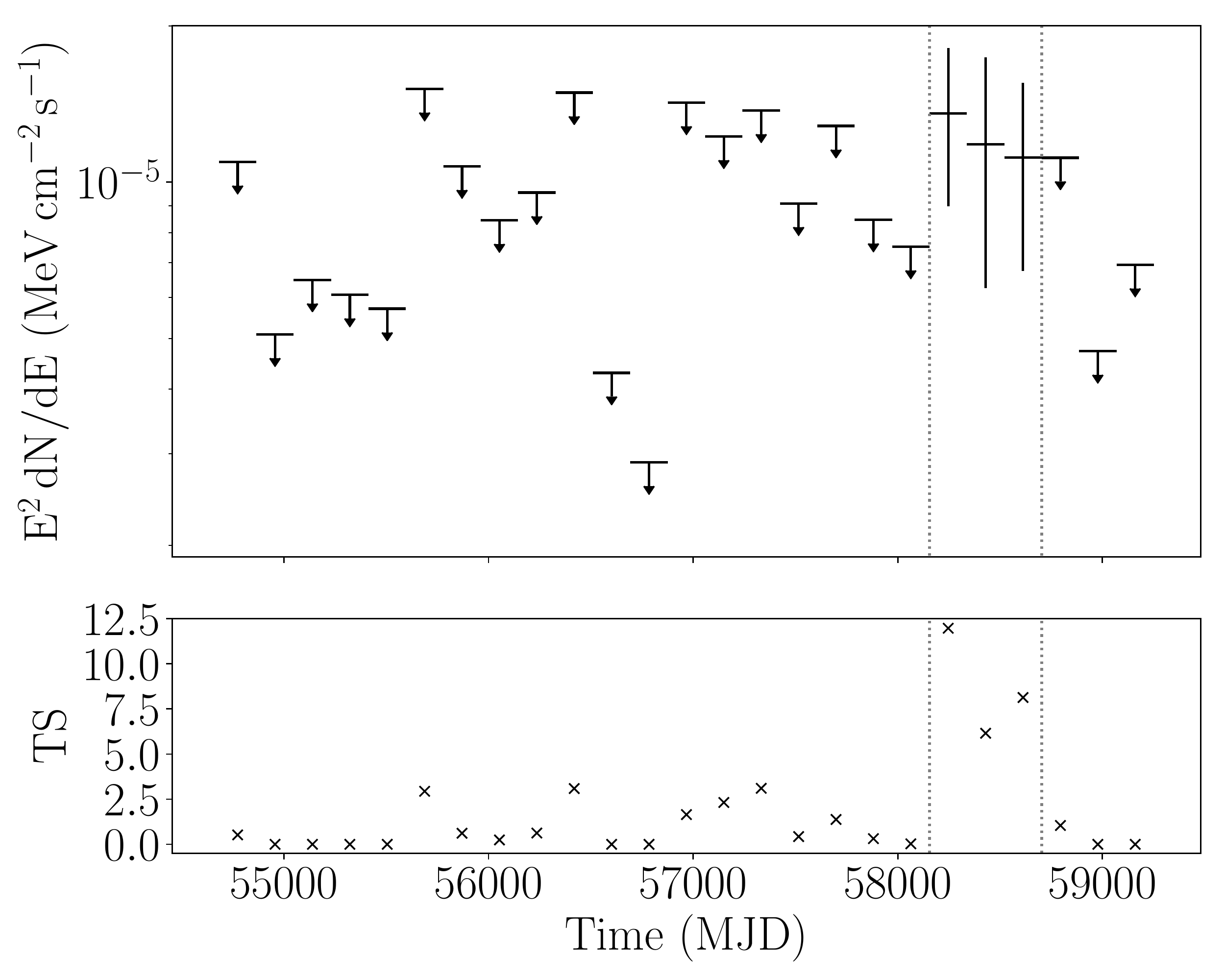}
    \caption{The \textit{Fermi}-LAT 6-monthly binned light-curve of SAX\,J1324.4-6200, showing the energy flux in the top panel, and the corresponding TS of each bin in the bottom panel. For bins with $\mathrm{TS} < 4$ we calculate a $95 \%$ confidence upper limit on flux. The vertical dotted lines indicate the beginning and the end of the $\gamma$-ray excess.}
    \label{fig:sax13_lc}
\end{figure}

SAX\,J1324-6200 (henceforth SAX13) is an X-ray pulsar, thought to be an accreting high-mass neutron star in orbit with a Be star (\citealt{angelini_discovery_1998}, \citealt{mereghetti_swift_2008}, \citealt{kaur_chandra_2009}). No orbital period is known for this system. We report a persistent TS of 12.8 over the full 12.5 year dataset; however, there is some evidence for sustained $\gamma$-ray emission from the position of SAX13 over an 18 month period throughout 2018 and 2019 (MJD 57972 - 58520), at the $2 \sigma$ to $3 \sigma$ level, suggesting this emission is likely transient. The $\gamma$-ray light-curve for this source is shown in Figure \ref{fig:sax13_lc}. There is no \textit{Swift}-BAT or MAXI light-curve for SAX13, nor are there any optical photometry measurements in the AAVSO database for the time period in question.

There are several catalogued sources near SAX13. The closest of these are 4FGL\,J1328.4-6231 ($\mathrm{TS} = 69.7$ at an angular offset from SAX13 of $0.690 \degree$), 4FGL\,J1321.1-6239 ($\mathrm{TS} = 116$, offset: $0.749 \degree$), 4FGL\,J1320.5-6256c ($\mathrm{TS} = 22.4$, offset: $1.033 \degree$) and 4FGL\,J1329.9-6108 ($\mathrm{TS} = 545$, offset: $1.093 \degree$). Additionally, we add a new source to our model with the \texttt{gta.find\_sources} algorithm, PS J1317.8-6157 ($\mathrm{TS} = 20.0$, offset: $0.779 \degree$). None of these have a 4FGL variability index high enough to indicate variability on monthly timescales. We generate light-curves for each of these sources (with the exception of the faint 4FGL\,J1320.5-6256c and PS J1317.8-6157 because these sources are sufficiently faint and have a large enough angular distance from SAX13 that we can be confident that they are not causing source confusion at the position of SAX13) using identical binning to the SAX13 light-curve, and do not see any significant enhancement in these light-curves at the time of the 18 month apparent SAX13 $\gamma$-ray excess, meaning that it is likely that this excess is independent of these sources. 

\begin{figure}
    \centering
    \includegraphics[width=240pt]{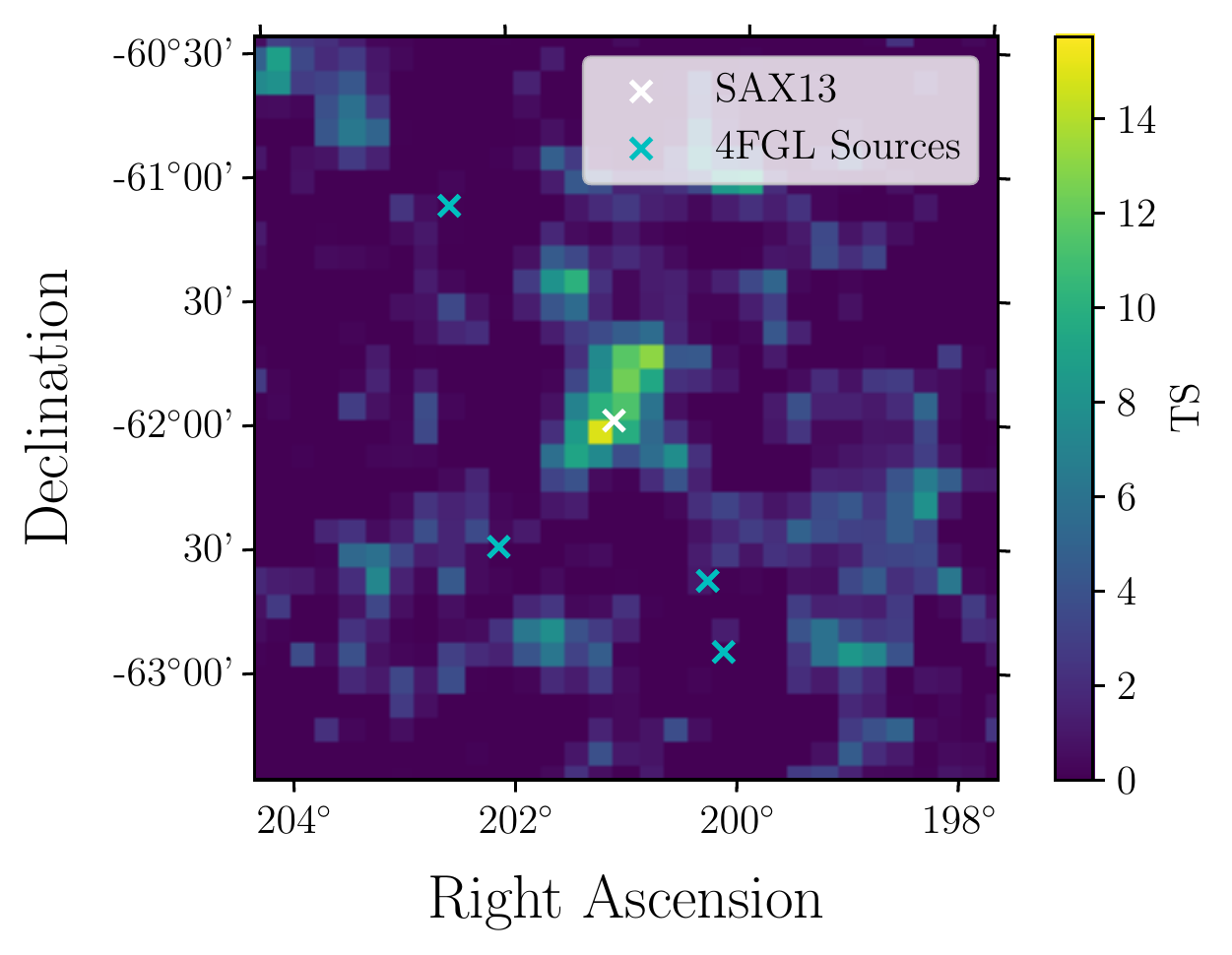}
    \caption{The TS map of the central $3\degree$ of the SAX\,J1324.4-6200 ROI during the 18 month period during which the $\gamma$-ray excess is observed. Here, the positions of the closest 4FGL sources are indicated by blue crosses, whilst the position of SAX\,J1324.4-6200 is indicated by a white cross. This TS map is generated after our ROI optimization and fit, but before a point source for SAX\,J1324.4-6200 is fitted to the model to highlight the spatial coincidence between the excess and the position of SAX\,J1324.4-6200. Bin widths are $0.1 \degree$. }
    \label{fig:sax13_TS}
\end{figure}

Considering only the photons detected within the 18 month excess we carry out an independent analysis of the same ROI over this 18 month period (using the same parameters, other than observation time, as in Table \ref{tbl:params}). We generate a TS map (Figure \ref{fig:sax13_TS}) of the centre of the ROI, and find that the peak of this excess is approximately spatially coincident with the position of SAX13. Fitting a power-law point source to the position of SAX13, we free this source and those within $1 \degree$ of the central position of SAX13 and execute a likelihood fit. We then run the \texttt{gta.localize} algorithm on the added SAX13 source, and find that the optimal position of the added source is $\mathrm{LII} = 306.8362 \degree  \pm 0.0699 \degree $, $\mathrm{BII} = 0.5534 \degree  \pm 0.0826 \degree$, offset from SAX13 by $0.0707 \degree$. Considering that this offset is less than the $95\%$ and $68\%$ containment radii of the added source ($0.1859 \degree$ and $0.1152 \degree$ respectively) this excess can be regarded as spatially coincident with the location of SAX13. With the point source at its optimal position, we calculate a TS of 28.7 over this 18 month period. Over the same period, there is no significant detection of 4FGL\,J1256.1-5919 or 4FGL\,J1320.5-6256c. 4FGL\,J1321.1-6239 is detected with a TS of 24.9 and 4FGL\,J1329.9-6108 with a TS of 49.4, making the SAX13 source comparable to these objects in terms of statistical significance during this time. 

\begin{figure}
    \centering
    \includegraphics[width=240pt]{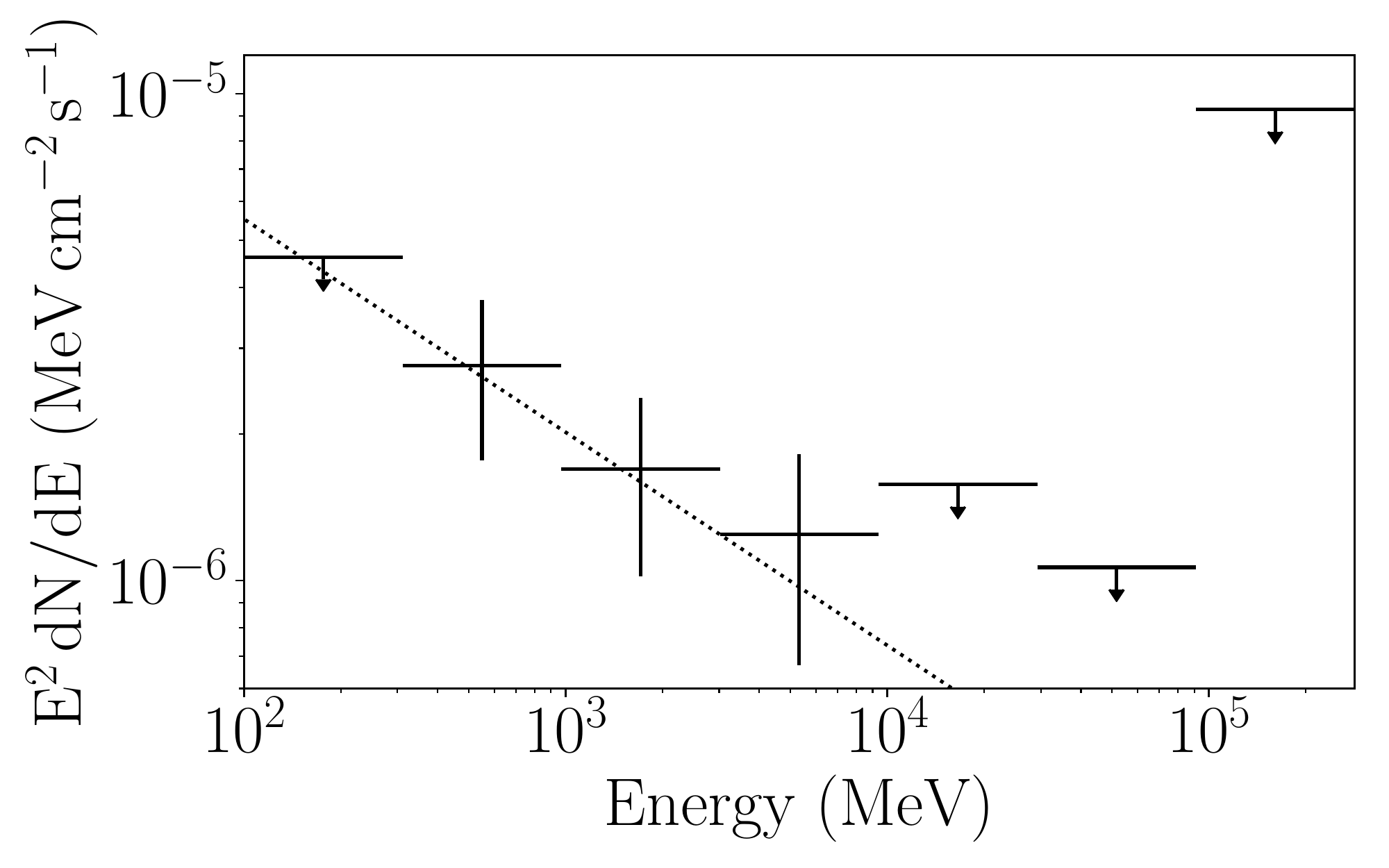}
    \caption{The Spectral Energy Distribution of the SAX\,J1324.4-6200 coincident source during the 18 month excess, with our power law fit indicated by the black dotted line. We place upper limits on any bin with $\mathrm{TS} < 4$. Whilst our baseline analysis uses 8 energy bins per decade, here we use 2 energy bins per decade to ensure sufficient photons for an accurate flux measurement in each bin.}
    \label{fig:sax13_SED}
\end{figure}

Given that the TS of the SAX13 source exceeds 25, we have sufficient photon statistics to carry out a spectral fit, shown in Figure \ref{fig:sax13_SED}. We find the best fit spectrum is a power-law with normalisation $N_{0} = 1.95 \times 10^{-12}$, spectral index $\Gamma = -2.43$ and scale energy $E_{0} = 1000 \, \mathrm{MeV}$. 

It is entirely possible that this $\gamma$-ray emission is associated with the system. The power law spectral fit and calculated spectral index are consistent with those known HMXB systems in the 4FGL with lower detection significances (the more significantly-detected systems are best fit by a log parabola spectral model). These are the Cyg X-1 with $\Gamma = -2.13$ and $z=8.55 \sigma$, HESS J0632+057 with $\Gamma = -2.17$ and $z=4.62 \sigma$ and PSR B1259-63 with $\Gamma = -2.75$ and $z=5.64 \sigma$, compared to SAX13 which we observe to have $\Gamma = -2.43$ and $z = 5.36 \sigma$. Considering that the accretor in this system is known to be a pulsar, and SAX13 is not a known microquasar, if this $\gamma$-ray emission does indeed come from SAX13 then it is likely to be from a pulsar wind interaction as seen in the $\gamma$-ray binary population. 
However, without any corroborating multi-wavelength data, it is difficult to be certain that this is indeed from SAX13 and not another undetected source nearby. Additionally, if this $\gamma$-ray emission is indeed from a pulsar wind interaction, the fact that we see only one emission episode during the lifetime of \textit{Fermi}-LAT suggests that the period of the system may be so long that it would be difficult to obtain the frequency of these interactions. 

\begin{figure}
    \centering
    \includegraphics[width=240pt]{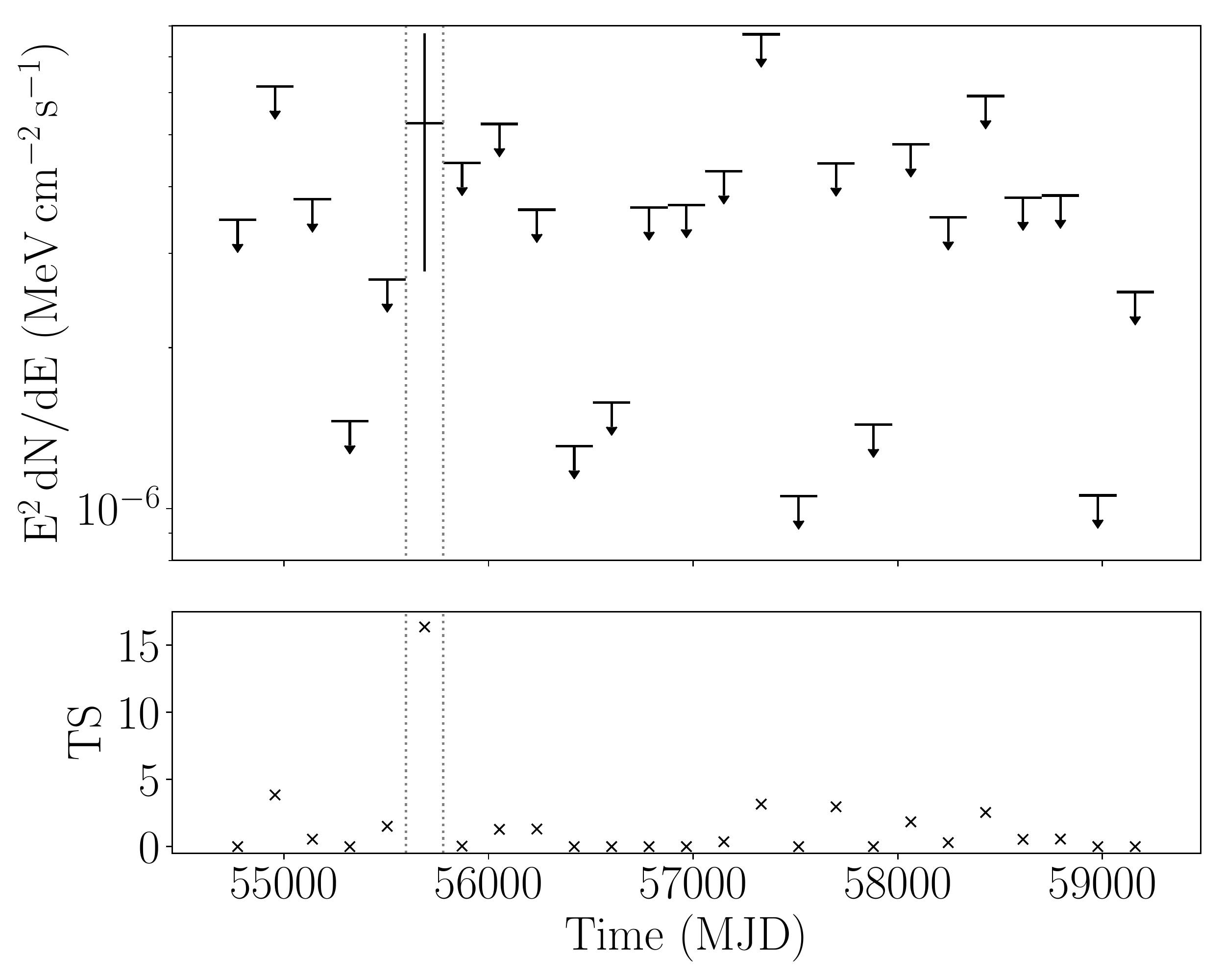}
    \caption{The \textit{Fermi}-LAT light-curve of the $\gamma$-ray excess coincident with 1H\,0749-600 shown in 6-month time bins. Upper limits are placed on any time bin where the TS of that bin is less than 4. The vertical dotted lines indicate the beginning and end of the apparent $\gamma$-ray excess.}
    \label{fig:1H07_lc}
\end{figure}

\subsection{1H\,0749-600}
\label{bin:1H07}
1H\,0749-600 (henceforth 1H07) is a HMXB with an unknown accretor and a Be companion star \citep{apparao_x-ray_1994}. Over the 12.5 year observation period we find a spatially-coincident excess with a TS of 14.4, which we assign the name PS~J0750.5-6116, and with a slight angular offset from the position of 1H07 of $0.170 \degree$. The nearest 4FGL sources are 4FGL~J0807.0-6102, which is the blazar PMN J0806-6101 ($\mathrm{TS} = 69.7$ at an angular offset from 1H07 of $2.028 \degree$) and 4FGL J0756.3-6431, which is the BL Lac blazar SUMSS J075625-643031 ($\mathrm{TS} = 52.53$, offset $3.489 \degree$). As neither of these blazars is particularly luminous, it is unlikely that source confusion is the cause of the persistent $\gamma$-ray excess PS~J0750.5-6116. 

We calculate the light-curve of PS~J0750.5-6116, shown in Figure \ref{fig:1H07_lc}, and find a $\gamma$-ray excess in one bin with $\mathrm{TS} = 16.4$ ($4.05 \sigma$) (MJD 55597 - 55779), and upper limits in the other 24 bins. As blazars are a generally variable class of $\gamma$-ray emitters, we produce light-curves of the two nearest 4FGL sources, but see no enhancement in the same bin as the PS J0750.5-6116 excess. This suggests this $4 \sigma$ time bin is independent of the $\gamma$-ray emission of the blazars.  It is likely that this single 6-month period is responsible for the majority of the $\gamma$-ray emission observed from the position of PS~J0750.5-6116. A re-analysis of this 6-month period results in a slightly increased TS of 17.2, having used the \texttt{gta.localize} algorithm to obtain a best fit position of LII $ = 273.8571 \degree \, \pm \, 0.0816 \degree$, BII $ = -16.8787 \degree \,\pm\, 0.0688\degree$. This gives an angular offset of $0.1581 \degree$ from the (IR) position of 1H07. Given that the $95\%$ positional uncertainty of PS\,J0750.5-6116 is $0.1814 \degree$, this is still spatially coincident with 1H07. 

This low-significance excess cannot be firmly associated with 1H07. Furthermore, as the nature of the accretor in this system is unknown, and no microquasar-like behaviour or pulsations have been observed, the physical mechanisms behind any $\gamma$-ray emission from this system are unclear. No orbital period is known for this system, so examining the $\gamma$-ray emission by orbital phase is not possible. We conclude that whilst it is entirely possible that PS~J0750.5-6116 represents faint $\gamma$-ray emission from 1H07, a lack of information makes a firm detection claim impossible.

\subsection{1H\,1238-599}
\label{bin:1H12}
\begin{figure}
    \centering
    \includegraphics[width=240pt]{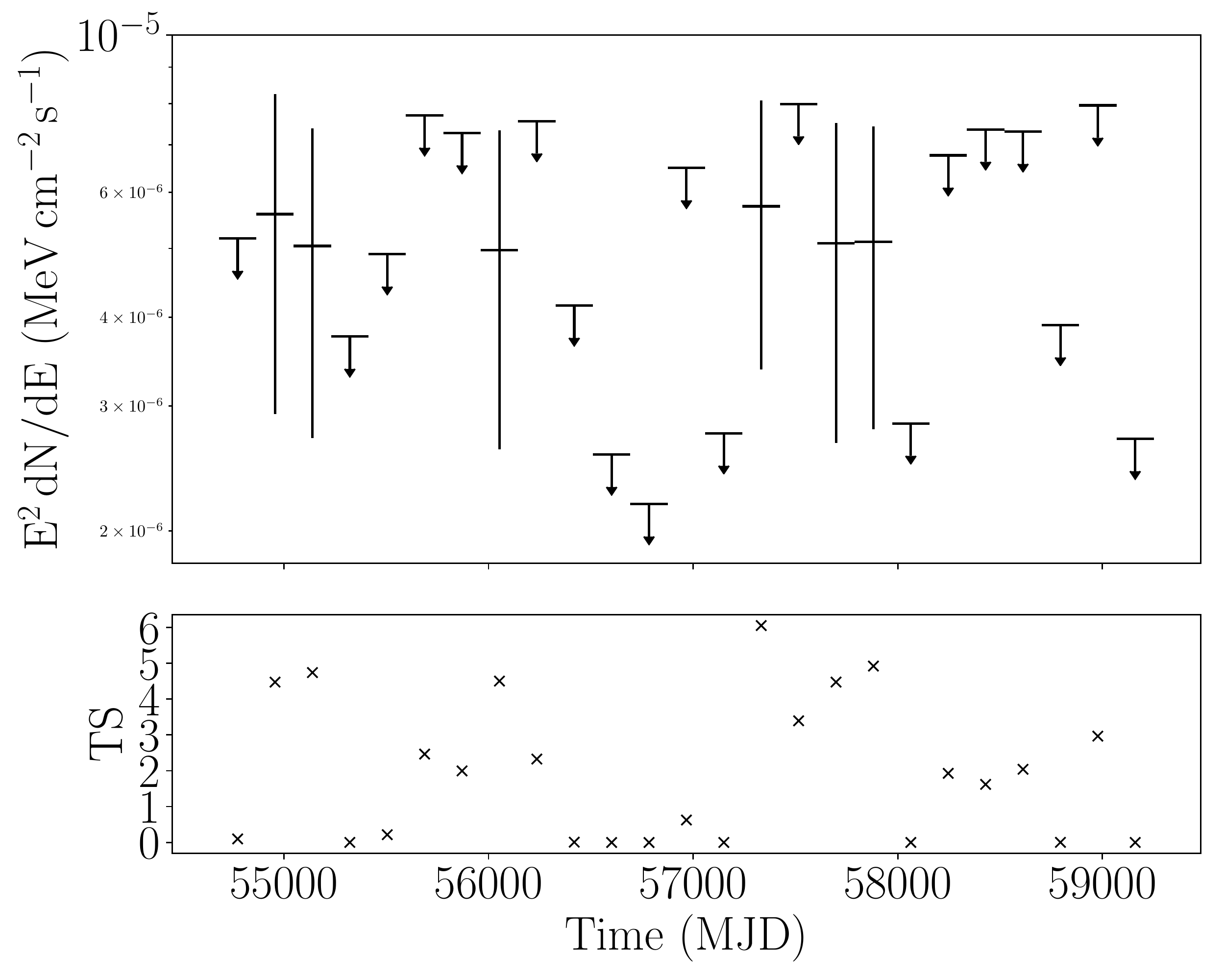}
    \caption{The \textit{Fermi}-LAT light-curve of the $\gamma$-ray excess coincident with 1H\,1238-599 with time bins of 6 month width. Upper limits are placed on any time bin where the TS of that bin is less than 4.}
    \label{fig:1H12_lc}
\end{figure}

1H\,1238-599 (henceforth 1H12) is an X-ray pulsar HMXB system \citep{huckle_discovery_1977}. Over the full 12.5 year dataset, the TS is 10.6, with the light-curve (Figure \ref{fig:1H12_lc}) showing borderline significance (approximately $2 \sigma$) $\gamma$-ray excesses across 6 of the 25 bins (MJD 54865 - 55231, MJD 55962 - 56145, MJD 57241 - 57425, and MJD 57607 - 57973). The nearest catalogued sources are 4FGL\,J1256.1-5919 ($\mathrm{TS} = 174$ and an angular offset from 1H12 of $1.983 \degree$), which is the blazar PMN J1256-5919, 4FGL\,J1244.3-6233 ($\mathrm{TS} = 428$, offset: $2.370 \degree$) and 4FGL\,J1253.3-5816 ($\mathrm{TS} = 48.1$, offset: $2.404 \degree$), which is the pulsar PSR J1253-5820. None of these sources has a catalogue variability index which would indicate variability on monthly timescales. As these sources are at some distance from the position of 1H12, it is unlikely that any $\gamma$-ray signal from the position of 1H12 is due to source confusion with a 4FGL source. Similarly, no uncatalogued sources of $\gamma$-rays are detected close to 1H12 by \texttt{gta.find\_sources}, the closest such source being approximately $3 \degree$ away. 

Given that the bins in which we measure an apparent flux are all at the $2\sigma$ level, it is difficult to perform any detailed tests of emission (for example, localisation) at the position of 1H12. As multi-wavelength data for this HMXB are not available, we cannot cross correlate the $\gamma$-ray light-curve with other wavelengths. Finally, an orbital period for this binary has not been measured, therefore we cannot correlate the light-curve with the system period. As a result, while it is possible that the light-curve of 1H12 is showing a very faint signal that could be from the binary, we could equally be measuring fluctuations in the Galactic diffuse background. 

\subsection{GRO\,J1008-57}
\label{bin:GRO10}
\begin{figure}
    \centering
    \includegraphics[width=240pt]{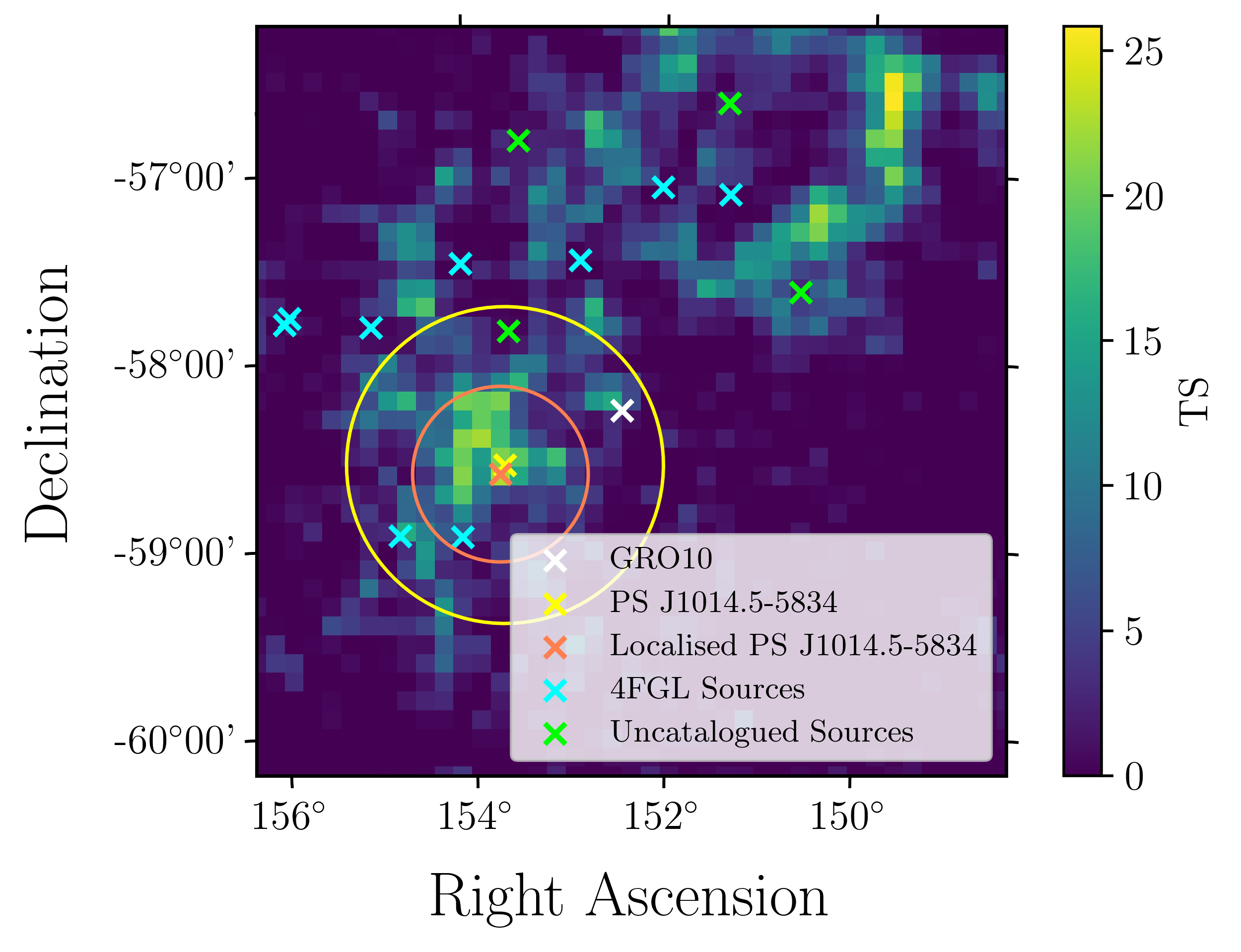}
    \caption{The TS map of the central $4\degree$ of the GRO\,J1008-57 ROI, after our likelihood fit and the \texttt{gta.find\_sources} algorithm, but with the coincident source, PS\,J1014.5-5834, removed from the model to reveal the TS of the coincident $\gamma$-ray emission. Here, the yellow and orange crosses refer to the positions of PS\,J1014.5-5834 both before and after the \texttt{gta.localize} algorithm, which the corresponding circles referring to the $95\%$ positional uncertainty of the source before and after localisation. The white cross indicates the catalogued location of GRO\,J1008-57, the blue crosses indicate the positions of other 4FGL sources, and the green crosses indicate the positions of the other uncatalogued sources added to the model by \texttt{gta.find\_sources}.}
    \label{fig:GRO10_TS}
\end{figure}

GRO J1008-57 (henceforth GRO10) is an X-ray pulsar/Be star HMXB system \citep{petre_gro_1993}, with an orbital period of 135.0 days. Over the 12.5 year observation time, we find a source, PS\,J1014.5-5834, that is spatially coincident with GRO10 with \texttt{gta.find\_sources}. The angular offset between GRO10 and PS\,J1014.5-5834 is less than the $95\%$ positional uncertainty of the source. However, PS\,J1014.5-5834 has an unusually large positional uncertainty of approximately $0.7 \degree$. Running the \texttt{gta.localize} algorithm, we find that, while the position of the source does not change significantly, the positional uncertainty decreases to the extent that this source is no longer coincident with the position of GRO10. Figure \ref{fig:GRO10_TS} shows the position of GRO10 together with the positional uncertainty of PS\,J1014.5-5834 both before and after localisation, and shows a somewhat extended $\gamma$-ray structure around PS\,J1014.5-5834.This is likely the cause of the large positional uncertainty, as \texttt{gta.find\_sources} does not account for spatially-extended $\gamma$-ray structures. 

\begin{figure}
    \centering
    \includegraphics[width=240pt]{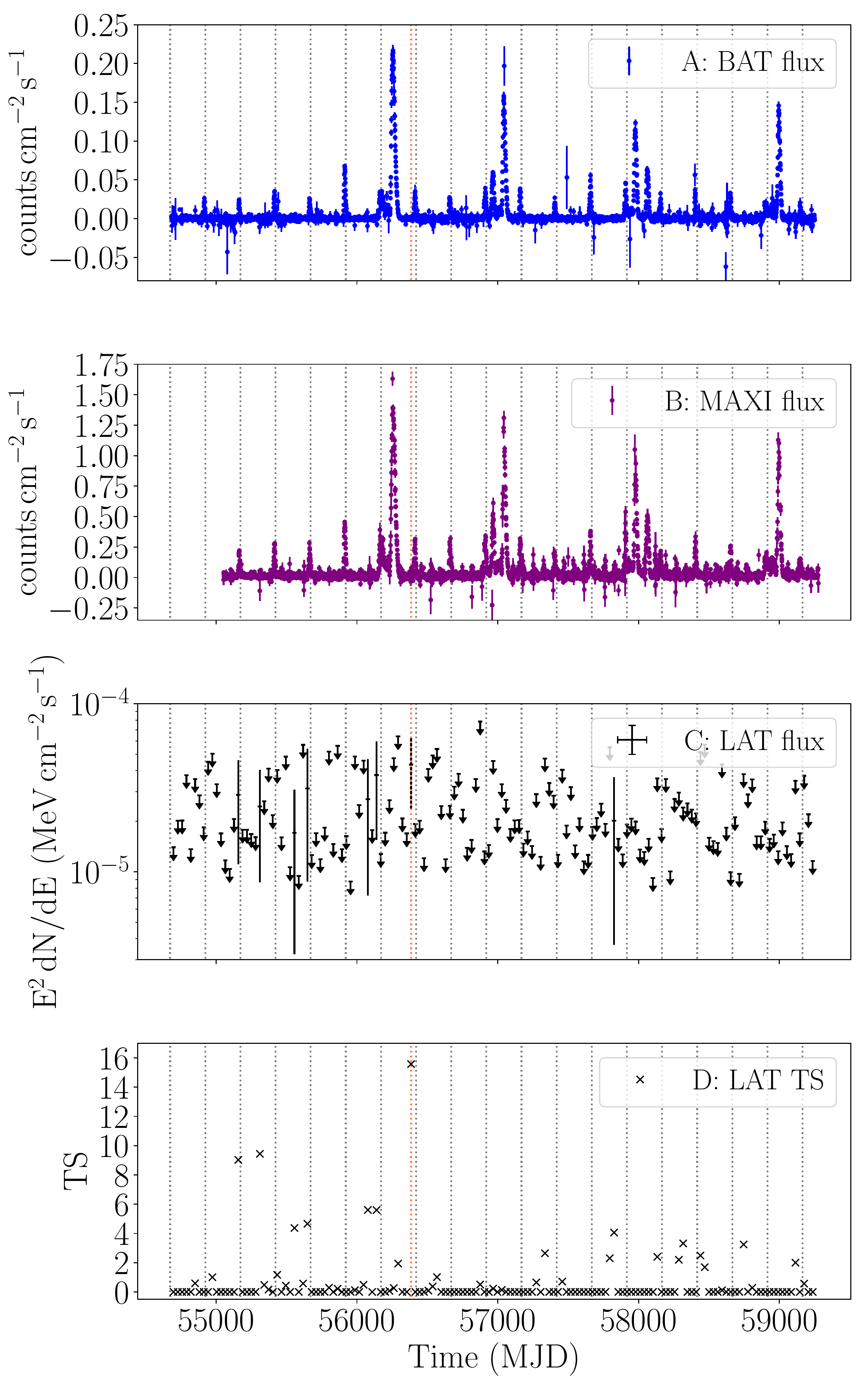}
    \caption{The daily binned light-curve of GRO\,J1008-57 from \textit{Swift}-BAT (Panel A) and MAXI (Panel B). The calculated monthly binned\textit{Fermi}-LAT light-curve for a source fitted to the position of GRO\,J1008-57 is shown below in Panel C, and the corresponding TS values of these bins shown in Panel D. We place 95\% confidence limits on any \textit{Fermi}-LAT energy flux bins with $\mathrm{TS} < 4$. The vertical orange line reflects the centre of the bin with peak $\gamma$-ray emission, and the vertical grey lines mark each periastron passage of GRO\,J1008-57 over the observation time.}
    \label{fig:GRO10_lc}
\end{figure}

\begin{figure}
    \centering
    \includegraphics[width=240pt]{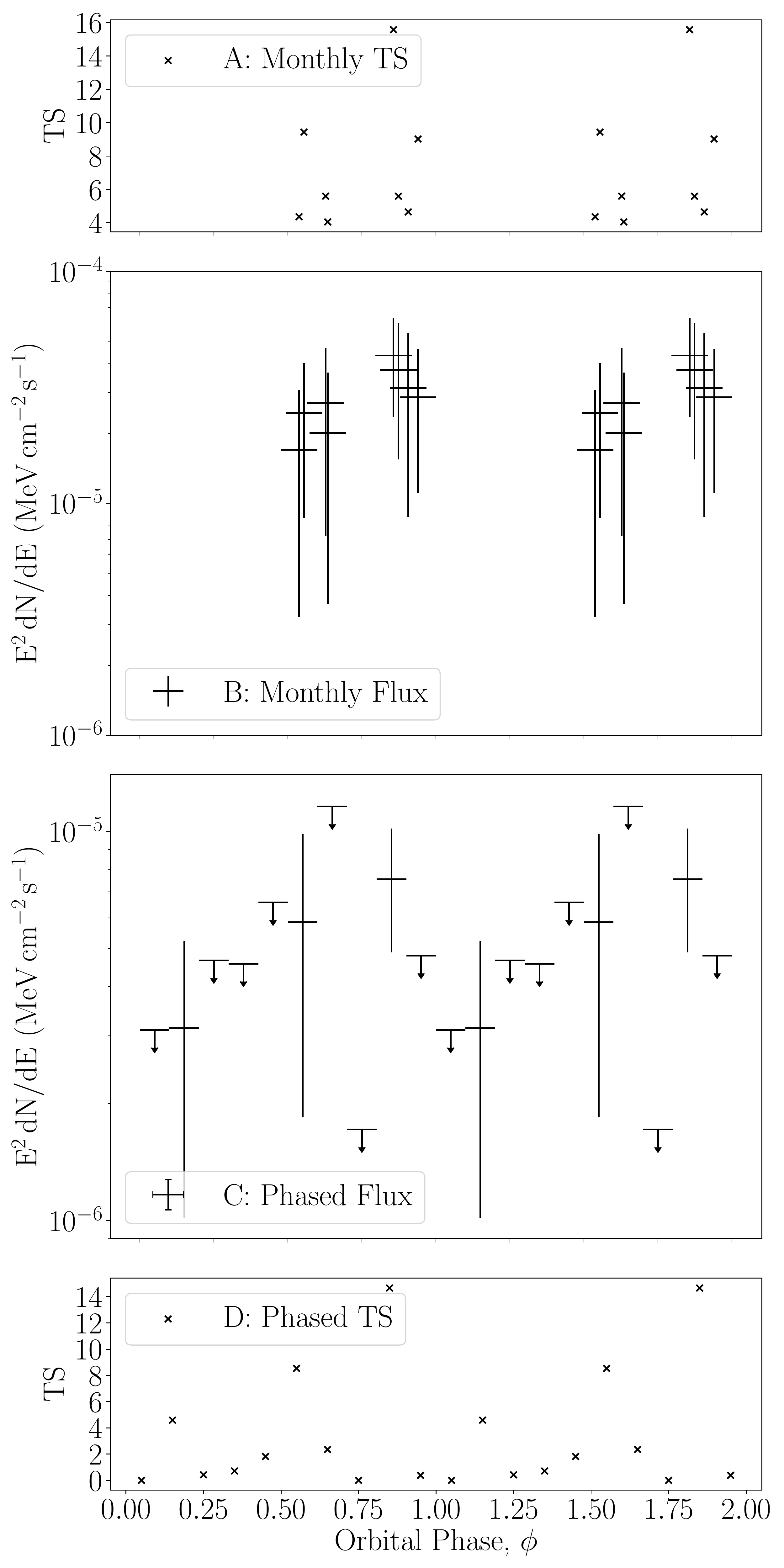}
    \caption{This figure displays the variability of GRO\,J1008-57 based on orbital phase, showing two full orbits. Panels A and B respectively show the TS and $\gamma$-ray flux of each monthly bin from Figure \ref{fig:GRO10_lc}, plotted by orbital phase. We do not include bins from the monthly light-curve where upper limits on flux were calculated. Panels C and D respectively show the $\gamma$-ray flux and TS of the phase folded light-curve of GRO\,J1008-57, with the orbit divided into 10 equal intervals. Here we fix upper limits on flux for any phase period where the $\mathrm{TS} < 4$. On the left vertical axis of the flux plots (B and C) we plot energy flux ($E^{2} \frac{dN}{dE}$). On the horizontal phase axis, phase is defined so that $0, \, 1, \, 2$ refer to periastron. } 
    \label{fig:GRO10_orb}
\end{figure}

GRO10 is a well-studied system, particularly in the X-ray waveband, with semi-predictable X-ray flares occurring at periastron \citep{kuhnel_gro_2013}, the most recent of which was observed in 2020 (\citealt{nakajima_maxigsc_2020} and \citealt{nakajima_maxigsc_2020-1}). \cite{xing_likely_2019} (henceforth \textit{Xing\,19}) studied the possibility of $\gamma$-ray emission using approximately 9 years of \textit{Fermi}-LAT observations but employing the 4 year LAT catalogue and correspondingly older background models. We use the 10 year 4FGL-DR2 with the corresponding Galactic and isotropic diffuse models, enabling more accurate modelling of the region around GRO10 than was possible with the older models and catalogue. \textit{Xing\,19} observed a $\gamma$-ray excess at the position of GRO10 to a TS of 7 ($z = 2.65 \sigma$) by adding a point source to the model, and carrying out a likelihood fit. They then carried out a stacked temporal analysis binned by orbital phase, dividing each orbit into 10 equal time bins and summing the bins from each orbit. Through this method, \textit{Xing\,19} identified 3 excesses, two around the middle of GRO10's orbit with $\mathrm{TS} \approx 5$  ($z \approx 2\sigma$), and one in the penultimate orbital phase bin preceding periastron with $\mathrm{TS} \approx 20$  ($z \approx 4.8\sigma$) Additionally, \textit{Xing\,19} identified 3 excesses by deriving a light-curve for their entire observation time and found $\mathrm{TS} \approx 9$ excesses in the bins centered on MJD 55135 and MJD 55559 and a $TS \approx 17.5$ excess in the bin centered on MJD 56383. Given the significance of these excesses, and the lack of other emission, it is likely these dominate their stacked orbital analysis and are primarily responsible for the TS values seen in the phased light-curve.

As we reject the hypothesis that PS\,J1014.5-5834 represents $\gamma$-ray emission from GRO10, we manually add a point source to our model and fit to it (after we have localised PS\,J1014.5-5834). We find a total TS of 7.9 over the 12.5 year observation time of this source, consistent with the TS of 7 found by \textit{Xing\,19} given the increased observation time used in this work. We take a slightly different approach to \textit{Xing\,19} by using monthly time bins (rather than dividing each orbit into 10 phases), so that there are approximately 8 time bins per orbit, and 149 bins in total, one per month. Figure \ref{fig:GRO10_lc} shows our \textit{Fermi}-LAT light-curve together with the \textit{Swift}-BAT and MAXI light-curves of this source, where four significant outbursts are seen mid-orbit, along with periodic brightening events which correspond to the periastron of GRO10's orbit. With respect to the 3 excesses observed by \textit{Xing\,19}, we identify the MJD 55135 ($\mathrm{TS} = 9.7$) excess to $\mathrm{TS} = 9.0$, the MJD 55559 ($\mathrm{TS} = 9.1$) excess to $\mathrm{TS} = 4.4$, and the MJD 56383 ($\mathrm{TS} = 17.5$) excess to $\mathrm{TS} = 15.6$, although our bins are approximately $20\%$ longer than those of \textit{Xing\,19}, and are not perfectly contemporaneous. In total, we observe one bin with approximately $4 \sigma$ $\gamma$-ray emission, two with approximately $3 \sigma$ emission and five with approximately $2 \sigma$ emission, although it must be stressed that the $2 \sigma$ bins are very marginal, with an approximately $5 \%$ chance that these individually arise by coincidence, and given the 25 bins present in the light-curve, one would expect 1.25 $2\sigma$ bins to appear by chance. 

Using the \texttt{Fermitools} algorithm \texttt{GTOPHASE}, we are able to assign a phase to each photon in our analysis. Whilst \texttt{GTOPHASE} is typically used for assigning phases to pulsars, \cite{rasul_gamma-rays_2019} demonstrate its suitability for dealing with the orbital phases of binary systems, and presumably this is the method that \textit{Xing\,19} employed, although this is not clear in the paper. In assigning the photon phases, we take the orbital ephemeris and period from \cite{bissinger_gro_2013} and assume that the first and second derivatives of period are zero. Given that the period of the orbit is of the order of hundreds of days, we do not expect the period to change significantly over the \textit{Fermi}-LAT mission time. Having assigned phases to each photon, we execute a likelihood analysis in 10 evenly-spaced phase bins to produce a phase-folded light-curve, analysing $\gamma$-ray emission by orbital phase in the same way as \textit{Xing\,19}.

Figure \ref{fig:GRO10_orb} shows the phase-folded light-curve of GRO10, alongside the flux points from the monthly-binned light-curve (Figure \ref{fig:GRO10_lc}) with a phase calculated for each bin. We see that of the 10 bins across the orbit, a $\gamma$-ray flux is apparent in 3 of these, the 2nd ($\mathrm{TS} = 4.6$), 6th ($\mathrm{TS} = 8.5$), and 9th bins ($\mathrm{TS} = 14.7$). The latter two bins are consistent with the results of \textit{Xing\,19}. However, where they measured a $\gamma$-ray flux in the 7th bin, we find only an upper limit. The monthly bins appear to cluster into two groups, with the first being coincident in phase with the 6th phase-folded bin (~$2.9 \sigma$), and the second being coincident with the 9th bin (~$3.8 \sigma$). We see no monthly flux points coincident with the third phase-folded flux measurement, but given the result in this bin is marginal in significance, this is unsurprising.

Considering the flux measurements in each bin are all below $5 \sigma$, we do not claim detection of any $\gamma$-rays from this system. While \textit{Xing\,19} establish that the most significant flux point (in the ninth phase bin) precedes periastron by a bin, the lack of detectable emission in either the first or 10th bin (immediately following and preceding periastron) casts some doubt on these being due to emission from GRO10. That said, the fact that every detectable monthly bin in the light-curve clusters around one of the two points indicates that there is likely some pattern to the apparent $\gamma$-ray excesses in this system, as it is unlikely\footnote{We find a $4.13 \times 10^{-9}$ chance of these two clusters occurring by chance.} that these flux points would cluster by chance in phase space, were they random background fluctuations.

\subsection{IGR\,J17544-2619}
\label{bin:IGR17}
IGR\,J17544-2619 (henceforth IGR17) is a HMXB system and the prototypical super-fast X-ray transient consisting of a likely pulsar in an unusually short $4.926 \pm 0.001$ day orbit with a massive (likely O-type) donor star \citep{bozzo_multi-wavelength_2016}. Over the 12.5 year \textit{Fermi}-LAT observation period,  we detect a $\gamma$-ray excess coincident with the position of IGR17 with $\mathrm{TS} = 19.7$ ($4.4 \sigma$), at a slight angular offset from the position of IGR17 of $0.151 \degree$. Using the \texttt{gta.localize} algorithm, we find a best fit position for this excess of $\mathrm{LII} = 3.3742  \degree \pm 0.0402 \degree$, $\mathrm{BII} = -0.2747 \degree \pm 0.0441 \degree$. At this best fit position, the new angular offset from the position of IGR17 is $0.0372 \degree$ and the TS of the excess increases slightly to 23.7. 

\begin{figure}
    \centering
    \includegraphics[width=240pt]{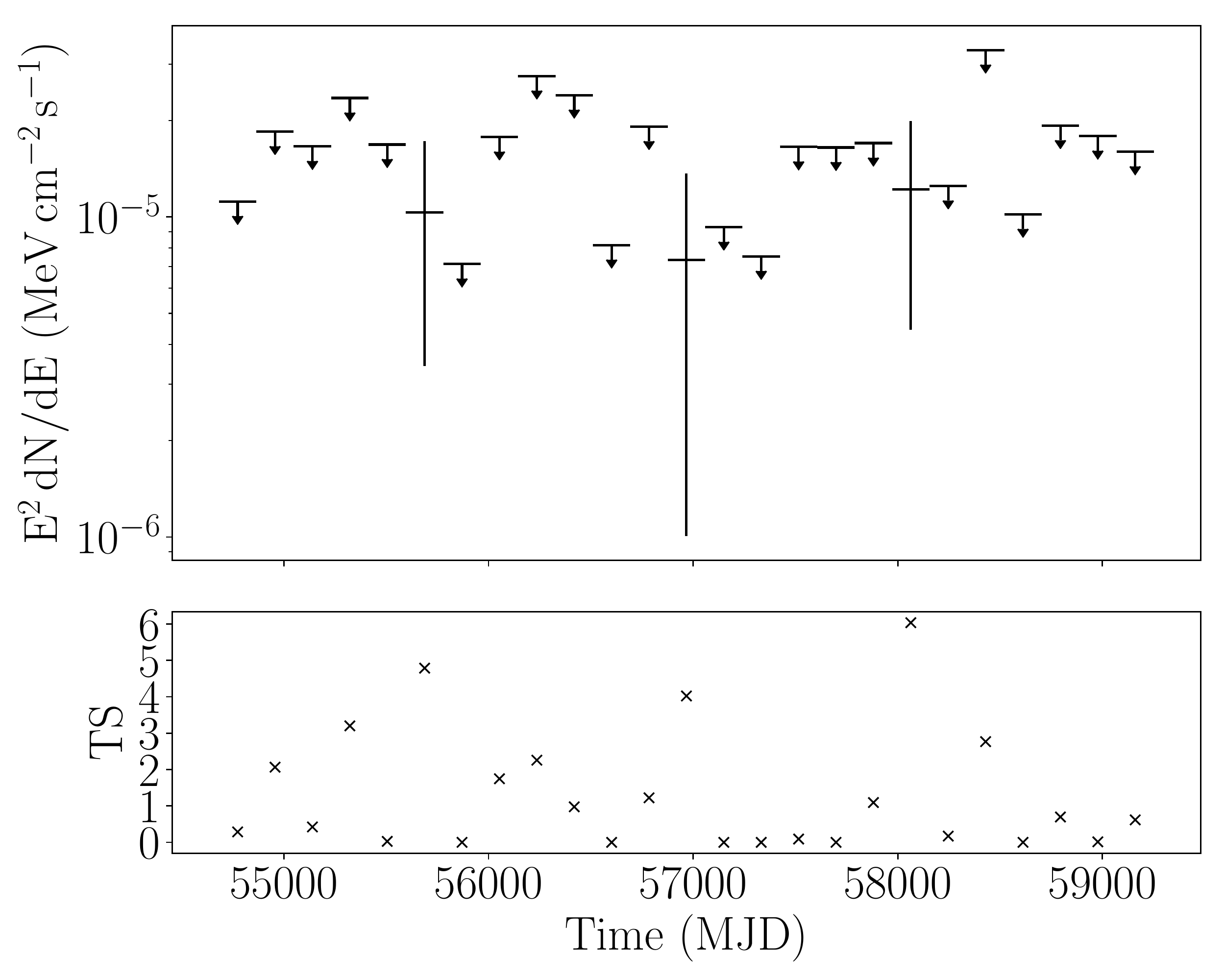}
    \caption{The \textit{Fermi}-LAT light-curve of the $\gamma$-ray excess coincident with IGR\,17544-2619 with 6-month time bins. Upper limits are placed on any time bin in which the TS is less than 4.}
    \label{fig:IGR17_lc}
\end{figure}

The nearest 4FGL sources in the sky are a source of unknown type, 4FGL J1754.4-2649 (angular offset from IGR17 of $0.499 \degree$, $\mathrm{TS}= 93.8$), 4FGL J1755.4-2552 (SNR G003.7-00.2, offset: $0.506 \degree$, $\mathrm{TS}= 157$), and the luminous unassociated $\gamma$-ray source 4FGL J1753.8-2538 ($\mathrm{TS}= 1500$, offset: $0.703 \degree$). There is no detectable variability in the 6-monthly binned light-curve (Figure \ref{fig:IGR17_lc}) of the excess, with 3 bins having a TS in the $2 \sigma \leq z < 3 \sigma$ range. These 3 bins do not correlate with any significant enhancements in the 6 monthly binned light-curves of the 3 closest $\gamma$-ray neighbours in the sky. This suggests that the $\gamma$-ray emission is unlikely to be due to confusion with a flare from a nearby object. 

We are unable to fit a model reliably to the SED of the IGR17 coincident excess due to limited photon statistics, and are thus unable to compare the spectrum of the excess with those of nearby sources. We cannot conclusively ascribe this excess to source confusion with the brightest nearby catalogued 4FGL source, 4FGL J1753.8-2538, and we cannot associate any features of the excess with IGR17 itself. However, as 4FGL J1753.8-2538 is very luminous ($\mathrm{TS} = 1500$), we cannot rule this out either. 

\subsection{IGR\,J19140+0951}
\label{bin:IGR19}
\begin{figure}
    \centering
    \includegraphics[width=240pt]{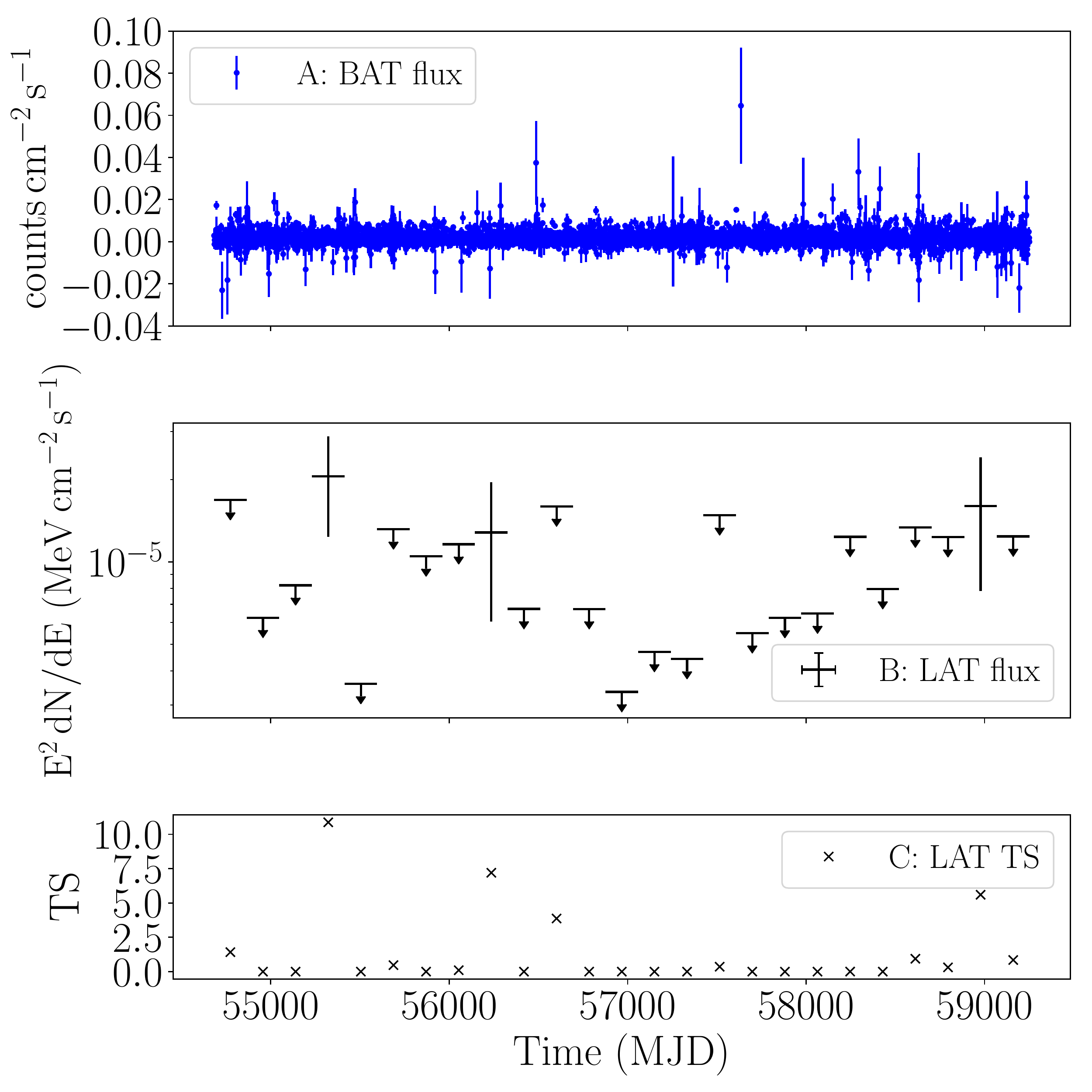}
    \caption{The daily binned light-curve of IGR\,J19140+0951 taken with \textit{Swift}-BAT shown in Panel A, with the 6-monthly binned \textit{Fermi}-LAT light-curve for a source fitted to the position of IGR\,J19140+0951 shown below in Panel B, and the corresponding TS values of these bins shown in Panel C. We place 95\% confidence limits on any \textit{Fermi}-LAT energy flux bins with $\mathrm{TS} < 4$. }
    \label{fig:IGR19_lc}
\end{figure}

IGR\,J19140+0951 (henceforth referred to as IGR19) is a HMXB with a likely neutron star accretor \citep{hannikainen_discovery_2004} and a supergiant B star donor (\citealt{hannikainen_nature_2007} and \citealt{zand_optical_2006}). Unlike most cases discussed in this paper, no persistent $\gamma$-ray excess is identified coincident with the position of IGR19. However, we identify 3 bins with a $\mathrm{TS} > 4$ in the 6 monthly binned light-curve shown in Figure \ref{fig:IGR19_lc}. None of these 3 bins corresponds to any significant enhancement in the X-ray waveband indicated by the \textit{Swift}-BAT daily light-curve, also shown in Figure \ref{fig:IGR19_lc}.

Three catalogued sources lie within a $0.5 \degree$ angular separation from the position of IGR19. These are 4FGL\,J1912.7+0957 ($\mathrm{TS} = 188$ and an angular offset of $0.335 \degree$), 4FGL\,J1914.7+1012c ($\mathrm{TS} = 110$, offset: $0.369 \degree$) and 4FGL\,J1913.3+1019 ($\mathrm{TS} = 137$, offset: $0.476 \degree$), confirmed to be the pulsar PSR\,J1913+1011. The three flux points do not correlate with any enhancements in the light-curve of any of the three closest sources, so it is unlikely that source confusion is responsible for these $\gamma$-ray excesses. As there is no known orbital information for IGR19 it is not possible to perform a phased analysis for this system, and the lack of a persistent $\gamma$-ray excess means that neither spectral analysis nor source localisation are possible. Although the small excesses at the position of IGR19 are independent of nearby sources, they are not significant enough to claim a detection, nor is there evidence to associate them with IGR19.  

\subsection{1A\,0535+262}
\label{bin:1A05}
1A\,0535+262 (henceforth 1A05) is a well-studied pulsar-Be star binary system with an orbital period of 110.3 days \citep{finger_quasi-periodic_1996}. 1A05 has been the target of previous searches for $\gamma$-ray emission (\citealt{acciari_gamma-ray_2011} \& \citealt{lundy_tev_2021}) and is well known for its giant X-ray outbursts, the most recent of which was in November 2020 (\citealt{bernardini_trigger_2020} \& \citealt{jaisawal_nicer_2020}) in addition to being a known source of non-thermal radio emission \citep{van_den_eijnden_vla_2020}. We find a $\gamma$-ray excess at the position of 1A05 with $\mathrm{TS}=12.4$, with the binary system itself being located roughly at the edge of the extended $\gamma$-ray source 4FGL\,J0540.3+2756e: the supernova remnant S\,147, which has an extension radius of $1.5 \degree$ \citep{abdollahi_fermi_2020}. The centroid of S\,147 is offset from 1A05 by $1.625 \degree$, and S\,147 has a TS of 1080. The closest nearby 4FGL sources lie within S\,147, the most significant of which is the unattributed point source 4FGL\,J0533.9+2838 ($\mathrm{TS} = 146$, angular offset from 1A05 of $2.572 \degree$). Given the large (several degree) separation of the nearest $\gamma$-ray point sources and the position of 1A05, if source confusion is responsible for the 1A05 coincident $\gamma$-ray excess, the confusion is likely with S\,147, which is a steady source\footnote{Supernova remnants are a non-variable class of $\gamma$-ray emitter, and S\,147 has a variability index of 6.7 in the 4FGL-DR2 which supports the hypothesis that no variability is observed on monthly timescales.}.

\begin{figure}
    \centering
    \includegraphics[width=240pt]{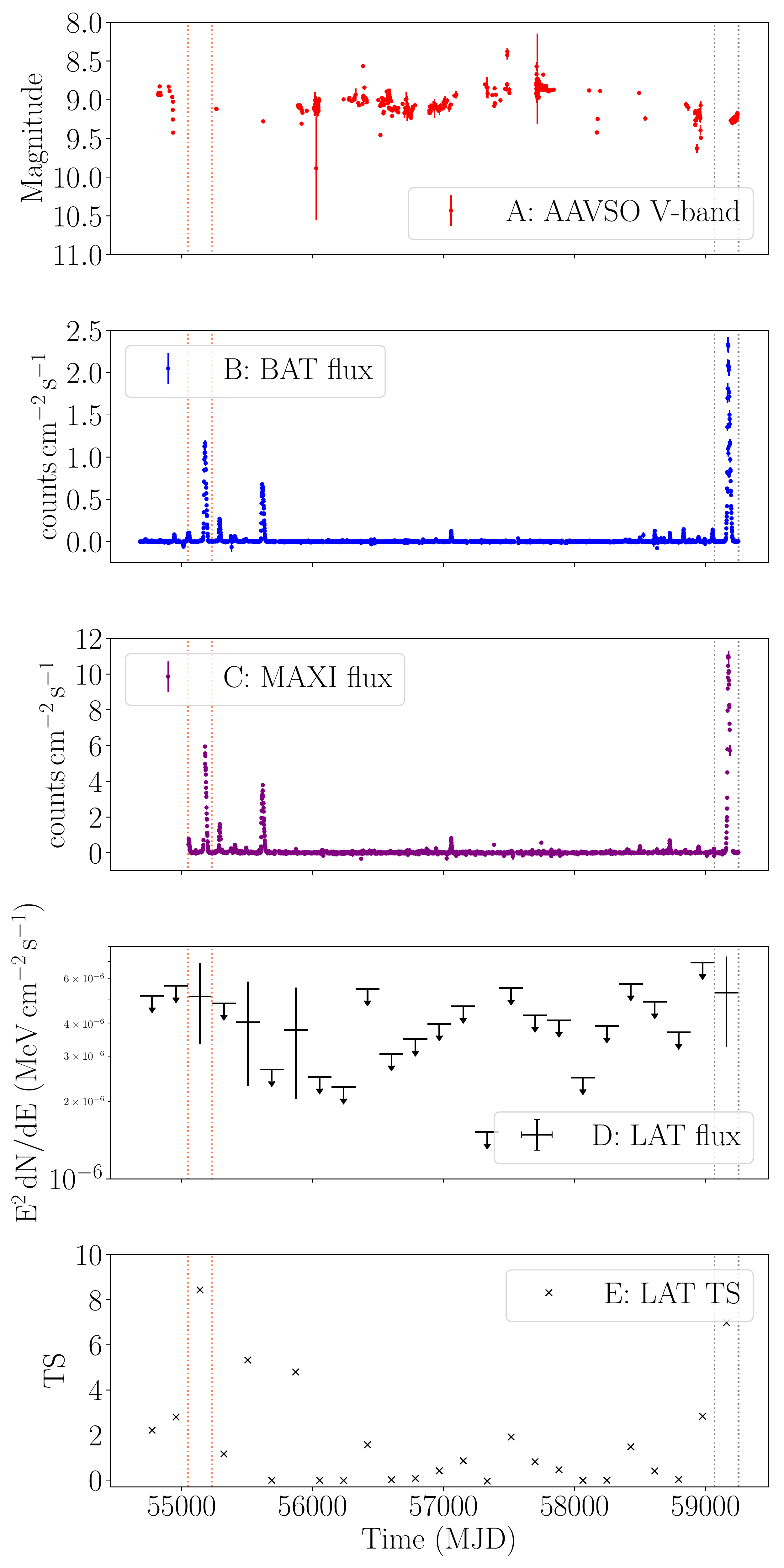}
    \caption{Light curves for 1A\,0535+262 - Panel A: AAVSO V-band optical light-curve; Panel B: \textit{Swift}-BAT daily light-curve; Panel C:  MAXI daily X-ray light-curve. Panel D shows the $\gamma$-ray energy flux measurements of the excess coincident with the position of 1A\,0535+262 with approximately 6 month bins, and Panel E shows the respective TS values of these bins. Upper limits on energy flux are calculated for any bin where $\mathrm{TS} < 4$. The vertical dotted orange lines indicate the start and end times of the $\gamma$-ray flux bin which is temporally coincident with the December 2009 Type II X-ray outburst, and the vertical grey dotted lines indicate the time interval of the $\gamma$-ray flux bin which is coincident with the November 2020 Type II X-ray outburst.}
    \label{fig:1A05_lc}
\end{figure}

\begin{figure}
    \centering
    \includegraphics[width=240pt]{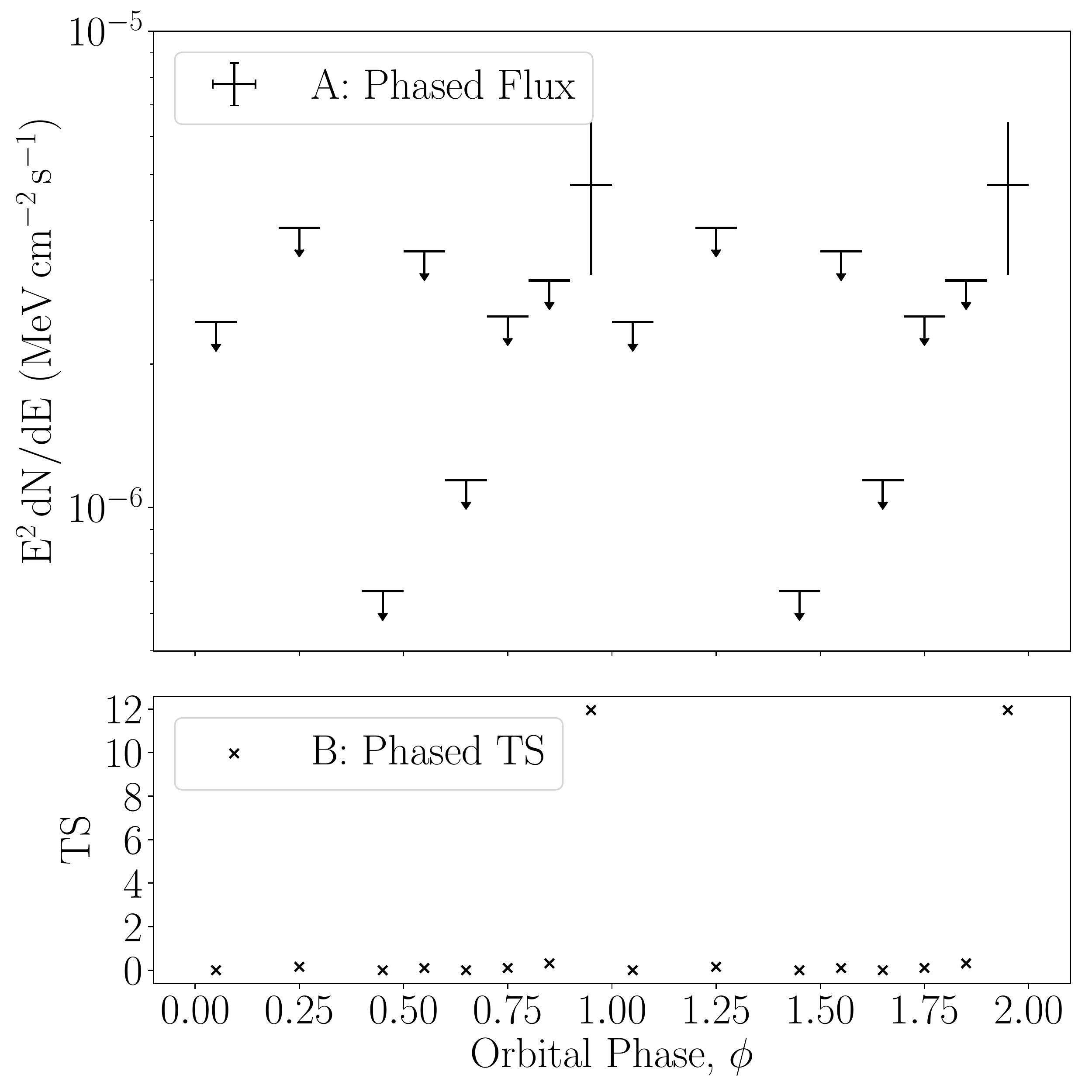}
    \caption{The orbital phase-folded light-curve of the $\gamma$-ray excess coincident with 1A\,0535+262 over the phase range $0 \leq \phi < 2$, with 10 phase bins per orbit. Panel A shows the phase folded energy flux of 1A\,0535+262, and Panel B shows the respective TS values of these phase bins, where upper limits are placed on any bins where $\mathrm{TS} < 4$. We note that our likelihood fit fails to identify a point source in the second and fourth orbital phase bins, thus no upper limit or TS is calculated for these bins. We define $\phi = 0, 1, 2$ as periastron and $\phi = 0.5, 1.5$ as apastron; the entire $\gamma$-ray excess is distributed in the phase bin immediately preceding orbital periastron.}
    \label{fig:1A05_phased}
\end{figure}

\begin{figure}
    \centering
    \includegraphics[width=240pt]{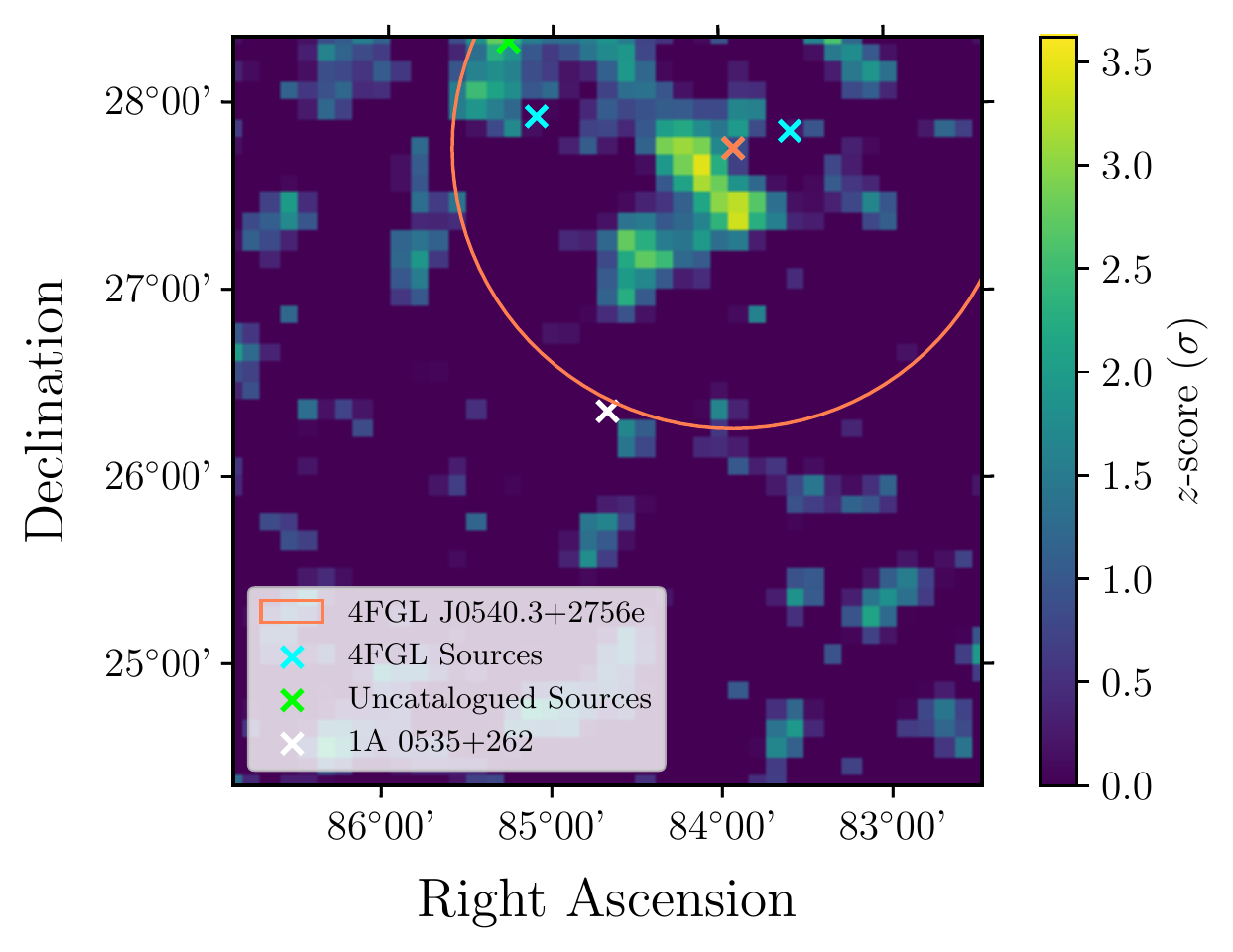}
    \caption{The significance map of the central $5\degree$ of the 1A\,0535+262 ROI, after our likelihood fit and the \texttt{gta.find\_sources} algorithm, in the phase range $0.9 \leq \phi < 1.0$. The blue crosses refer to the positions of 4FGL sources and the green crosses refer to the positions of uncatalogued sources. The orange circle and cross refer to the extent and centroid of the supernova remnant S\,147. The white cross indicates the catalogued location of 1A\,0535+262. The $\gamma$-ray excess appears to be slightly offset from the position of 1A\,0535+262, however such offsets are common with low significance excesses. Furthermore, the combination of several lower significance bins causes the $3.5 \sigma$ excess we observe from 1A\,0535+262.}
    \label{fig:1A05_TS}
\end{figure}

1A05's giant Type II X-ray outbursts peak at several times the brightness of the Crab Nebula. Figure \ref{fig:1A05_lc} shows the multi-wavelength light-curve of 1A05, with three very bright X-ray outbursts and numerous smaller outbursts seen with \textit{Swift}-BAT. There is no obvious correlation between these outbursts and the AAVSO optical measurements, although these observations don't cover the entire \textit{Fermi}-LAT mission.

The two brightest outbursts occurring during the \textit{Fermi}-LAT observations analysed in this work occur during December 2009, with a peak X-ray flux of $1.2 \, \mathrm{counts \, cm^{-2} \, s^{-1}}$, and during November 2020, with a peak X-ray flux of $2.4 \, \mathrm{counts \, cm^{-2} \, s^{-1}}$. We measure a $\gamma$-ray flux in the $2 \sigma \leq z < 3 \sigma$ significance range in the 6 month bin contemporaneous with both of these outbursts, with these two bins being the most significant in the entire light-curve. In addition to these two bins, two additional $\gamma$-ray flux measurements are made with slightly lower significance, one of which immediately precedes the third largest observed outburst in the \textit{Swift}-BAT light-curve. We observe that for the majority of our \textit{Fermi}-LAT observation time (the majority of the 2010s) 1A05 appears to be in relative quiescence, and that our $\gamma$-ray flux points are broadly concentrated around the active periods near the December 2009 and November 2020 outbursts. Although the bins are all of low significance (and thus have limited photon statistics), and longer than the X-ray outbursts themselves, there does appear to be some correlation between the evidence for $\gamma$-ray emission and X-ray activity.

1A05 has a known orbital period of 110.3 days, which enables us to phase-fold the $\gamma$-ray data using \texttt{GTOPHASE}. Figure \ref{fig:1A05_phased} shows the phase-folded light-curve of the excess coincident with 1A05. This shows that the only measurable $\gamma$-ray emission occurs in the range $0.9 \leq \phi < 1.0$, immediately preceding periastron. This flux bin has a TS of approximately 12 ($z = 3.5 \sigma$), which is comparable to the significance of the excess over the total \textit{Fermi}-LAT observation time; a significance map for this flux bin is shown in Figure \ref{fig:1A05_TS}. All other bins have a TS of approximately 0, and in two bins it was not possible to fit a point source at the position of 1A05 at all. This indicates that essentially all of the $\gamma$-ray flux from the excess coincident with 1A05 is concentrated in the one phase bin preceding periastron. Whilst it is possible for a phase folded light-curve to be dominated by a short, single, significant event, Figure \ref{fig:1A05_lc} shows that the flux is spread across several bins, each with comparable significance, so this is not the case here. 

Given the 1A05 $\gamma$-ray excess has only a $3.5 \sigma$ significance, we lack the photon statistics to generate an SED of the source. A combination of this with the fact that 1A05 lies on the edge of the diffuse emission of S\,147 also makes positional localisation impossible. Nevertheless, the evidence (if only at the $3.5 \sigma$ level) suggests that 1A05 could be a very faint $\gamma$-ray binary fueled by wind-wind interactions, or neutron star accretion. This is further supported by the fact that there are no other variable $\gamma$-ray sources near 1A05. Finally, the $\gamma$-ray flux from the mission-long light-curve of the 1A05 excess shows a weak correlation between $\gamma$-ray flux and X-ray activity, with measurable $\gamma$-ray fluxes generally corresponding to periods when 1A05 was in outburst, suggesting that neutron star accretion outbursts could be responsible for the $\gamma$-ray emission.

In order to reach the $5 \sigma$ threshold required for a typical claim of discovery, another 12.5 years of all-sky observations would be needed with \textit{Fermi}-LAT, assuming the object's emission characteristics do not change. It is unlikely that \textit{Fermi}-LAT will continue to operate for this long, but future observatories (for instance, AMEGO \citep{mcenery_all-sky_2019}) which will operate in the MeV gap where the peak $\gamma$-ray emission of many XRBs may be located could detect the emission from 1A05 more significantly.

\subsection{GRO\,J2058+42}
\label{bin:GRO20}
GRO\,J2058+42 (henceforth GRO20) is a pulsar-Be star HMXB \citep{wilson_discovery_2005} with a 55 day orbital period \citep{wilson_gro_2000}, discovered with the Compton observatory during a Type II outburst in 1995 (\citealt{wilson_gro_1995} \& \citealt{grove_gro_1995}). The most recent outburst of GRO20 was in March 2019, with triggers from both \textit{Swift}-BAT \citep{barthelmy_swift_2019} and \textit{Fermi}-GBM, and additional follow up observations from \textit{AstroSat} \citep{mukerjee_astrosat_2020}. 

\begin{figure}
    \centering
    \includegraphics[width=240pt]{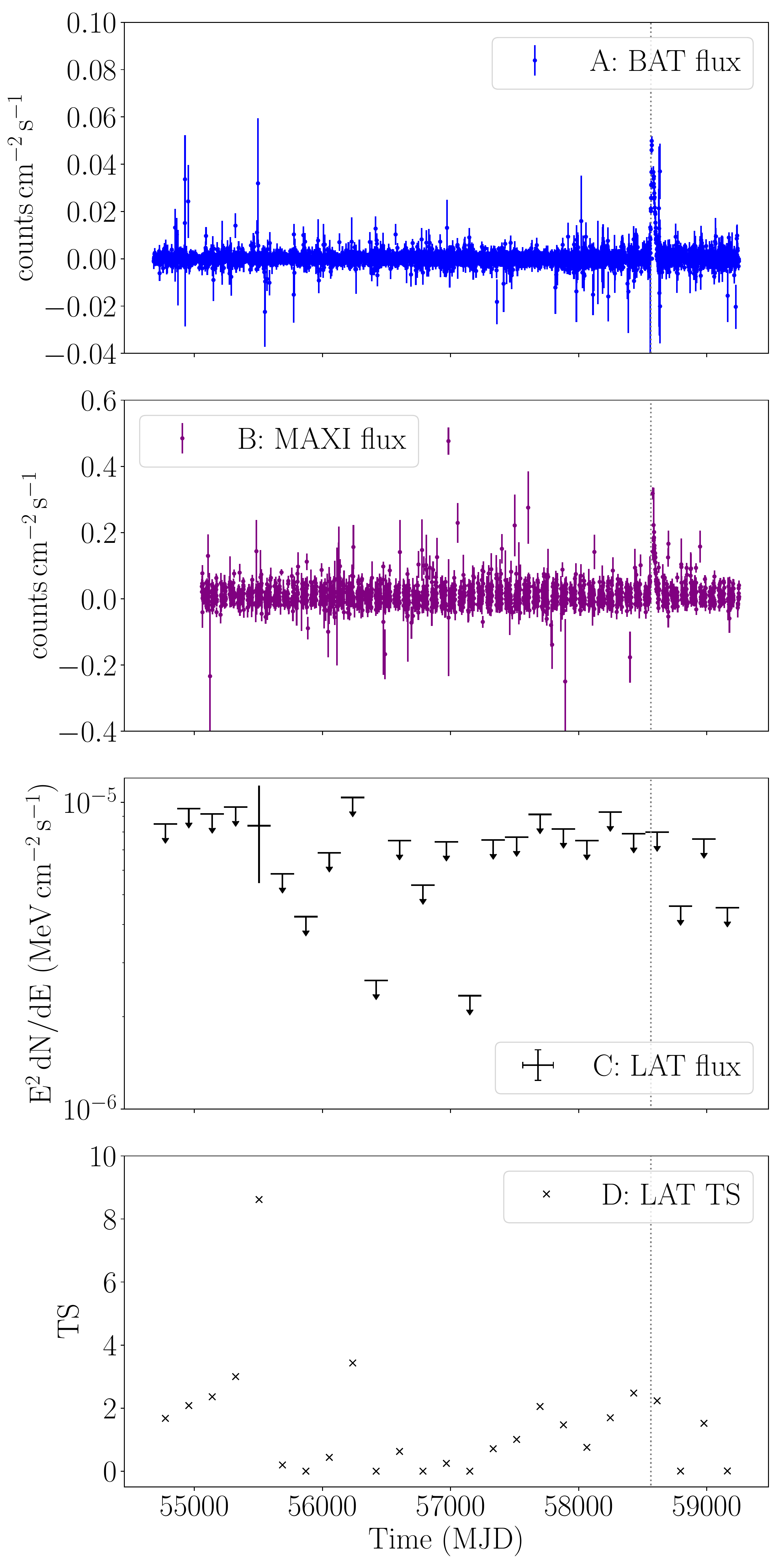}
    \caption{The \textit{Swift}-BAT and MAXI daily binned light-curves of GRO\,J2058+42 are shown in Panels A and B respectively, with the 6-month energy flux measurements and respective TS values of the coincident $\gamma$-ray excess shown in Panels C and D respectively. We place 95\% confidence limits on any \textit{Fermi}-LAT energy flux bins with $\mathrm{TS} < 4$. There is only one flux measurement from the light-curve of GRO\,J2058+42; this is not coincident with the March 2019 X-ray enhancement, the beginning of which is indicated by the vertical grey dotted line.}
    \label{fig:GRO20_lc}
\end{figure}

There is a small $\gamma$-ray excess coincident with the position of GRO20 with a TS of 16.3 ($z = 4.0 \sigma$), with a single flux measurement (MJD 55414 - 55596) in the 6-month binned light-curve (Figure \ref{fig:GRO20_lc}) and upper limits otherwise. This measurement is not coincident with the March 2019 X-ray enhancement, which is the only known outburst during the mission time of \textit{Fermi}-LAT\footnote{The outburst in May 2008 \citep{krimm_transient_2008} occurred several months before the beginning of \textit{Fermi}-LAT observations.}. Additionally, given that the most significant bin in the light-curve of the excess reaches only $\mathrm{TS} = 8.61$, evidence for long-term variability is very weak.

\begin{figure}
    \centering
    \includegraphics[width=240pt]{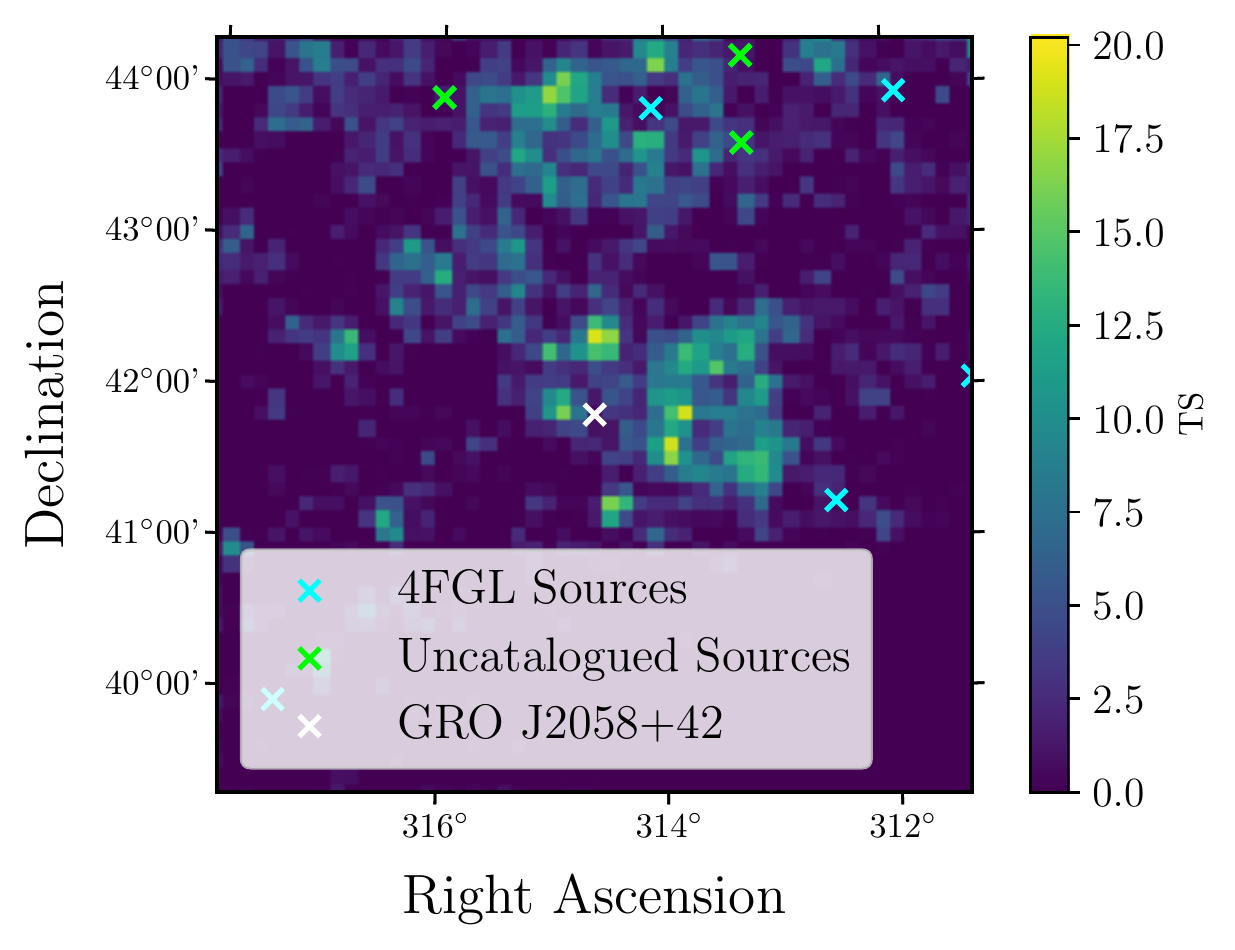}
    \caption{The TS map of the central $3\degree$ of the GRO\,J2058+42 ROI, after our likelihood fit and the \texttt{gta.find\_sources} algorithm. The blue crosses refer to the positions of 4FGL sources and the green crosses refer to the positions of uncatalogued sources. The white cross indicates the catalogued location of GRO\,J2058+42. Our spatial bins have an angular width of $0.1 \degree$.}
    \label{fig:GRO20_TS}
\end{figure}

Figure \ref{fig:GRO20_TS} shows the TS map of the region around GRO20, with the HMXB located within a wider $\gamma$-ray excess. There are no catalogued $\gamma$-ray sources within the immediate vicinity of GRO20, the closest sources being 4FGL\,J2050.0+4114c ($\mathrm{TS} = 34.0$ and an angular offset of $1.729 \degree$) and 4FGL\,J2056.4+4351c ($\mathrm{TS} = 297$, offset $2.122 \degree$) associated with the X-ray source 1RXS\,J205549.4+435216. Neither of these sources displays any variability according to their variability indices in the 4FGL-DR2. 

\begin{figure}
    \centering
    \includegraphics[width=240pt]{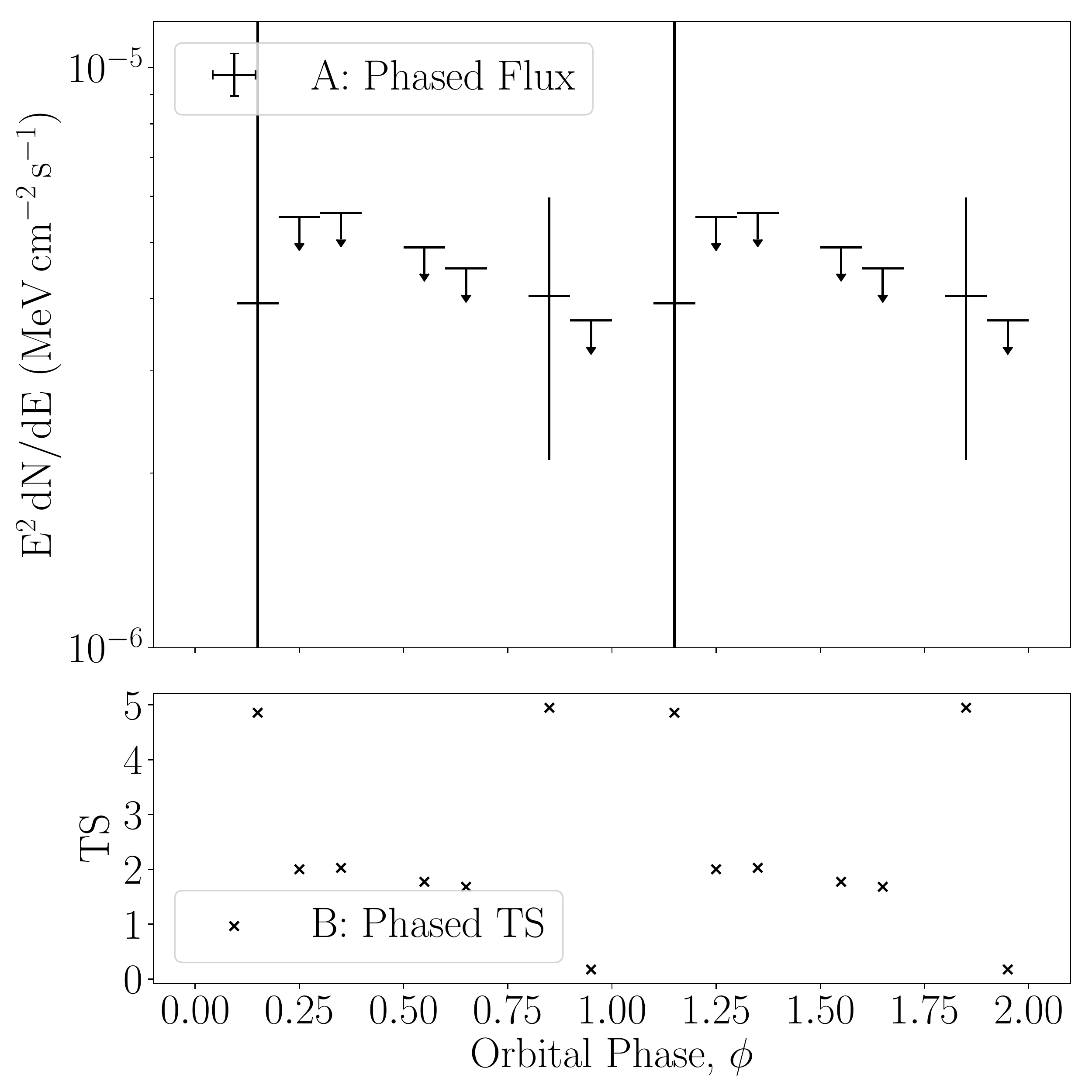}
    \caption{The orbital phase-folded light-curve of the $\gamma$-ray excess coincident with GRO\,J2058+42 over the phase range $0 \leq \phi < 2$, with 10 phase bins per orbit. Panel A shows the phase folded energy flux of GRO\,J2058+42, and Panel B shows the respective TS values of these phase bins, where upper limits are placed on any bins where $\mathrm{TS} < 4$. We note that our likelihood fit fails to identify a point source in the first, fifth and eighth orbital phase bins, thus no upper limit or TS is calculated for these bins. We define $\phi = 1$ as periastron and $\phi = 0.5$ as apastron. We note that there is a very large uncertainty ($\pm \, 4.18 \times 10^{-5} \, \mathrm{MeV} \, \mathrm{cm}^{-2} \mathrm{s}^{-1}$) associated with the flux measurement in the second and twelfth bins due to the very limited photon statistics.}
    \label{fig:GRO20_phased}
\end{figure}

Given the low ($z < 5 \sigma$) significance of the observed excess, we cannot carry out a spectral analysis or localisation. However, we do have an orbital period and ephemeris for this system \citep{wilson_gro_2000} so are able to produce a phase-folded light-curve with phase steps of 0.1. Given that GRO20 is a pulsar system with a Be companion star, one might expect $\gamma$-ray emission to peak around periastron ($\phi = 0,1,2 $), where the shocks between the pulsar wind and stellar wind are most intense, although if the neutron star is accreting during outburst it is likely that $\gamma$-ray emission could be fuelled by the accretion processes rather than a wind-wind interaction at periastron. Our phase folded light-curve of the GRO20 excess is shown in Figure \ref{fig:GRO20_phased}.There are weak ($z \approx 2 \sigma$) indications of $\gamma$-ray emission in the phase ranges $0.2 \leq \phi < 0.3$ and $0.8 \leq \phi < 0.9$. 
We conclude that there is likely no orbital modulation in the weak $\gamma$-ray excess we observe from the position of GRO20. 

There is no evidence for any significant variability over the mission time of \textit{Fermi}-LAT, nor is there evidence for any significant orbital modulation of the putative $\gamma$-ray flux. We conclude that there is no evidence for $\gamma$-ray emission from GRO20. As the immediate area around GRO20 appears to contain diffuse $\gamma$-ray emission, it is possible that a weak, unknown, extended source could be causing source confusion at the position of GRO20. There could also be one, or multiple, unresolved $\gamma$-ray point sources. 

\begin{figure}
    \centering
    \includegraphics[width=240pt]{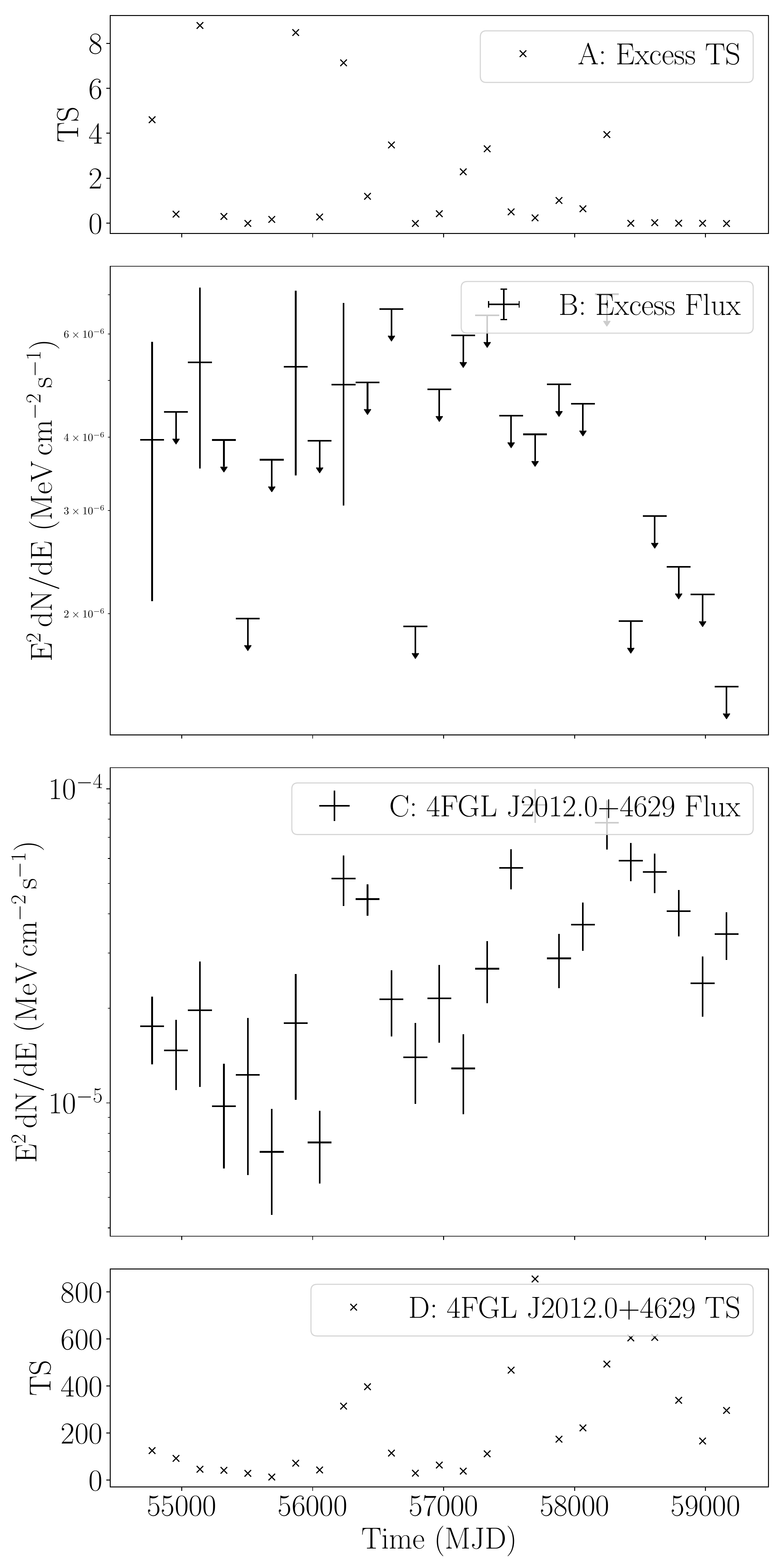}
    \caption{Panels B and A show the $\gamma$-ray fluxes and associated TS values for the excess coincident with the optical position of W63\,X-1. Panels C and D show the $\gamma$-ray fluxes and associated TS values for the nearby Bl Lac 4FGL\,J2012.0+4629, without the excess in the model. We use approximately 6 month bins in each of these light-curves, and 95\% confidence upper limits on flux are used for any bin where the corresponding TS value is less than 4.} 
    \label{fig:W63_com_lc}
\end{figure}

\subsection{W63\,X-1}
\label{bin:W63}
W63\,X-1 is a pulsar X-ray binary system, likely with a Be or OB star companion \citep{rho_discovery_2004} and located within the W63 supernova remnant, itself located within the Cygnus\,X star forming region \citep{sabbadin_elliptical_1976}. We observe a persistent $\gamma$-ray excess coincident with the position of W63\,X-1 with $\mathrm{TS} = 13.2$ ($z = 3.6 \sigma$). W63\,X-1 is a poorly-studied X-ray binary system; no orbital period is known and there is no recorded flux variability in any waveband. 

The closest $\gamma$-ray neighbour to the W63\,X-1 excess is the highly variable BL Lac type blazar 4FGL\,J2012.0+4629 also known as 7C\,2010+4619. We detect 4FGL\,J2012.0+4629 to a significance of $\mathrm{TS} = 4710$, and with an angular offset from the position of W63\,X-1 of $1.435 \degree$. Although it is unlikely, given the separation between the W63\,X-1 excess and the blazar, the highly variable and luminous nature of this source means we must test for source confusion, which we do by generating comparative light-curve of the blazar using the same binning scheme which we use at the binary position. 

Figure \ref{fig:W63_com_lc} shows the light-curves of both the W63\,X-1 coincident excess, and 4FGL J2012.0+4629. There is weak evidence ($2 \sigma \leq z < 3 \sigma$) for emission from the position of W63\,X-1 in 4 time bins, all spread across the first half of the \textit{Fermi}-LAT mission. At this time, 4FGL J2012.0+4629 appears to be in a lower flux state before a year-long flux enhancement, the beginning of which corresponds with the last $\gamma$-ray bin in the light-curve of the apparent W63\,X-1 excess. Confusion with the blazar is therefore unlikely to be the source of the $\gamma$-ray excess.

Given the marginal nature of all of the $\gamma$-ray flux measurements in the light-curve, the lack of measurable variability of the excess and a lack of multi-wavelength data, it is impossible to identify any correlations between wavebands. Poor photon statistics preclude spectral analysis or localisation. We cannot associate the $\gamma$-ray excess with W63\,X-1, but nor can we exclude the possibility that the excess is caused by W63\,X-1. 
Given that both the excess and W63\,X-1 lie within the larger supernova remnant W63 itself, it is possible that the small $\gamma$-ray excess represents very faint $\gamma$-ray emission from the supernova remnant rather than the binary system.

\subsection{RX\,J2030.5+4751}
\label{bin:RX20}
RX\,J2030.5+4751 (henceforth referred to as RX20) is a HMXB system consisting of a neutron star or black hole and a Be star \citep{belczynski_apparent_2009}. The orbital period of this system is unknown, but the 100 year optical light-curve indicates long-term variability on the timescale of decades \citep{servillat_dasch_2013}. We identify a $\gamma$-ray excess spatially coincident with the position of RX20 with a TS of 30.81. The 6-monthly binned light-curve of the source indicates that this excess seems to be largely dominated by one bin from MJD 56145-56328 (which reaches approximately $5 \sigma$). A measurable flux is observed in 7 other time bins, although at a lower level and with larger uncertainties, so we place upper limits on those bins. There is no enhancement contemporaneous with this $\gamma$-ray enhancement in the optical AAVSO V-band. There are no X-ray light-curves for this source available from either \textit{Swift}-BAT or MAXI.

\begin{figure}
    \centering
    \includegraphics[width=240pt]{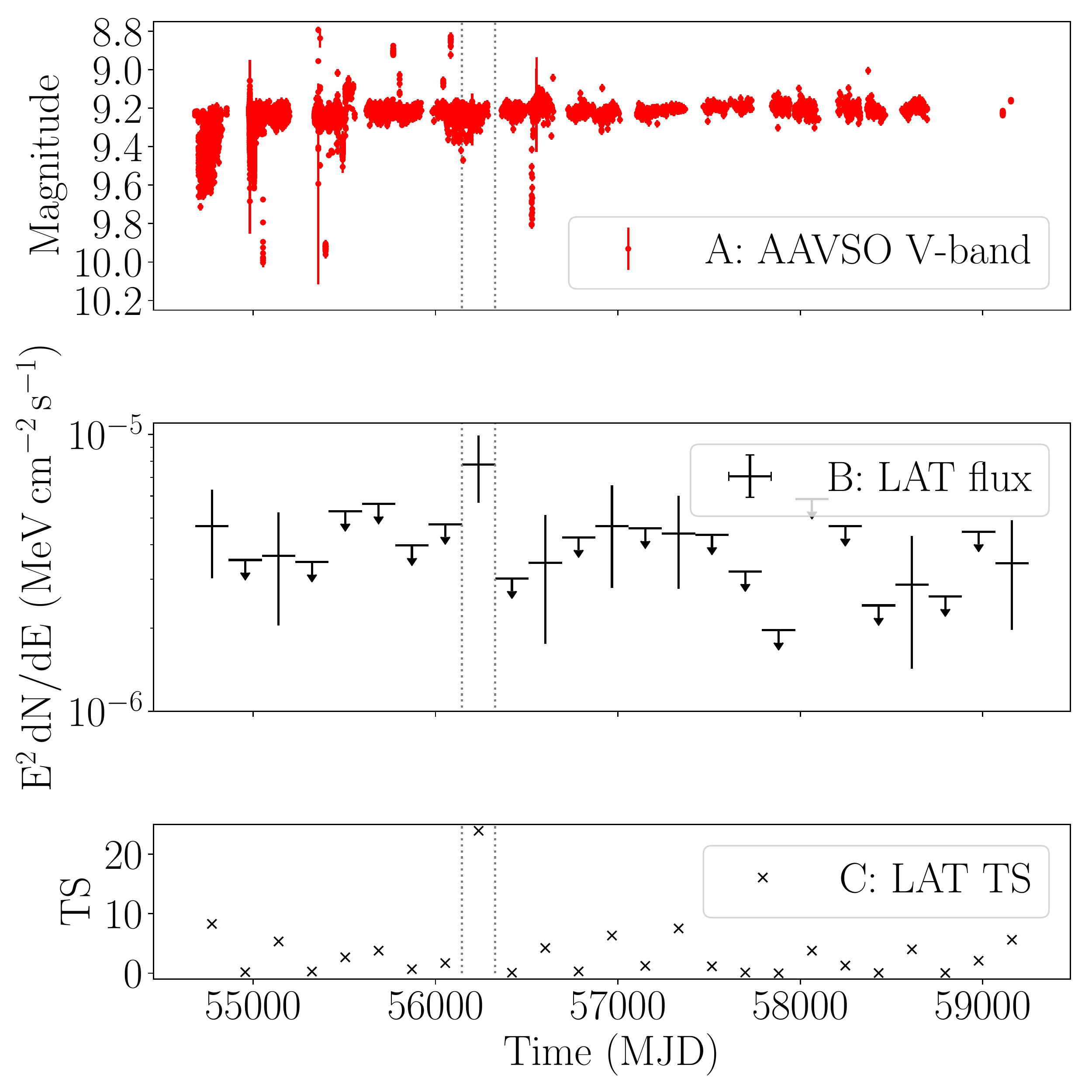}
    \caption{The AAVSO V-band optical light-curve of RX\,J2030.5+4751 is shown in Panel A, with the 6-monthly binned \textit{Fermi}-LAT light-curve for a source fitted to the position of RX\,J2030.5+4751 shown below in Panel B, and the corresponding TS values shown in Panel C. We place 95\% confidence limits on any \textit{Fermi}-LAT energy flux bins with $\mathrm{TS} < 4$. The vertical dashed lines indicate the beginning and end of the 6-month period in which there is a significant enhancement of the excess. }
    \label{fig:RX20_lc}
\end{figure}

The nearest $\gamma$-ray sources to RX20 are 4FGL J2026.0+4718 ($\mathrm{TS} = 25.4$ and an angular offset of $0.942 \degree$), 4FGL\,J2035.9+4901, associated with the blazar 2MASS\,J20355146+4901490 ($\mathrm{TS} = 25.4$, offset: $0.942 \degree$) and 4FGL\,J2029.5+4925, associated with the BL Lac type blazar MG4\,J202932+4925 ($\mathrm{TS} = 454$, offset: $1.567 \degree$). In addition to these catalogued sources, we also identify a second $\gamma$-ray excess which we name PS\,J2027.4+4728 ($\mathrm{TS} = 25.4$, offset: $0.942 \degree$).

\begin{figure}
    \centering
    \includegraphics[width=240pt]{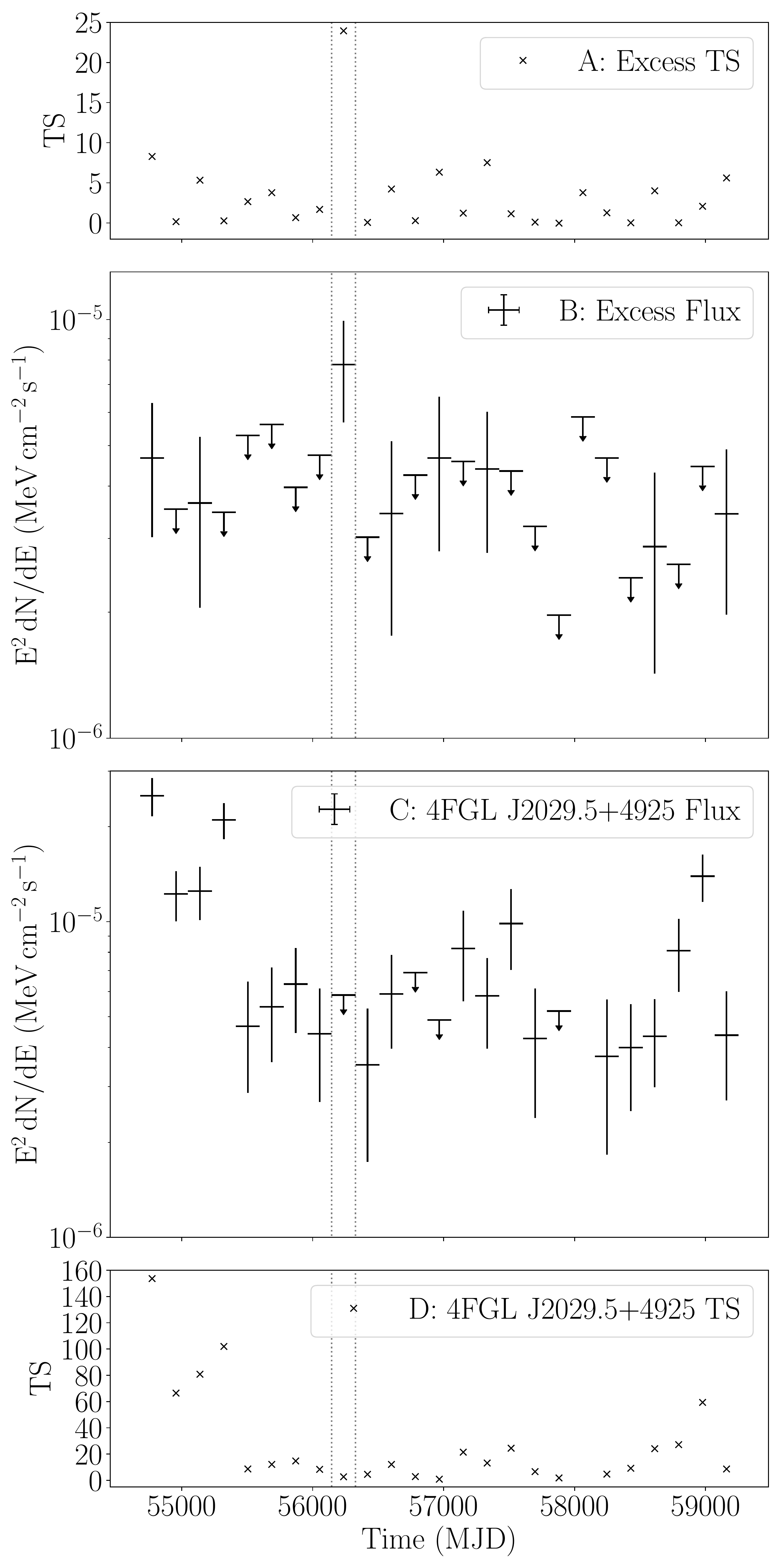}
    \caption{Panels B and A show the $\gamma$-ray fluxes and associated TS values for the excess coincident with the optical position of RX\,J2030.5+4751. Panels C and D show the $\gamma$-ray flux and associated TS values for the nearby $\gamma$-ray source 4FGL\,J2029.5+4925, without the excess in the model. We use 6-month bins in each of these light-curves, and 95\% confidence upper limits on flux are used for any bin where the corresponding TS value is less than 4. The dotted lines indicate the duration of the observed soft $\gamma$-ray flare.} 
    \label{fig:RX20_com_lc}
\end{figure}

Whilst the observed excess exceeds $5 \sigma$ in significance, we note that the photons almost entirely lie at just above 100 MeV, meaning any SED of the excess would not provide any meaningful observation (as a $\gamma$-ray flux would only be measured in one bin). We note that this is also the case during the time bin of the flare. The very soft nature of the $\gamma$-ray excess coincident with RX20 is somewhat problematic, as \textit{Fermi}-LAT's angular resolution in the MeV range is several degrees. Coupled with the relatively low photon count (a few thousand), this means the \texttt{gta.localize} algorithm is unable to properly converge in this case and we cannot unambiguously associate the excess with RX20. As two of the three nearest 4FGL sources to RX20 are blazars, which are usually variable, and the third source is of unknown nature and may also be variable, we generate light-curves of each source in order to examine whether a flare from one of these sources is causing source confusion. Whilst 4FGL\,J2026.0+4718 and 4FGL\,J2035.9+4901 are not significantly variable, the BL Lac object, 4FGL\,J2029.5+4925, is. Figure \ref{fig:RX20_com_lc} shows the light-curve of 4FGL\,J2029.5+4925, and clearly shows that for approximately the first two years of the \textit{Fermi}-LAT mission the blazar is in an enhanced flux state relative to the rest of the mission. However, during the time where we observe the flare coincident with RX20, there are only flux upper limits from 4FGL\,J2029.5+4925. Hence we are confident that the $\gamma$-ray excess observed at the position of RX20 is independent of nearby 4FGL $\gamma$-ray sources. 

The orbital period and the nature of the accretor in RX20 are unknown, and the system is not a known microquasar which makes $\gamma$-ray emission from a jet unlikely. Given that most Be star HMXBs have a neutron star accretor \citep{belczynski_apparent_2009}, it is likely that this is also the case for RX20, and it is possible that the soft $\gamma$-ray flare we observe coincident with RX20 is representative of either a wind driven interaction at the periastron of the system, as observed in the known $\gamma$-ray binary population, or a neutron star accretion outburst. However, without X-ray data for this time we cannot be certain.

\subsection{4U\,2206+543}
\label{bin:4U22}

\begin{figure}
    \centering
    \includegraphics[width=240pt]{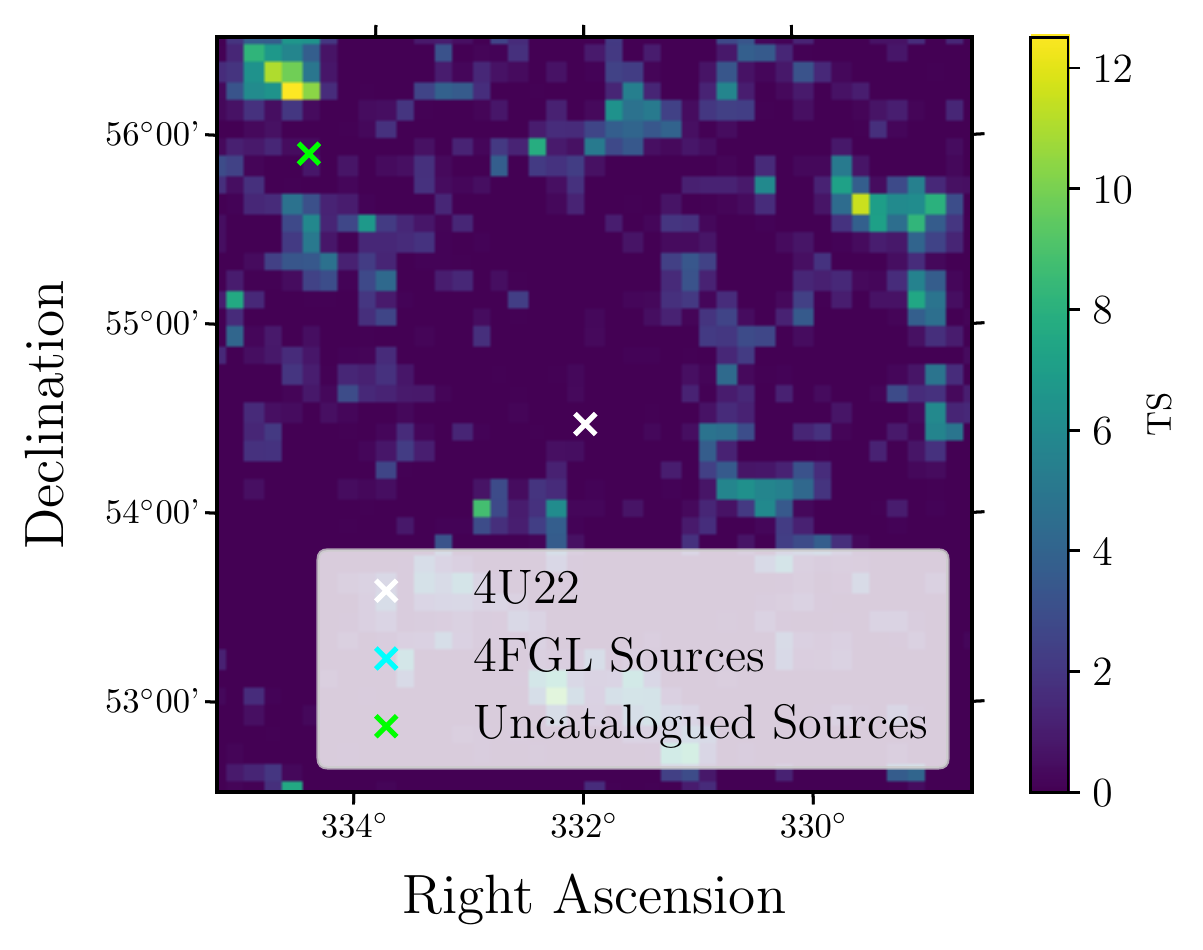}
    \caption{The TS map of the central $3\degree$ of the 4U\,2206+543 ROI across the full 12.5 year observation time. The positions of the closest 4FGL sources are indicated by blue crosses, whilst the positions of sources identified with the \texttt{gta.find\_sources} algorithm are indicated by green crosses. This TS map is generated after our ROI optimization and fit, but before a point source for 4U\,2206+543 is fitted to the model, to highlight the spatial coincidence between the excess and the position of 4U\,2206+543. Bins are $0.1 \degree$ across. We note that given the soft nature of this excess, it is possible that this is a product of very soft photons from the Galactic plane itself, rather than a genuine signature of $\gamma$-ray emission from a HMXB. This would explain why no excess is observed at the position of the white cross, as this apparent `source' is a product of background photons.}
    \label{fig:4U22_TS}
\end{figure}

\begin{figure}
    \centering
    \includegraphics[width=240pt]{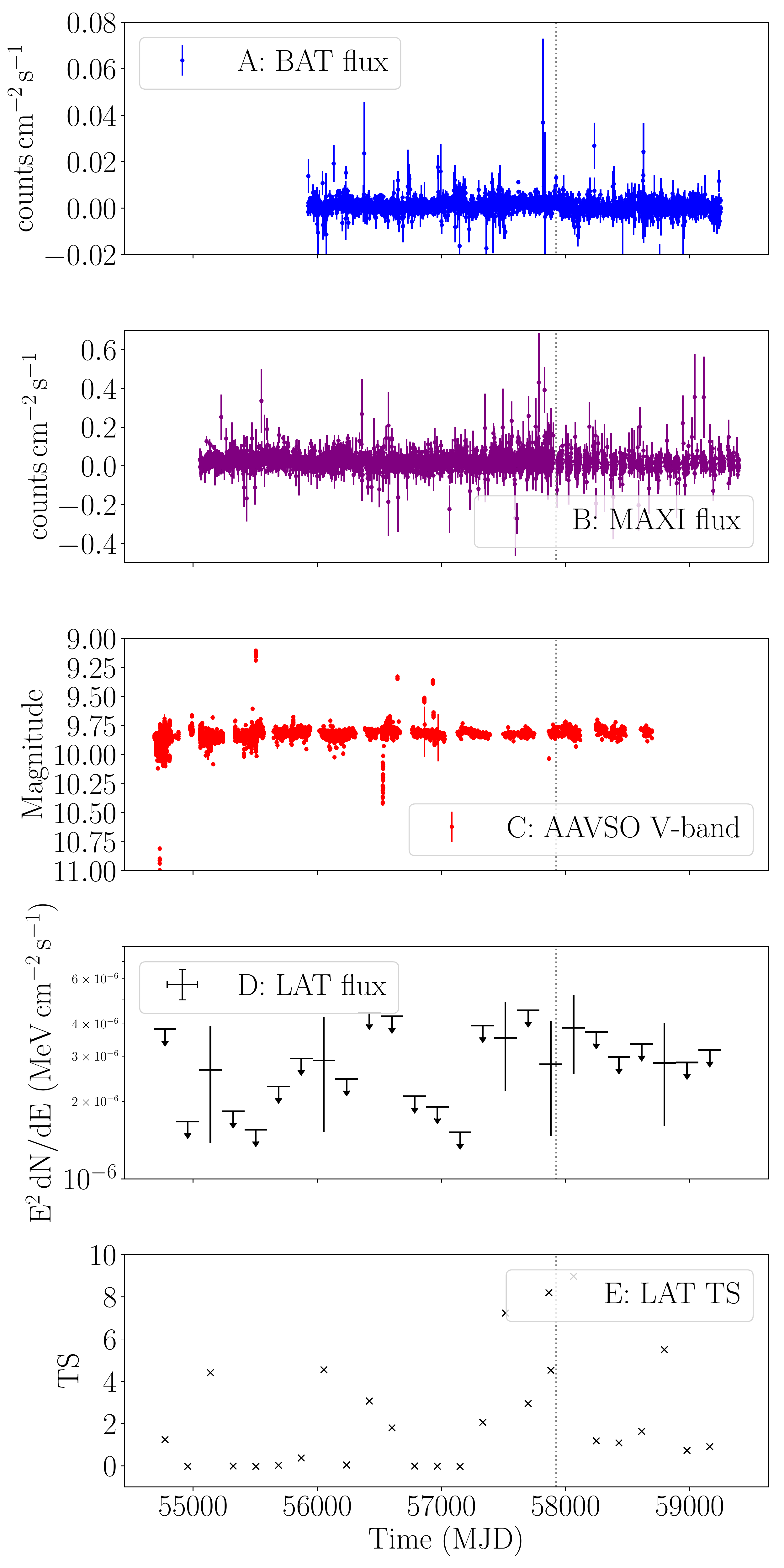}
    \caption{The multi-wavelength light-curve of 4U\,2206+543, and the coincident $\gamma$-ray excess. Panel A shows the \textit{Swift}-BAT X-ray data, which does not cover the entirety of the LAT observations, and Panel B shows the MAXI daily X-ray light-curve. The bin sizes for each X-ray light-curve are the same, with 1-day bin widths. Panel C shows the available AAVSO V-band optical photometry observations of 4U\,2206+543 for the duration of the LAT mission. Panel's D and E show the energy flux and respective TS of 4U\,2206+543 from the \textit{Fermi}-LAT light-curve, where upper limits are fixed to any flux bin where $\mathrm{TS} < 4$. The vertical grey dotted line indicates the beginning of the INTEGRAL period where an enhancement in X-ray data is observed.}
    \label{fig:4U22_lc}
\end{figure}

4U\,2206+543 (henceforth 4U22) is a HMXB system with a Be star companion, a pulsar accretor (\citealt{negueruela_search_2007}, \citealt{finger_spin-down_2009} \& \citealt{wang_spin_2013}) and a  9.57 day orbital period. We find a $\gamma$-ray excess coincident with the position of 4U22 with a TS of 30.53. With a Galactic latitude of $\mathrm{BII} = -1.11 \degree$, 4U22 is on the Galactic plane; however, it is relatively isolated from other $\gamma$-ray point sources. The nearest catalogued source is the pulsar 4FGL\,J2215.6+5135 (PSR J2215+5135), with a TS of 1940 and an angular offset from the position of 4U22 of $3.149 \degree$. This is a highly significant source, but given the separation between the pulsar and the position of 4U22, it is unlikely that source confusion explains the excess at 4U22's position. We also identify 4 uncatalogued, sub-threshold $\gamma$-ray excesses between $2 \degree$ and $3 \degree$ angular offset from 4U22; these are also unlikely to cause source confusion with 4U22 given that they are all less significant than the excess coincident with 4U22, and are $>2 \degree$ from 4U22.

Similar to RX20 (Section \ref{bin:RX20}), 4U22's spectrum is extremely soft with the entirety of the $\gamma$-ray flux being concentrated at just above 100 MeV, making any meaningful spectral analysis impossible for this source. Given the very soft nature of this apparent excess, the localisation fit fails and as shown in the TS map (Figure \ref{fig:4U22_TS}) there is no visually obvious excess centered on the position of 4U22. 

Figure \ref{fig:4U22_lc} shows the multi-wavelength light-curve of 4U22, with daily X-ray data from both MAXI and \textit{Swift}-BAT, optical AAVSO V-band photometry and the 6-month binned \textit{Fermi}-LAT energy flux and associated TS of the spatially coincident excess. The $\gamma$-ray flux of the excess coincident with the position of 4U22 is generally consistent for the bins where a measurement is made, and the upper limits consistent otherwise. The statistical significance of all bins is relatively low, with a maximum measured TS of approximately 9 (3$\sigma$). Due to the short orbital period of 4U22, we cannot identify any regular Type I X-ray outbursts, nor it is possible to identify orbital periodicity from the \textit{Fermi}-LAT data. 

\begin{figure}
    \centering
    \includegraphics[width=240pt]{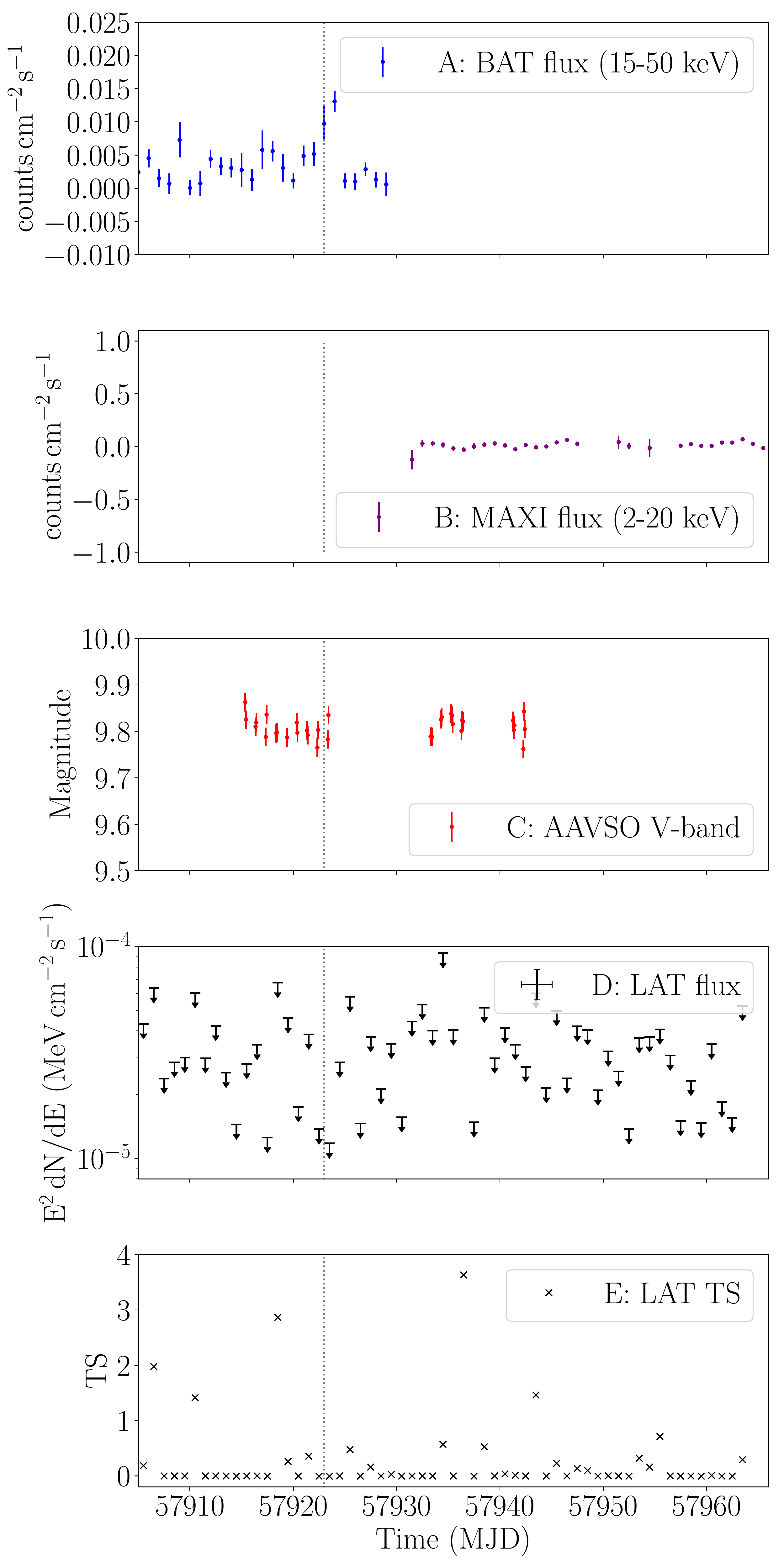}
    \caption{The multi-wavelength light-curve of 4U\,2206+543, and the coincident $\gamma$-ray excess during June and July 2017, when the INTEGRAL enhancement was detected (the start of which is denoted by the grey dotted line). Panel A shows the \textit{Swift}-BAT X-ray data, which covers the time of the enhancement and Panel B shows the MAXI X-ray data, available only for the time period after the enhancement measured with \textit{Swift}-BAT. The bin sizes for each X-ray light-curve are the same, with 1 day bin widths. Panel C shows the available AAVSO V-band photometric observations of 4U\,2206+543; no significant optical enhancement is seen. Panels D and E show the energy flux and respective TS of 4U\,2206+543 from the \textit{Fermi}-LAT daily light-curve, where upper limits are fixed to all flux bins, as no bin has a TS greater than 4. }
    \label{fig:4U22_day_lc}
\end{figure}

An enhancement of emission between 20 and 100 keV  was observed from 4U22 in June 2017 with INTEGRAL \citep{di_gesu_integral_2017}, together with a small enhancement we observe in the \textit{Swift}-BAT data. No MAXI data are available during the INTEGRAL observation period. There is a measurable $\gamma$-ray flux in the 6-month bin coincident with the 2017 enhancement (denoted by the grey vertical line in Figure \ref{fig:4U22_lc}), but such a flux is not unique to this time. As the  hard X-ray/soft $\gamma$-ray  enhancement lasted only days, we also generate a daily $\gamma$-ray light-curve to establish whether any $\gamma$-ray emission exists on the timescale of this enhancement. This light-curve is shown in Figure \ref{fig:4U22_day_lc}. No significant $\gamma$-ray emission is detected on daily timescales during June and July 2017, with only upper limits measured. We conclude that the enhancement reported by \cite{di_gesu_integral_2017} produced no measurable, contemporaneous, high-energy $\gamma$-ray flux. 

A lack of variability from the excess coincident with the position of 4U22 and a lack of information regarding the true position of this very soft excess (which has a PSF of $3.5 \degree$) makes it impossible to associate this excess with 4U22. Given the soft nature of this excess, it is possible that this apparent source is a product of excess very soft photons from the Galactic plane itself, which is very difficult to model, rather than a genuine signature of $\gamma$-ray emission from a HMXB. This hypothesis is further supported by the apparent lack of any point source excess shown in the TS map (Figure \ref{fig:4U22_TS}), coincident with the position of 4U22, despite a $5.5 \sigma$ point source being fitted to this position.

\subsection{IGR\,J00370+6122}
\label{bin:IGR00}
\begin{figure}
    \centering
    \includegraphics[width=240pt]{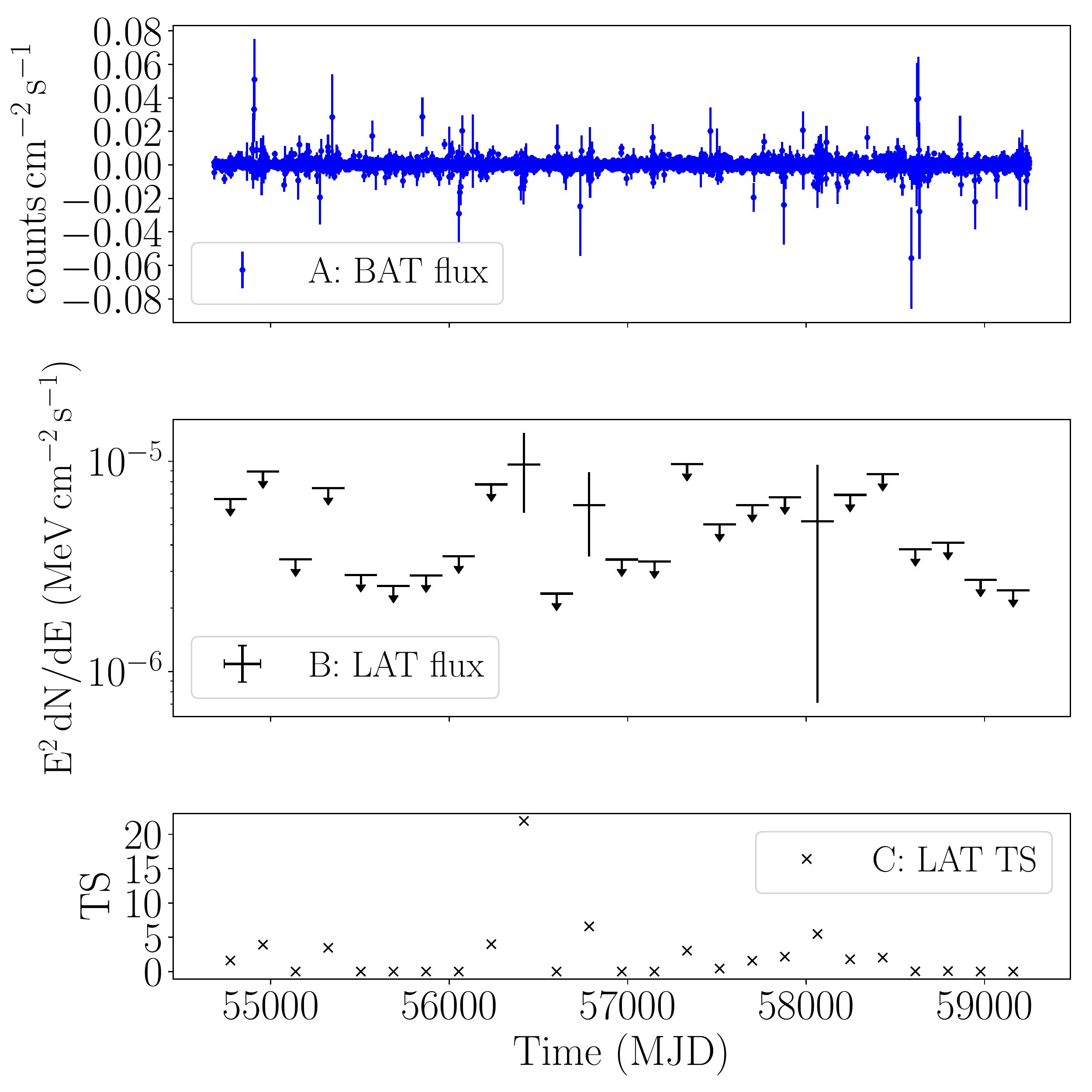}
    \caption{The \textit{Swift}-BAT daily binned light-curve of IGR\,J00370+6122 is shown in Panel A, with the 6 month energy flux measurements and respective TS values of the coincident $\gamma$-ray excess shown in Panels B and C respectively. As is consistent with the other light-curves we produce, we place 95\% confidence limits on any \textit{Fermi}-LAT energy flux bins with $\mathrm{TS} < 4$.}
    \label{fig:IGR00_lc}
\end{figure}

IGR\,J00370+6122 \citep{den_hartog_igr_2004} (henceforth IGR00) is an X-ray binary system with a pulsar accretor \citep{int_zand_probable_2007} and a B1Ib class companion star \citep{negueruela_bd_2004} with an orbital period of 15.7 days \citep{grunhut_orbit_2014}. There is a very marginal persistent $\gamma$-ray excess coincident with the position of IGR00, with a $\mathrm{TS} = 7.30$ ($z = 2.7 \sigma$), however in the six monthly binned light-curve we see evidence for emission at $\mathrm{TS}=21.9$ ($z = 4.7 \sigma$) in one of the bins (MJD 56328-56511), and \}{see weak ($2 \sigma$) evidence for emission} in two other bins (MJD 56693 - 56876 and MJD 58155 - 58338). This suggests that there may be transient $\gamma$-ray emission at IGR00's position. The $\gamma$-ray light-curve of the excess and the IGR00 X-ray light-curve are shown in Figure \ref{fig:IGR00_lc}. There is no apparent correlation between the X-ray light-curve of IGR00 and the $\gamma$-ray light-curve of the excess. 

\begin{figure}
    \centering
    \includegraphics[width=240pt]{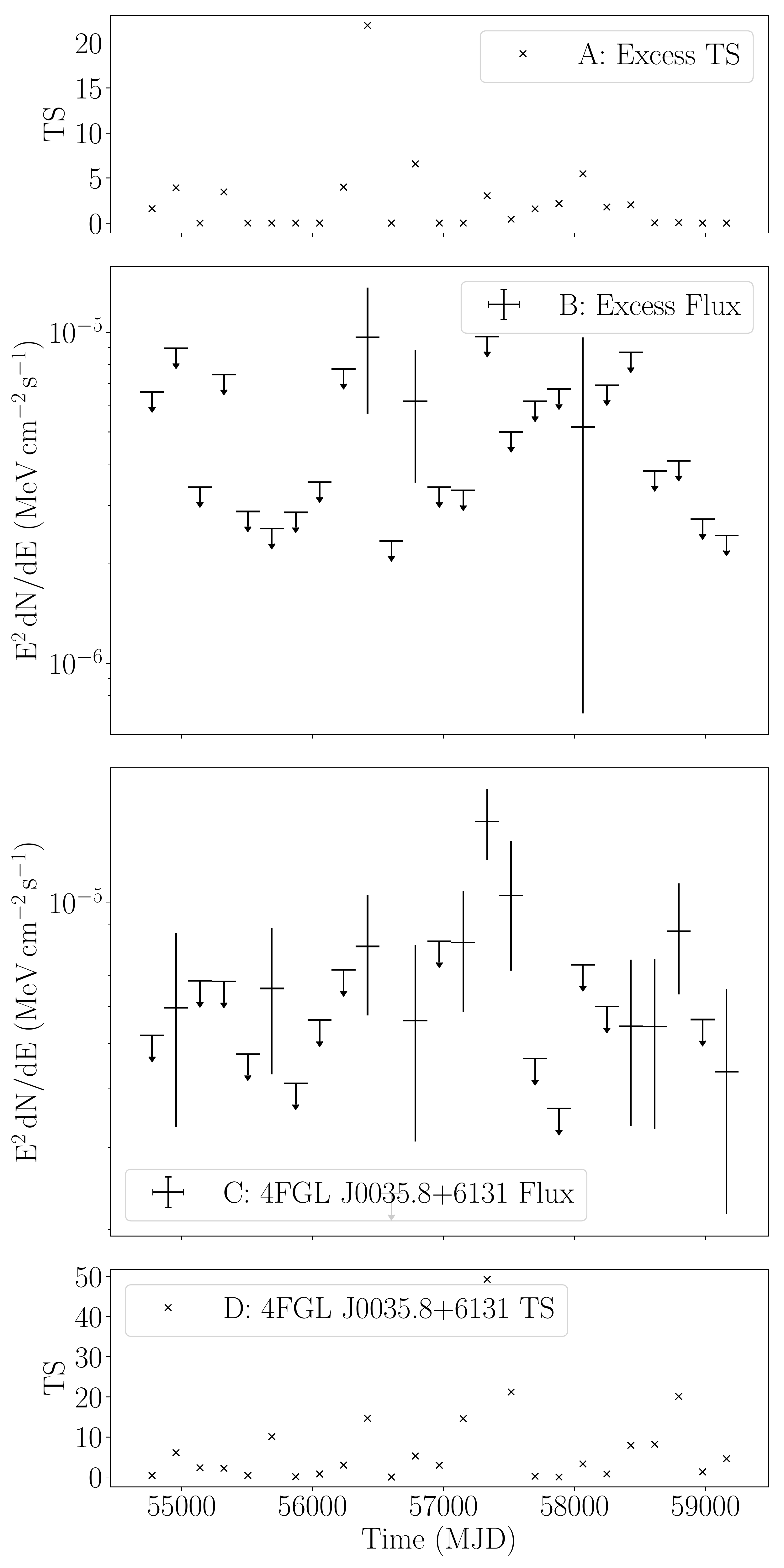}
    \caption{Panels B and A show the $\gamma$-ray flux and associated TS values of these flux points for the excess coincident with the position of IGR\,J00370+6122. Panels C and D show the $\gamma$-ray flux and associated TS values of these flux values for the nearby $\gamma$-ray blazar 4FGL J0035.8+6131, without the excess in the model. We use approximately 6 month bins in each of these light-curves, and 95\% confidence upper limits on flux are used for any bin where the corresponding TS value is less than 4.}
    \label{fig:IGR00_com_lc}
\end{figure}

There is a variable blazar 4FGL\,J0035.8+6131 (also known as LQAC\,008+061, $\mathrm{TS} = 71.3$ at an angular offset of $0.225 \degree$) close to the position of IGR00. Given the small angular offset between the blazar and position of IGR00, it is likely that source confusion is responsible for the excess, rather than it being a genuine signature of $\gamma$-ray emission from a HMXB. We generate a light-curve of 4FGL\,J0035.8+6131 using the same 6-month binning scheme used for the light-curve of the excess (Figure \ref{fig:IGR00_lc}), which is compared with that of the excess coincident with IGR00 in Figure \ref{fig:IGR00_com_lc}. Where we see apparent emission in the excess light-curve, a similar flux is observed in the light-curve of the blazar. Furthermore, the two proceeding bins (an upper limit and a flux measurement) are also similar in value to those of the blazar. A third, lower significance, flux measurement is observed later in the light-curve of the IGR00 excess where only an upper limit is observed from the blazar. This is likely a chance fluctuation. 

The light-curve alone suggests that source confusion is likely the cause of the excess at the position of IGR00. Nonetheless, we reanalyse the ROI using the same parameters as described in Table \ref{tbl:params} over the time range of the most significant $\gamma$-ray bin in Figure \ref{fig:IGR00_lc} (MJD 56328 - MJD 56510). Following this reanalysis, we add a point source to the position of IGR00, and perform a localisation fit in order to optimise the positional fit of the excess. 

\begin{figure}
    \centering
    \includegraphics[width=240pt]{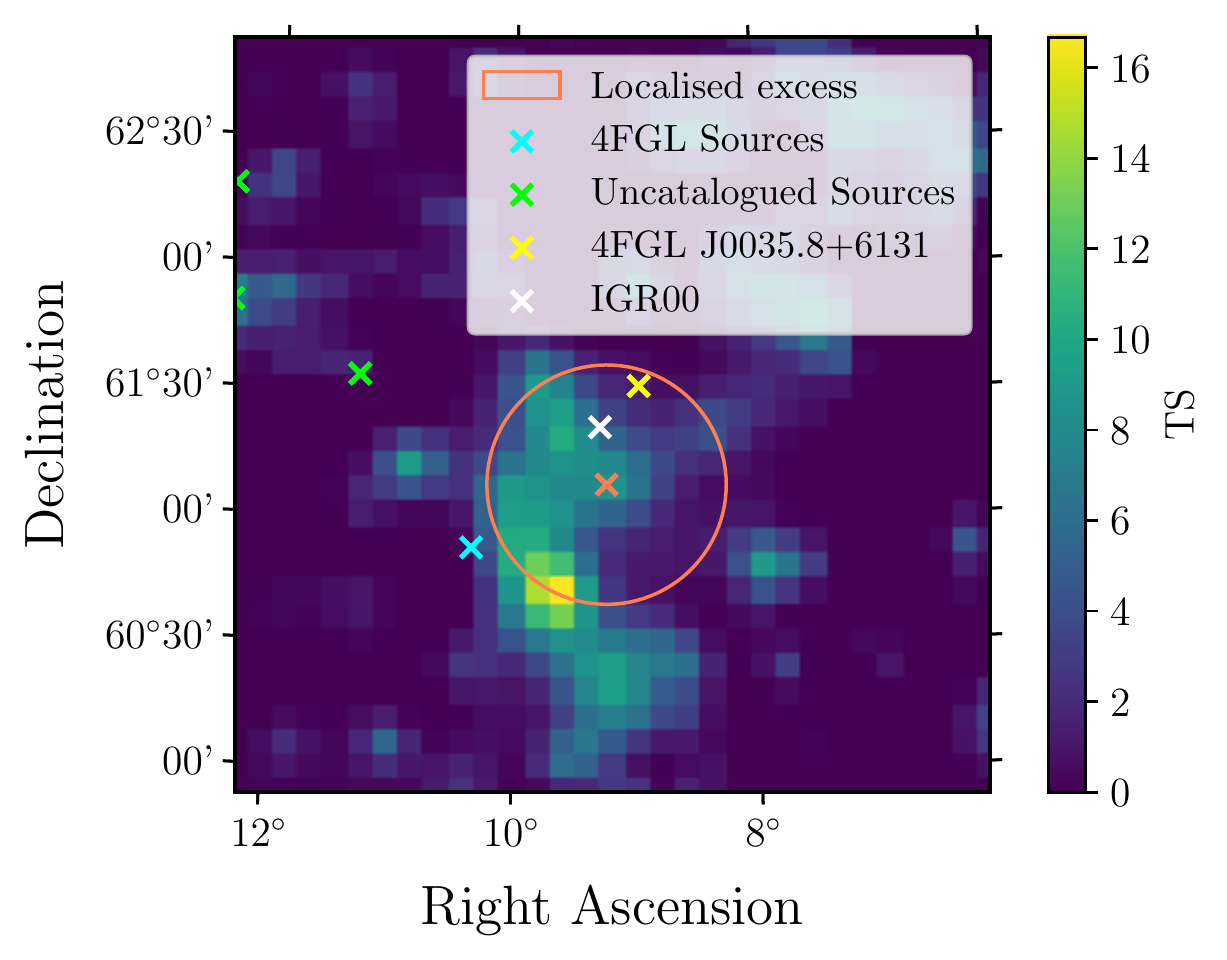}
    \caption{The TS map of the central $3\degree$ of the IGR\,J00370+6122 ROI across the MJD 56328 - MJD 56510 period. Here, the positions of the closest 4FGL sources are indicated by blue crosses, whilst the positions of sources identified with the \texttt{gta.find\_sources} algorithm are indicated by green crosses. The position of IGR\,J00370+6122 is indicated by a white cross, the position of the blazar is indicated by a yellow cross and the localized excess is indicated by an orange cross, with the 95\% positional uncertainty on the excess indicated by the orange circle. As both the blazar and the binary lie within the positional uncertainty of the excess, we conclude it is not possible to determine which object is the cause of the excess.  Bin widths are $0.1 \degree$. }
    \label{fig:IGR00_TS}
\end{figure}

Figure \ref{fig:IGR00_TS} shows the position of the blazar, 4FGL\,J0035.8+6131, together with the position of IGR00 and the positional uncertainty of the localised excess. As both the position of the blazar and the binary lie within the 95\% uncertainty bound, it is impossible to determine which of these is the cause of the excess, or indeed if there is an unresolved source causing it. Therefore we conclude that this $\gamma$-ray excess is unlikely to represent $\gamma$-ray emission from IGR00.

\subsection{Confirmed False Positives}
Of the 20 binaries where we detect either a persistent or transient $\gamma$-ray excess, we determine that 8 of these have significant evidence which suggest that this is a false positive result. Discussions of each of these false positive excesses are included in Appendix \ref{appendix:false_positive}. The binaries with confirmed false positive results are: IGR\,J16320-4751, IGR\,J16358-4726, IGR\,J16465-4507, 1WGA\,J0648.0-4419, AX\,J1740.1-2847, H\,1833-076, GS\,1839-04 and SAX J2103.5+4545. 

The false positives can be broken down into two categories; the first is where we see a significant excess which appears to be coincident with the position of the HMXB in question. Given the significant photon statistics available, we perform a localisation of this excess, and upon examining the new positional fit find that the excess is no longer spatially coincident with the position of the binary, greatly decreasing the likelihood of association between the $\gamma$-ray excess and the HMXB. The second group of false positive excesses are those which are definitively due to source confusion with another source. In this case, this excess cannot be proven to be independent, and therefore cannot be associated with the HMXB in question.

\section{Conclusions}
We survey the positions of 114 high mass X-ray binaries using \textit{Fermi}-LAT, searching for $\gamma$-ray emission in the MeV-GeV energy range. Our survey includes 4 known $\gamma$-ray emitting X-ray binaries, LS 5039, Cygnus X-1, Cygnus X-3 and LS I +61 303. These are all detected more significantly than in the most recent \textit{Fermi}-LAT point source catalogue, the 4FGL-DR2. 

We employ a binned likelihood method, using the \texttt{Fermipy} and \texttt{Fermitools} software together with the catalogues of \citealt{liu_catalogue_2006} in order to carry out our survey, where we test for both persistent $\gamma$-ray emission, and transient $\gamma$-ray emission using a 6 month binning scheme. Where a $\gamma$-ray excess is identified a more detailed analysis is performed. We identify some evidence for $\gamma$-ray emission from the positions of 20 HMXBs where no previous emission has been observed. The significances of the $\gamma$-ray excesses observed varied considerably: from $3 \sigma$ over 12.5 years of observations in the case of IGR\,J16358-4726, to almost $8 \sigma$ in a single 6-month bin in the case of SAX\,J2103.5+4545, without any evidence for persistent emission.

\subsection{Persistent Emission}

Many cases where an excess is observed are likely to be false positives, where a detailed analysis establishes that the excess at the position of the binary is likely to be caused by something else. In particular, in the case of 4 HMXBs with a coincident excess (IGR\,J16358-4726, H\,1833-076, GS\,1839-04 and SAX\,J2103.5+4545), localisation of this excess caused its best fit position to move, rendering the excess no longer spatially coincident with the position of the HMXB.

Given the PSF of the LAT instrument, it is possible for photon contamination to occur between sources within $~ 1 \degree$ of each other, particularly at lower photon energies. Such source confusion is especially prevalent on the Galactic plane. We establish that false positive $\gamma$-ray excesses due to source confusion occur in 4 of 20 cases. These are IGR\,J16320-4751, 1WGA\,J0648.0-4419, AX\,J1740.1-2847 and IGR\,J00370+6122. Additionally, we suspect that source confusion is also responsible for the excesses observed coincident with IGR J16465-4507 and 
IGR J17544-2619, although further evidence is needed to establish whether this is the case. Finally, in the case of 4U\,2206+543 we observe a very soft, isolated excess which displays no variability and which cannot be associated with the binary itself. In this particular case (given that there are no nearby point sources) the excess may be caused by source confusion with the Galactic $\gamma$-ray background. 

In the case of 5 of the excesses, there is a lack of evidence to associate the excess to the binary itself, yet it remains spatially coincident when localised (if photon statistics enable localisation) and there is no evidence for source confusion. In these cases, the $\gamma$-ray excess \textit{may} represent $\gamma$-ray emission from the binary. However, further evidence is needed, particularly evidence of time-variability, which may be associated with orbital phase. Systems where there is a lack of evidence to determine whether the excess is caused by the binary are 1H\,0749-600, 1H\,1238-599, IGR\,J19140+0951, GRO\,J2058+42, W63\,X-1 and also 4U\,2206+543, although as noted above, source confusion with the Galactic background may be a factor here.

\subsection{Transient Emission}

In 2 of the systems, SAX\,J1324.4-6200 (SAX13) and RX\,J2030.5+4751 (RX20), we see evidence for transient $\gamma$-ray emission across timescales of months that cannot be attributed to the $\gamma$-ray background or any known source. In the case of RX20, there is roughly $5 \sigma$ evidence for emission in a 6-month time bin, and furthermore a persistent $\gamma$-ray excess of $5.6 \sigma$. We rule out source confusion as the cause of this excess, and the variability seen in the light-curve suggests that the excess observed comes from a genuine point source of $\gamma$-rays.  Given that the orbit of this system is unknown, the nature of the light-curve and companion star suggests that this system could be a long-orbit $\gamma$-ray binary with the light-curve enhancement caused by the periastron of a likely neutron star and the Be companion star. 

In the case of SAX13, we observe a $5.4 \sigma$ $\gamma$-ray excess which is visible over a continuous 18 month period. This excess is not caused by source confusion with any known source, and localizing the emission reveals that it is consistent with the position of SAX13; it has a power-law spectrum with $\Gamma = -2.43$, which is broadly consistent with the spectral indices of other power-law HMXBs. Given that SAX13 is likely a pulsar-Be star system, and has an unknown orbital period, it is entirely possible that this excess represents $\gamma$-ray emission from around the time of the binary's periastron, which would make SAX13 a $\gamma$-ray binary with a long orbital period. Further, long-term, study and monitoring of this system in both the X-ray and $\gamma$-ray wavebands are needed to test this hypothesis.

In the final 2 systems, GRO\,J1008-57 and 1A\,0535+262 there is tentative evidence for $\gamma$-ray emission which varies by orbital phase. With GRO10, we see a flux measurement in the MJD 58886 - 59069 bin where $z = 3.8 \sigma$, and a less significant bin $z = 3.0 \sigma$ immediately after apastron. A previous study \citealt{xing_likely_2019} suggested that this represents tentative evidence of orbital modulation and that the $\gamma$-ray excess is caused by GRO10. While we find evidence for some modulation with the orbital phase of GRO10, the lack of any measurable $\gamma$-ray emission in the $0.9 < \phi < 1.1$ range makes it harder to reconcile this with a wind-wind collision emission model, where GeV $\gamma$-ray emission would be expected to occur at periastron, although emission peaks are seen at times other than periastron in LSI\,+61\,303.

The case of 1A\,0535+262 presents the strongest evidence for new $\gamma$-ray emission from a HMXB from our survey. 1A\,0535+262 is a pulsar-Be star system, and not a known microquasar, hence any $\gamma$-ray activity would be expected to originate from either wind-wind interactions or, given that 1A05 is strongly accreting during outburst, a novel accretion related method. We observe a marginal persistent excess coincident with 1A05 with $\mathrm{TS} = 12.4$ ($3.5 \sigma$), and find evidence that $\gamma$-ray activity may be coincident with the giant X-ray flares the system undergoes from time-to-time. The `smoking gun' for the $\gamma$-ray excess originating from this binary is that all of the $\gamma$-ray flux is concentrated in the phase bin immediately preceding periastron, with a $3.5 \sigma$ measurement in this bin, and $\approx 0 \sigma$ in all other phase bins. This suggests that the sustained $\gamma$-ray excess we observe across the 12.5 year dataset is all occurring at periastron. The chances of another undiscovered system with this exact periodic behaviour being within the source confusion radius of 1A05 is exceptionally small. Therefore, whilst the significance of the persistent flux is only $3.5 \sigma$, we are reasonably confident that this represents a sub-threshold hint of $\gamma$-ray emission from 1A05.

\subsection{Summary}

Eight HMXBs have confirmed $\gamma$-ray emission and are listed in the 4FGL-DR2, with several other HMXBs being confirmed as $\gamma$-ray emitters but not included in the 4FGL-DR2. In this paper we identify a promising hint of emission from 1A\,0535+262, tentative evidence of $\gamma$-ray excesses from a further 3 HMXBs (SAX\,J1324.4-6200, GRO\,J1008-57 and RX\,J2030.5+4545) and excesses coincident with 5 HMXBs (1H\,0749-600, 1H\,1238-599, IGR\,J19140+0951, GRO\,J2058+42, and W63\,X-1), although there is a lack of evidence to establish these 5 $\gamma$-ray excesses as being products of processes occurring in their respective binary systems. Where we are able to produce spectra of these excesses, all appear to be soft, with a spectrum likely reaching a maximum below the \textit{Fermi}-LAT energy range. This makes these HMXBs ideal targets for instruments with lower energy (i.e. covering the MeV gap), for example the forthcoming mission AMEGO \citep{mcenery_all-sky_2019}. Furthermore, if these excesses represent the dip between the synchrotron and inverse Compton peaks, then these binaries may be detected by instruments with a higher energy detection threshold in the GeV - TeV energy range.

\section*{Acknowledgements}

The authors would like to acknowledge the excellent data and analysis tools provided by the NASA \textit{Fermi} collaboration, without which this work could not be done.  In addition, this work has made use of the SIMBAD database, operated at CDS, Strasbourg, France.  We would like to thank Matthew Capewell for useful discussions, and the anonymous referee for timely and constructive advice. 

MH acknowledges funding from Science and Technology Facilities Council (STFC) PhD Studentship ST/S505365/1, and PMC and CBR acknowledge funding from STFC consolidated grant ST/P000541/1. 

\section*{Data Accessibility}
The \textit{Fermi}-LAT data are all publicly accessible at the NASA LAT data server, located at https://fermi.gsfc.nasa.gov/ssc/data/access/ . For our LAT data analysis we use \texttt{Fermitools v1.2.23} available at https://fermi.gsfc.nasa.gov/ssc/data/analysis/software/ and \texttt{Python 2.7} package \texttt{Fermipy v0.19.0} available at https://fermipy.readthedocs.io/en/latest/install.html . We use the pre-computed \textit{Swift}-BAT daily light-curves available here https://swift.gsfc.nasa.gov/results/transients/ and the AAVSO photometry data available here https://www.aavso.org/data-download . The MAXI data are located at http://maxi.riken.jp/top/lc.html .





\bibliographystyle{mnras}
\bibliography{references.bib}


\appendix
\newpage
\onecolumn
\section{High Mass X-Ray Binary Sample}
\label{sec:appendixA}
\begin{table}[!ht]
\centering
\begin{tabular}{ccccccc}
\hline \hline
Binary Name & Alt. Name & LII & BII & Position Type & Period & Flux U.L. ($\mathrm{MeV}\,\mathrm{cm}^{-2} \, \mathrm{s}^-1$)\\
\hline
1H 1253-761 & HD 109857 & 302.14353 & -12.51748 & O & 0.0 & 4.19$\times$10$^{-8}$ \\
IGR J12349-6434 & RT Cru & 301.15792 & -1.75063 & X & 0.0 & 9.60$\times$10$^{-8}$ \\
2RXP J130159.6-635806 & & 304.08824 & -1.12109 & IR & 0.0 & 8.66$\times$10$^{-8}$ \\
1H 1249-637 & BZ Cru & 301.95802 & -0.20313 & O & 0.0 & 1.36$\times$10$^{-7}$\\
4U 1223-624 & BP CRU ? & 300.09815 & -0.03512 & O & 41.59 & 2.68 $\times$10$^{-7}$\\
2S 1417-624 &  & 313.02125 & -1.59848 & O & 42.12 & 2.59$\times$10$^{-8}$ \\
2S 1145-619 & V801 CEN & 295.61107 & -0.24028 & O & 187.5 & 1.52$\times$10$^{-7}$\\
SAX J1324.4-6200 &  & 306.79301 & 0.60938 & X & 0.0 & N.A. \\
1E 1145.1-6141 & V830 CEN & 295.48987 & -0.00984 & O & 14.4 &  7.37$\times$10$^{-8}$\\
1A 1118-615 & HEN 3-640 & 292.49858 & -0.89174 & O & 0.0 & 2.42$\times$10$^{-7}$\\
4U 1258-61 & V850 CEN & 304.10272 & 1.24742 & IR & 133.0 & 1.32$\times$10$^{-7}$ \\
IGR J11435-6109 &  & 294.8992 & 0.70118 & O & 52.46 & 6.57$\times$10$^{-8}$ \\
1H 0749-600 & HD 65663 & 273.69327 & -16.85817 & IR & 0.0 & N.A \\
1A 1244-604 &  & 302.45915 & 2.22543 & X & 0.0 & 9.76 $\times$10$^{-8}$ \\
4U 1119-603 & V779 CEN & 292.09035 & 0.33556 & X & 2.09 & 4.74$\times$10$^{-8}$\\
1H 1238-599 &  & 301.76038 & 2.65032 & X & 0.0 & N.A.\\
IGR J11215-5952 & HD 306414 & 291.89319 & 1.07292 & O & 0.0 & 7.96$\times$10$^{-8}$\\
1A 1246-588 &  & 302.6956 & 3.74924 & X & 0.0 & 2.28$\times$10$^{-7}$\\
GRO J1008-57 & & 282.99992 & -1.82173 & O & 135.0 & N.A \\
1H 1255-567 & MU.02 CRU & 303.36468 & 5.70047 & O & 0.0 & 1.03$\times$10$^{-7}$ \\
RX J1037.5-5647 & LS 1698 & 285.35291 & 1.4326 & X & 0.0 & 2.32$\times$10$^{-8}$ \\
XTE J1543-568 &  & 324.95121 & -1.46121 & X & 75.56 & 6.93$\times$10$^{-8}$\\
1H 1555-552 & HD 141926 & 326.97618 & -1.23892 & O & 0.0 & 2.45$\times$10$^{-7}$ \\
2S 1553-542 &  & 327.94412 & -0.85703 & X & 30.6 & 2.83$\times$10$^{-8}$\\
1H 0739-529 & HD 63666 & 266.31268 & -13.72584 & O & 0.0 & 3.66$\times$10$^{-8}$ \\
4U 1538-52 & QV NOR & 327.41949 & 2.1637 & O & 3.73 & 4.37$\times$10$^{-8}$\\
IGR J16195-4945 & HD 146628? & 333.53932 & 0.33326 & X & 0.0 & 3.56$\times$10$^{-7}$ \\
IGR J16318-4848 & & 335.61599 & -0.44776 & O & 0.0 & 2.46$\times$10$^{-8}$\\
IGR J16283-4838 &  & 335.32671 & 0.10203 & X & 0.0 & 2.03$\times$10$^{-7}$\\
IGR J16320-4751 &  & 336.32997 & 0.16892 & X & 8.96 & N.A.\\
IGR J16358-4726 &  & 337.12311 & -0.00089 & X & 0.0 & N.A. \\
AX J1639.0-4642 &  & 338.00124 & 0.07508 & IR & 0.0 & 4.89$\times$10$^{-7}$\\
IGR J16418-4532 &  & 339.19309 & 0.51511 & X & 0.0 & 4.72$\times$10$^{-8}$ \\
IGR J16479-4514 &  & 340.16406 & -0.12466 & X & 0.0 & 6.75$\times$10$^{-7}$\\
IGR J16465-4507 &  & 340.05357 & 0.13505 & IR & 0.0 & N.A. \\
1WGA J0648.0-4419 & HD 49798 & 253.70642 & -19.1412 & O & 1.55 & N.A. \\
IGR J16493-4348 &  & 341.37079 & 0.60251 & X & 0.0 & 1.67$\times$10$^{-7}$\\
GS 0834-430 &  & 262.02096 & -1.51074 & O & 105.8 & 4.05$\times$10$^{-8}$\\
AX J1700-419 &  & 344.04452 & 0.23717 & X & 0.0 & 1.70$\times$10$^{-7}$\\
OAO 1657-415 &  & 344.36915 & 0.31918 & X & 10.4 & 1.01$\times$10$^{-7}$\\
4U 0900-40 & HD 77581 & 263.05839 & 3.92993 & O & 8.96 & 3.22$\times$10$^{-8}$\\
4U 1700-37 & HD 153919 & 347.75446 & 2.1734 & O & 3.41 & 1.10$\times$10$^{-7}$\\
IGR J17091-3624 &  & 349.52595 & 2.23569 & R & 0.0 & 1.50$\times$10$^{-7}$\\
EXO 1722-363 &  & 351.49727 & -0.35395 & IR & 9.74 & 8.00$\times$10$^{-8}$\\
RX J0812.4-3114 & LS 992 & 249.57107 & 1.5477 & IR & 81.3 & 1.81$\times$10$^{-8}$\\
XTE J1739-302 &  & 358.06784 & 0.44517 & X & 0.0 & 6.75$\times$10$^{-7}$\\
RX J1739.4-2942 &  & 358.64668 & 0.73046 & X & 0.0 & 2.37$\times$10$^{-7}$\\
AX J1740.1-2847 &  & 359.49377 & 1.08387 & X & 0.0 & 1.08$\times$10$^{-7}$\\
AX J1749.1-2733 &  & 1.5827 & 0.06234 & X & 0.0 & 6.49$\times$10$^{-7}$\\
AX J1749.2-2725 &  & 1.701 & 0.1157 & X & 0.0 & 4.06$\times$10$^{-7}$\\
RX J1744.7-2713 & V3892 SGR & 1.35781 & 1.05224 & O & 0.0 & 2.09$\times$10$^{-7}$\\
GRO J1750-27 &  & 2.37283 & 0.50774 & X & 29.8 & 1.24$\times$10$^{-7}$\\

\hline

\end{tabular}
\end{table}

\newpage

\begin{table}[!ht]
\centering
\begin{tabular}{ccccccc}
\hline \hline
Binary Name & Alt. Name & LII & BII & Position Type & Period & Flux U.L. ($\mathrm{MeV}\,\mathrm{cm}^{-2} \, \mathrm{s}^-1$)\\
\hline
IGR J17544-2619 & & 3.23599 & -0.33559 & IR & 0.0 & N.A.\\
3A 0726-260 & V441 PUP & 240.28165 & -4.05037 & O & 34.5 & 2.91$\times$10$^{-8}$\\
SAX J1819.3-2525 & V4641 SGR & 6.75638 & -4.79765 & O & 2.8 & 9.69$\times$10$^{-8}$\\
SAX J1802.7-2017 &  & 9.41747 & 1.04356 & X & 4.6 & 1.05$\times$10$^{-7}$\\
SAX J1818.6-1703 & HD 168078? & 14.07813 & -0.71028 & X & 0.0 & 1.04$\times$10$^{-7}$\\
RX J1826.2-1450 & LS 5039 & 16.88157 & -1.28923 & R & 3.9 & N.A.\\
AX J1820.5-1434 &  & 16.47185 & 0.06991 & X & 0.0 & 2.61$\times$10$^{-7}$\\
4U 1807-10 &  & 18.60547 & 3.85183 & X & 0.0 & 2.91$\times$10$^{-7}$\\
XTE J1829-098 &  & 21.69699 & 0.2786 & X & 0.0 & 3.42$\times$10$^{-7}$\\
H 1833-076 &  & 24.46252 & -0.16075 & X & 0.0 & N.A\\
XTE J0658-073 & [M81] I-33 & 220.12859 & -1.76725 & O & 0.0 & 6.26$\times$10$^{-8}$\\
AX J1838.0-0655 &  & 25.23678 & -0.19035 & X & 0.0 & 1.47$\times$10$^{-7}$\\
GS 1839-06 &  & 26.61754 & -0.50815 & X & 0.0 & 1.05$\times$10$^{-7}$\\
AX J1841.0-0536 &  & 26.76429 & -0.23879 & O & 0.0 & 1.52$\times$10$^{-7}$\\
AX 1845.0-0433 &  & 28.14552 & -0.65617 & X & 0.0 & 2.62$\times$10$^{-8}$\\
GS 1839-04 &  & 27.87383 & 0.11023 & X & 0.0 &N.A. \\
IGR J18483-0311 &  & 29.74117 & -0.75374 & X & 0.0 & 8.30$\times$10$^{-8}$\\
GS 1855-02 &  & 31.24518 & -2.70507 & X & 0.0 & 1.25$\times$10$^{-7}$\\
XTE J1855-026 &  & 31.07627 & -2.09629 & X & 6.067 & 5.84$\times$10$^{-8}$\\
2S 1845-024 &  & 30.42054 & -0.40562 & X & 241.0 & 7.22$\times$10$^{-7}$\\
GS 1843+009 &  & 33.03653 & 1.6896 & O & 0.0 & 5.24$\times$10$^{-8}$\\
XTE J1901+014 &  & 35.38393 & -1.61576 & O & 0.0 & 2.16$\times$10$^{-7}$\\
4U 1901+03 &  & 37.1618 & -1.25 & X & 22.58 & 4.05$\times$10$^{-8}$\\
XTE J1858+034 &  & 36.80574 & -0.02467 & O & 0.0 & 4.31$\times$10$^{-7}$\\
3A 1909+048 & SS 433 & 39.69421 & -2.2447 & O,R & 13.1 & 2.13$\times$10$^{-7}$\\
SAX J0635.2+0533 &  & 206.15276 & -1.04229 & O & 11.2 & 1.48$\times$10$^{-7}$\\
4U 1909+07 &  & 41.89715 & -0.81151 & IR & 4.4 & 1.04$\times$10$^{-7}$\\
XTE J1859+083 &  & 41.14664 & 2.06249 & X & 0.0 & 8.31$\times$10$^{-8}$\\
XTE J1906+09 & & 42.49725 & 1.17483 & O & 0.0 & 1.21$\times$10$^{-7}$\\
4U 1907+09 & & 43.74539 & 0.47045 & O & 8.38 & 1.04$\times$10$^{-7}$\\
IGR J19140+0951 &  & 44.29629 & -0.46868 & X & 13.558 & 3.68$\times$10$^{-7}$\\
IGR J06074+2205 &  & 188.38516 & 0.81378 & O & 0.0 & 1.43$\times$10$^{-7}$\\
1A 0535+262 & V725 TAU & 181.44505 & -2.64342 & O & 111.0 & N.A.\\
XTE J1946+274 &  & 63.20703 & 1.39573 & O & 169.2 & 1.57$\times$10$^{-7}$\\
1H 0556+286 & HD 249179 & 181.28416 & 1.85966 & O & 0.0 & 6.33$\times$10$^{-8}$\\
KS 1947+300 &  & 66.08802 & 2.09797 & O & 40.4 & 7.62$\times$10$^{-8}$\\
4U 0352+309 & X PER & 163.07842 & -17.1335 & O & 250.3 & 6.67$\times$10$^{-8}$\\
4U 1956+35 & Cyg X-1 & 71.33508 & 3.0668 & O,R & 5.0 & N.A.\\
EXO 2030+375 &  & 77.15175 & -1.24157 & O & 46.02 & 6.64$\times$10$^{-8}$\\
EXO 051910+3737.7 & V420 AUR & 170.05323 & 0.71029 & O & 0.0 & 2.16$\times$10$^{-7}$\\
4U 2030+40 & Cyg X-3 & 79.8426 & 0.69512 & R & 0.2 & N.A.\\
GRO J2058+42 &  & 83.56978 & -2.65543 & O & 55.03 & N.A.\\
RX J0440.9+4431 & LS V +44 17 & 159.84708 & -1.27013 & O & 0.0 & 4.30$\times$10$^{-8}$\\
W63 X-1 &  & 82.31659 & 5.42815 & X & 0.0 & N.A.\\
SAX J2103.5+4545 &  & 87.12993 & -0.68507 & O & 12.68 & 6.76$\times$10$^{-8}$\\
RX J2030.5+4751 & SAO 49725 & 85.2307 & 5.04764 & O & 0.0 & N.A.\\
1H 2202+501 & BD +49 3718 & 97.24782 & -4.04112 & O & 0.0 & 2.81$\times$10$^{-8}$\\
V 0332+53 & BQ CAM & 146.05195 & -2.19388 & O & 34.25 & 7.04$\times$10$^{-8}$\\
1H 1936+541 & DM +53 2262 & 85.84987 & 15.9024 & O & 0.0 & 3.60$\times$10$^{-8}$\\
4U 2206+543 & BD +53 2790 & 100.60312 & -1.10596 & O & 9.57 & N.A.\\
XTE J0421+560 & CI CAM & 149.17637 & 4.13342 & O & 19.41 & 5.95$\times$10$^{-8}$\\
1H 2138+579 & V490 CEP & 99.01253 & 3.31318 & O & 0.0 & 9.50$\times$10$^{-8}$\\
2S 0053+604 & GAMMA CAS & 123.57681 & -2.14848 & O & 203.59 & 1.01$\times$10$^{-7}$\\
1E 0236.6+6100 & LS I +61 303 & 135.67529 & 1.08626 & O,R & 26.496 & N.A.\\
SAX J2239.3+6116 &  & 107.73456 & 2.36233 & O & 262.0 & 8.39$\times$10$^{-8}$\\
RX J0146.9+6121 & LS I +61 235 & 129.54108 & -0.8001 & O & 0.0 & 1.19$\times$10$^{-7}$\\
IGR J00370+6122 & BD +60 73 & 121.22213 & -1.46464 & O & 15.665 & 1.48$\times$10$^{-7}$\\
\hline
\end{tabular}
\label{tbl:sample}
\end{table}

\newpage

\begin{table}[!ht]
\centering
\begin{tabular}{ccccccc}
\hline \hline
Binary Name & Alt. Name & LII & BII & Position Type & Period & Flux U.L. ($\mathrm{MeV}\,\mathrm{cm}^{-2} \, \mathrm{s}^-1$)\\
\hline

4U 0115+634 & V635 CAS & 125.92366 & 1.02574 & O & 24.3 & 6.47$\times$10$^{-8}$\\
2S 0114+650 & V662 CAS & 125.70998 & 2.56353 & O & 11.6 & 5.65$\times$10$^{-8}$\\
IGR J01363+6610 & EM* GGR 212 & 127.39482 & 3.7248 & O & 0.0 & 4.44$\times$10$^{-8}$\\
IGR J01583+6713 &  & 129.35216 & 5.18871 & O & 0.0 & 1.08$\times$10$^{-7}$\\
\hline
\end{tabular}

\caption{A table of each HMXB included in the survey; the binary name column gives the X-ray name of each X-ray binary included in the survey from the Liu et al. catalogue, and the alternate name column gives either the optical name of the system from Liu et al. or the most commonly used name for the system (i.e. Cyg X-1). LII and BII are the Galactic coordinates of the system used in our data analysis; the position type gives the waveband from which the coordinates were taken (for example, O corresponds to the coordinates of the optical counterpart, X to the X-ray, etc.). The period refers to the orbital period of the binary system measured in days; where no binary period is known a value of 0 is shown. The Flux U.L value refers to the 95\% upper confidence limit on energy flux (with units of $\mathrm{MeV}\,\mathrm{cm}^{-2} \, \mathrm{s}^-1$) obtained from the position of each HMXB where no measurement of a $\gamma$-ray flux is made. Where a measurement of energy flux is made, these are shown in the Survey Results section. }
\label{tbl:sample}
\end{table}

\newpage

\twocolumn

\section{False Positive $\gamma$-ray Excesses}
\label{appendix:false_positive}
\subsection{IGR\,J16320-4751}
\label{bin:IGR1632}
\begin{figure}[h!]
    \centering
    \includegraphics[width=240pt]{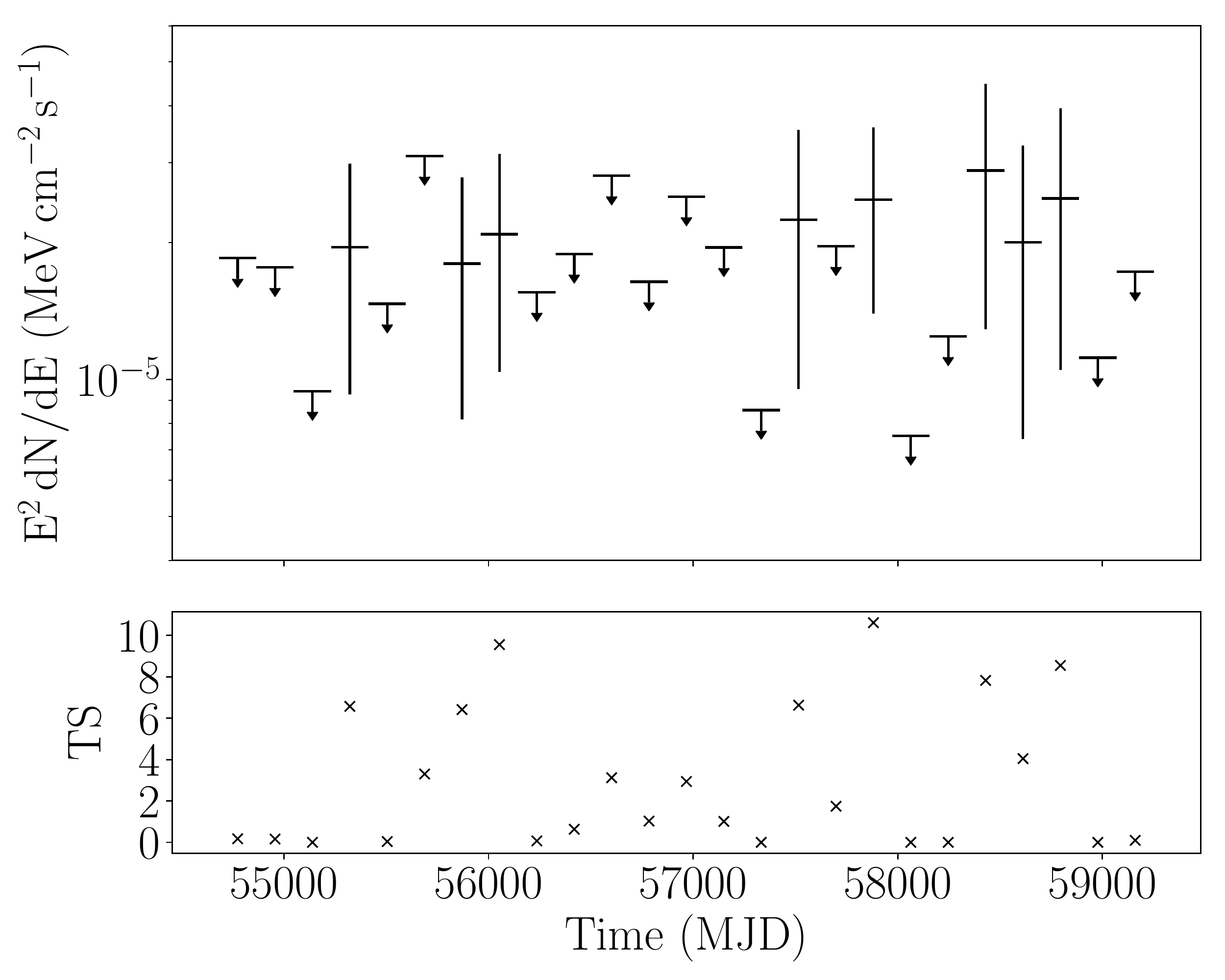}
    \caption{The \textit{Fermi}-LAT light-curve of the $\gamma$-ray excess coincident with IGR\,J16320-4751 with time bins of 6 month width. Upper limits are placed on any time bin where the TS of that bin is less than 4.}
    \label{fig:IGR1632_lc}
\end{figure}

IGR\,J16320-4751 (henceforth IGR1632) is a system thought to be a pulsar in orbit with either a giant K-type, or O/B, star \citep{rodriguez_xmm-newton_2003}. It is notable for its variable nature in the X-ray waveband and particularly short orbital period of approximately 9 days \citep{garcia_spectral_2018}. 
Over the 12.5 year observation time, we obtain a TS of 31.1 in support of the hypothesis that there is a $\gamma$-ray source at this position, with numerous bins in the 6-month binned light-curve (Figure \ref{fig:IGR1632_lc}) with $2 \sigma \leq z < 3 \sigma$, but no clear evidence for variability. 

\begin{figure}
    \centering
    \includegraphics[width=240pt]{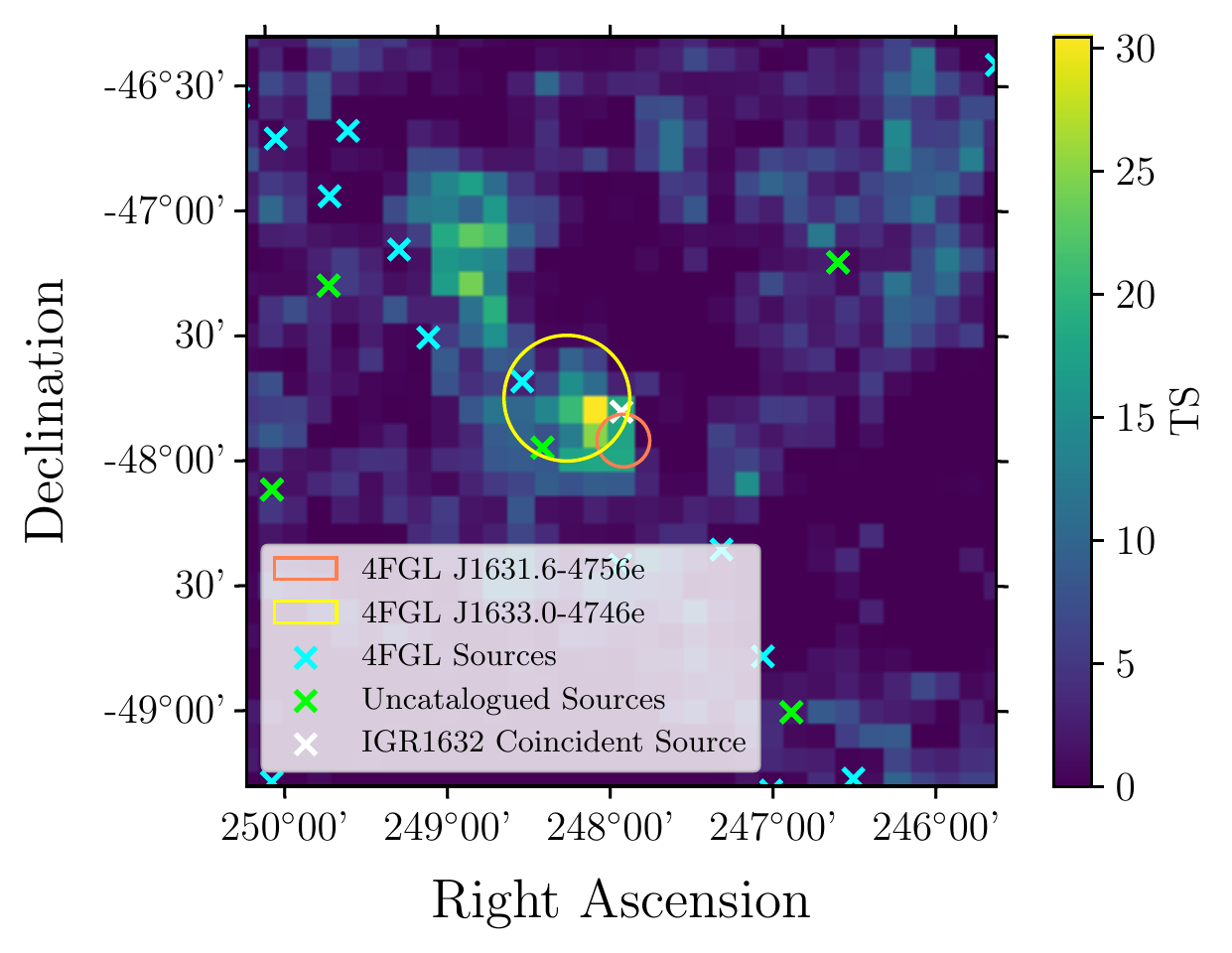}
    \caption{A TS map showing the excess coincident with IGR~J16320-4751 (white cross) and the position of the binary, the two extended sources, and other 4FGL sources (blue crosses), and additional sources identified by the \texttt{gta.find\_sources} algorithm (green crosses).} 
    \label{fig:IGR1632_TS}
\end{figure}

\begin{figure}
    \centering
    \includegraphics[width=240pt]{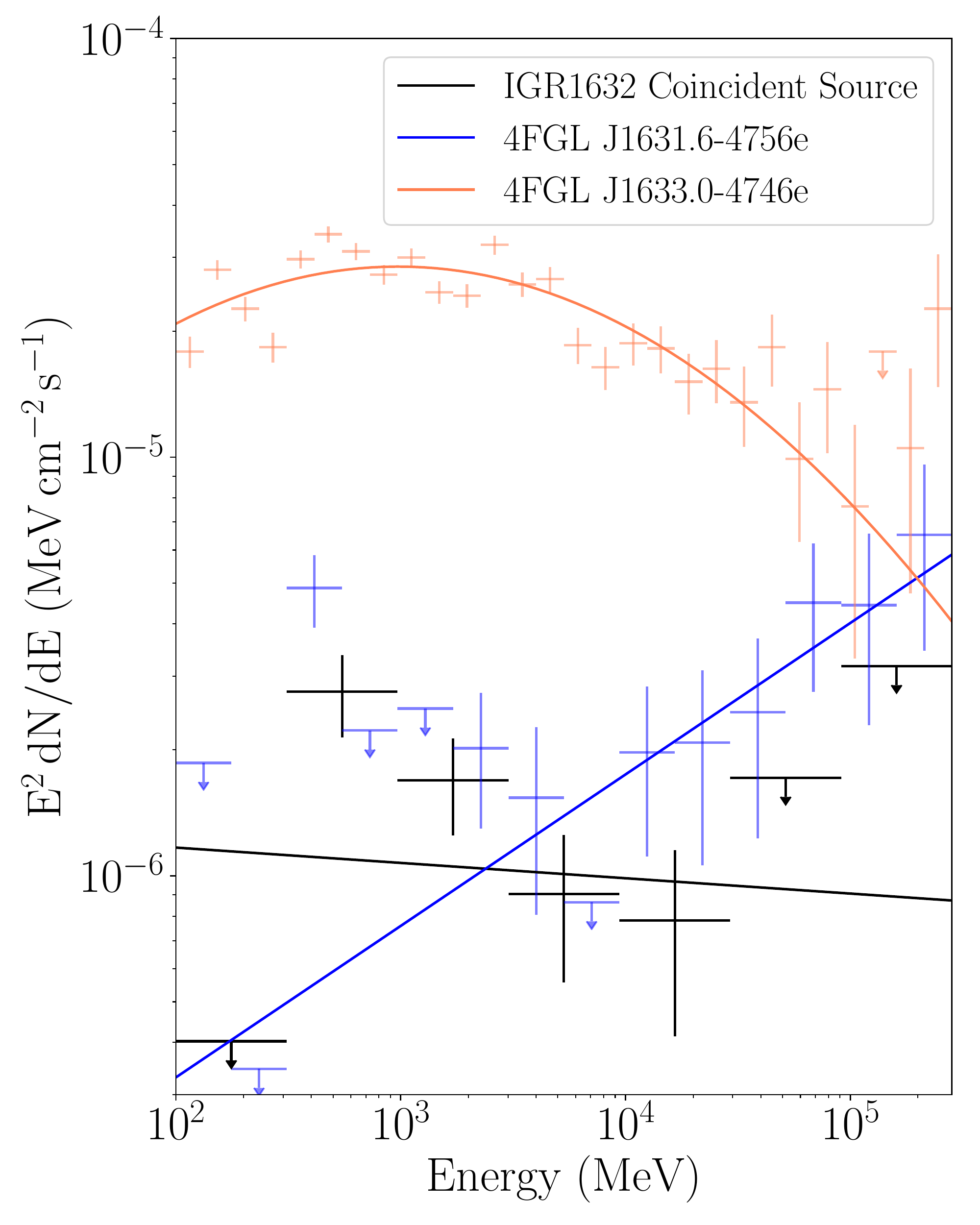}
    \caption{The SEDs of the added source coincident with IGR~J16320-4751 (black), and the two extended sources within which it lies: 4FGL\,J1631-4756e (blue) and 4FGL\,J1633.0-4746e (orange). The IGR~J16320-4751 coincident source and 4FGL\,J1631-4756e are both fitted with power-law spectral models, and 4FGL\,J1633.0-4746e is fitted with a log-parabola. The number of bins per decade is chosen based on the available photon statistics for each source and upper limits are fitted to any bin with $\mathrm{TS} < 4$.} 
    \label{fig:IGR1632_sed}
\end{figure}

Whilst it is possible that this coincident $\gamma$-ray point source is associated with IGR1632, we strongly suspect source confusion has caused a false detection, as IGR1632 lies $0.087 \degree$ away from 4FGL\,J1631.6-4756e ($\mathrm{TS} = 47.3$) and $0.198 \degree$ away from 4FGL\,J1633.0-4746e ($\mathrm{TS} = 4240$). Both sources are associated in the 4FGL with the TeV PWN HESS\,J1632-478 \citep{aharonian_hess_2006}. Such proximity between any two sources can see one source (usually the more significant) contaminating the second with photons. IGR1632 is extremely close to \textit{two} significant and extended sources rather than a single point source, and lies within our calculated radius of the larger and more significant of the two, 4FGL\,J1633.0-4746e. Figure \ref{fig:IGR1632_TS} shows the extent of the coincident $\gamma$-ray excess with the position of IGR1632 overlaid, and the extent of the two extended sources shown to highlight the positional coincidence between these and the IGR1632 source. 

We generate light-curves of both extended sources to compare against the IGR1632 light-curve to identify any correlated variability (a signature of source confusion), or differences in variability (a signature of source independence); however, neither of the extended sources is significantly variable (4FGL\,J1633.0-4746e is variable at the $2.00 \sigma$ level, 4FGL\,J1631.6-4756e at the $0.20 \sigma$ level), and no correlations are observed between these light-curves and the IGR1632 coincident source light-curve. Therefore, the light-curve calculation does neither supports nor rejects the hypothesis that the IGR1632 excess is the product of source confusion. 

We generate an SED of the IGR1632 coincident source across the entire energy range we employ, and find that the best fit to the data is a power law with spectral index $\Gamma = -2.04 \pm 0.09$, which indicates an almost flat spectrum source, although we note that this fit to the data appears to be poor. This SED is illustrated in Figure \ref{fig:IGR1632_sed}. We also generate SEDs of 4FGL\,J1633.0-4746e, which has a log-parabola spectral shape, and 4FGL\,J1631.6-4756e which is best modelled by a hard power law. The 4FGL indicates a power law spectral index of $\Gamma = 2.17$ for the more significant of the two sources, 4FGL\,J1633.0-4746e, marginally softer than that of the IGR1632 coincident source. It is highly likely that the spectrum for this coincident source is essentially that of 4FGL\,J1633.0-4746e with a smaller normalisation and a contaminating component from 4FGL\,J1631.6-4756e which slightly hardens the spectral index. Given this, and the fact that the predicted counts of both extended sources (in a reference model) decrease by roughly the same number of counts as is predicted for the IGR1632 coincident source, we conclude that this apparent source is a false positive, and the emission originates from the extended 4FGL sources.

\subsection{IGR\,J16358-4726}
\label{bin:IGR1635}
\begin{figure}
    \centering
    \includegraphics[width=240pt]{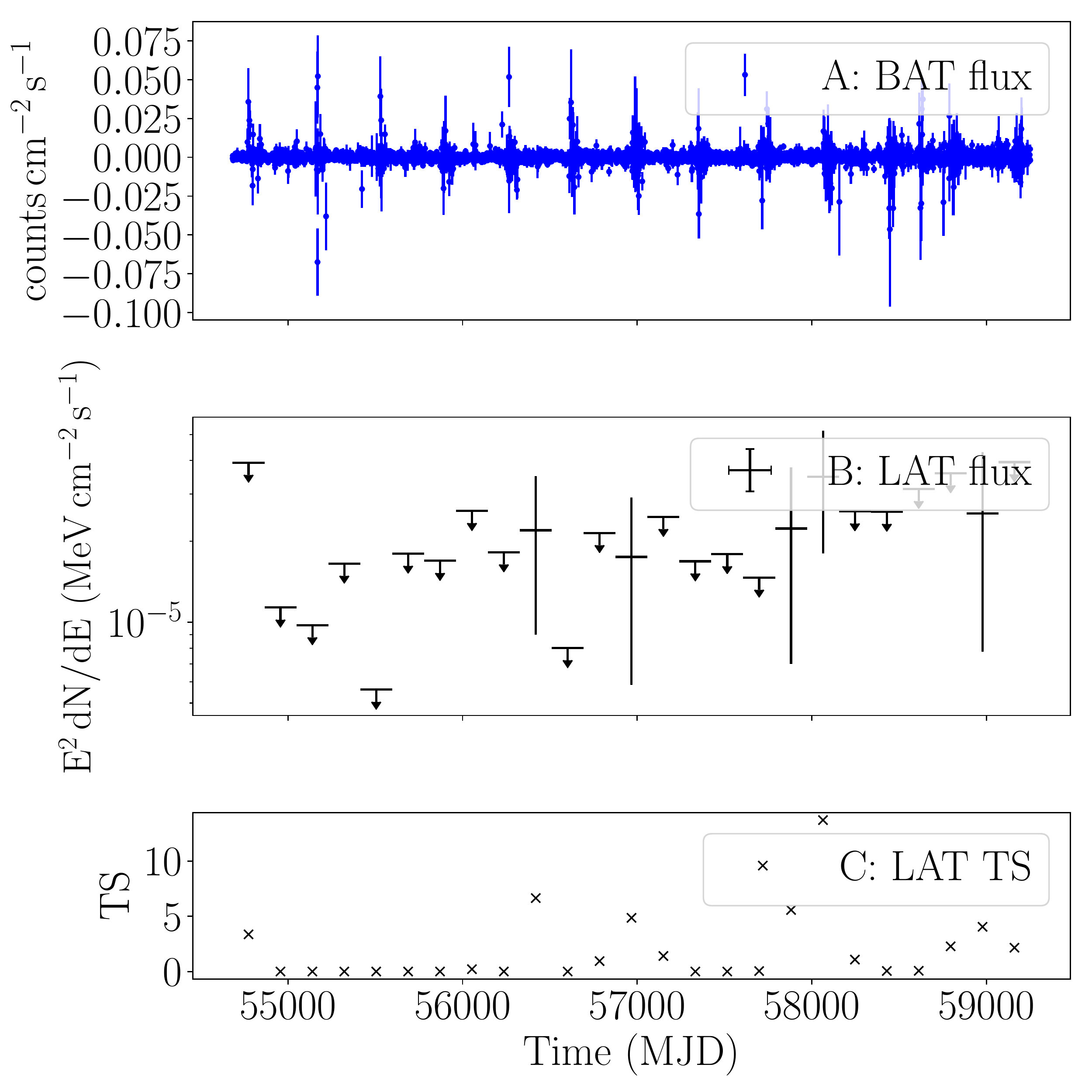}
    \caption{The daily binned light-curve of IGR\,J16358-4726 taken with \textit{Swift}-BAT shown in Panel A, with the calculated monthly-binned \textit{Fermi}-LAT light-curve for a source fitted to the position of IGR\,J16358-4726 shown below in Panel B, and the corresponding TS values of these bins shown in Panel C. We place 95\% confidence limits on any \textit{Fermi}-LAT energy flux bins with $\mathrm{TS} < 4$. Whilst some weak periodic activity is seen in the \textit{Swift}-BAT light-curve, this does not correspond to any known timescales for the system, nor does it correlate with the $\gamma$-ray light-curve.} 
    \label{fig:IGR1635_lc}
\end{figure}

IGR\,J16358-4728 (\citealt{revnivtsev_igr_2003} and \citealt{bodaghee_chandra_2012}) (henceforth IGR1635) is either a HMXB in which a pulsar orbits a super-giant Be star (\citealt{chaty_multi-wavelength_2008} and \citealt{rahoui_multi-wavelength_2008}), or, based on the presence of CO lines in the K-band spectrum making the companion star a KM giant, a symbiotic LMXB \citep{nespoli_k-band_2010}. Under this former hypothesis, one might expect $\gamma$-ray emission at periastron. We detect a $\gamma$-ray excess coincident with this source with $\mathrm{TS} = 9.5$ over the LAT observation period, and also measure a flux value where $\mathrm{TS} > 4$ in 5 light-curve bins. This light-curve is shown together with the daily \textit{Swift}-BAT light-curve in Figure \ref{fig:IGR1635_lc}. Of the 5 bins which exceed the $\mathrm{TS} > 4$ threshold in this light-curve, only one exceeds the $3\sigma$ level. This bin, and the preceding bin are the only adjacent bins, making it difficult to perform any subsequent analysis of this source. 

It seems likely that the origin of this weak $\gamma$-ray excess is source confusion with the nearby, significantly extended source 4FGL~J1636.3-4731e (SNR\,G337.0-00.1, $\mathrm{TS} = 1500$) as they are separated by an angular offset of $0.087 \degree$ and 4FGL J1636.3-4731e is extended by a radius of $0.11 \degree$, so the apparent excess is within the $\gamma$-ray extension of this SNR. We generate a light-curve for this source and find that it is not variable, with the exception of one particularly low flux point. Both the IGR1635 coincident source and the extended source are well-fitted by a steady source model.

We do not have sufficient photon statistics in the IGR1635 coincident excess for a comparative spectral analysis with 4FGL J1636.3-4731e. As we cannot effectively distinguish one source from the other with variability we cannot prove the independence of the IGR1635 excess from 4FGL J1636.3-4731e, nor associate this excess with the binary in question, hence we conclude it is likely a false positive caused by source confusion.

\subsection{IGR\,J16465-4507}
\label{bin:IGR1646}
\begin{figure}
    \centering
    \includegraphics[width=240pt]{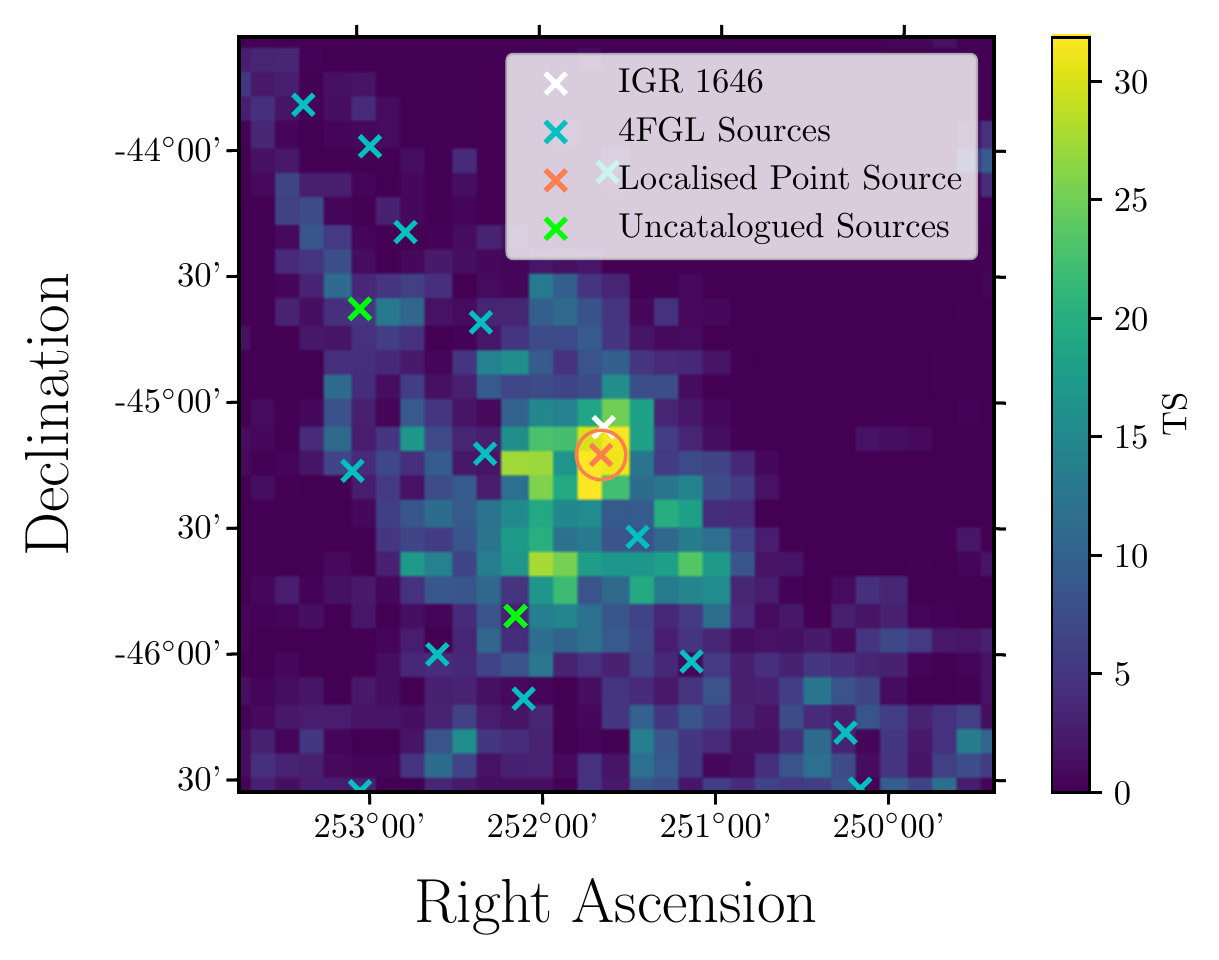}
    \caption{A TS map with $0.1 \degree$ bins, over the full 12.5 year observation time, showing the $\gamma$-ray excess and with the localised point source shown by the orange cross. The orange circle shows the 95\% positional uncertainty on the point source, after localisation. The white cross indicates the infrared position of IGR\,J16465-4507. The blue crosses show the position of sources from the 4FGL-DR2, and the green crosses show the position of uncatalogued sources identified by the \texttt{gta.find\_sources} algorithm. Note the presence of a 4FGL source (a blue cross) behind the legend of this figure.} 
    \label{fig:IGR1646_TS}
\end{figure}

IGR J16465-4507 (henceforth IGR1646) (\citealt{lutovinov_ntegral_2004} and \citealt{romano_100-month_2014}) is a supergiant HMXB system consisting of a Be star \citep{negueruela_search_2007} and a pulsar \citep{lutovinov_discovery_2005} orbiting each other with a period of 30.2 days \citep{la_parola_detection_2010}. We detect a point source coincident with the position of IGR1646 to $\mathrm{TS} = 50.8$, equivalent to $z = 7.1 \sigma$, the most significant of all the coincident sources and excesses in our survey. The nearest sources to the IGR1646 coincident source are 4FGL\,J1645.8-4533c (angular offset = $0.456 \degree$, $\mathrm{TS} = 417.93$) which is tentatively associated with the LMXB 4U 1642-45. Further sources near the IGR1646 coincident source include 4FGL J1649.2-4513c (angular offset = $0.456 \degree$, $\mathrm{TS} = 417.93$) and 4FGL J1649.3-4441 (angular offset = $0.642 \degree$, $\mathrm{TS} = 205.14$). Neither source has a multi-wavelength counterpart.

Figure \ref{fig:IGR1646_TS} shows the TS map of the central part of the IGR1646 ROI, centered on the position of IGR1646. Given the significance of the recorded point source, we are able to localise the emission with the \texttt{gta.localize} algorithm and refit our point source to the peak of the $\gamma$-ray emission. We find the best positional fit for the point source is $\mathrm{LII} = 339.9764 \degree  \pm 0.0324 \degree $, $\mathrm{BII} = 0.0557 \degree  \pm 0.0498 \degree$. There is thus an angular offset from the position of IGR1646 of $0.1107 \degree$, compared to a 95\% positional uncertainty of $0.0981 \degree$. The TS of the source also increases from 50.8 to 59.7 in its new position. Thus, following localisation we no longer consider this point source to be spatially coincident with the binary system and find it more likely that this new point source is either a product of source confusion with one of the nearby 4FGL sources, or an unknown new source which is unlikely to be associated with IGR1646.

\begin{figure}[h!]
    \centering
    \includegraphics[width=240pt]{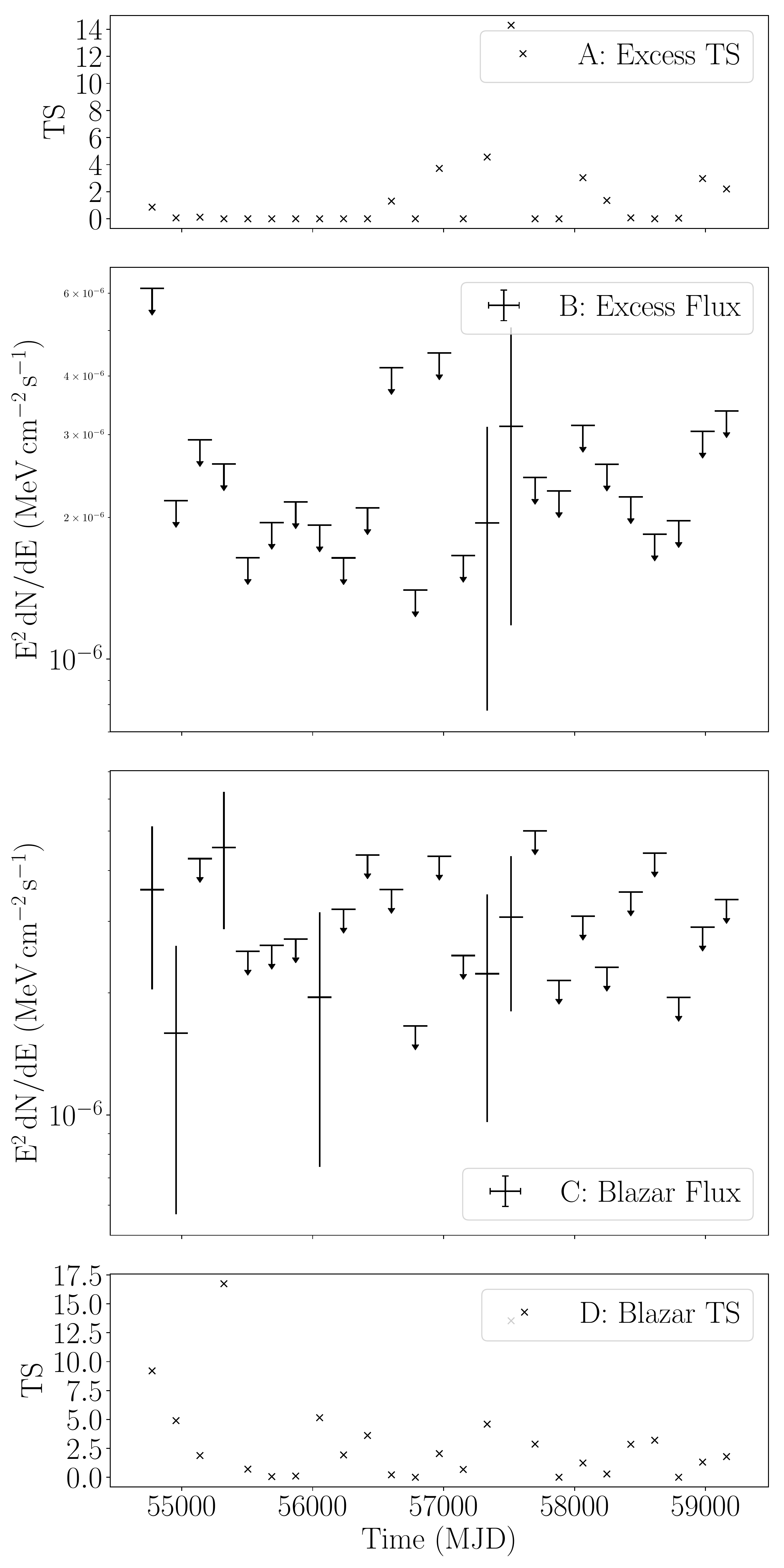}
    \caption{Panels B and A show the $\gamma$-ray flux and associated TS values of these flux points for the excess coincident with the optical position of 1WGA\,J0648.0-4419. Panels C and D show the $\gamma$-ray flux and associated TS values of these flux values for the nearby blazar 4FGL\,J0647.7-4418, without the excess in the model. We use approximately 6 month bins in each of these light-curves, and 95\% confidence upper limits on flux are used for any bin where the corresponding TS value is less than 4. } 
    \label{fig:1WGA06_lc}
\end{figure}

\subsection{1WGA\,J0648.0-4419}
\label{bin:1WGA06}
1WGA\,J0648.0-4419 (also known as HD 49798, and henceforth referred to as 1WGA06) is a binary system consisting of a pulsar \citep{israel_discovery_2009} and an O-type sub-dwarf star \citep{kupfer_first_2020}. We find a $\gamma$-ray excess coincident with the position of 1WGA06 with a TS of 18.5, giving a $z$-score of $4.3 \sigma$. The nearest 4FGL source to 1WGA06 is 4FGL\,J0647.7-4418 (associated with the blazar SUMSS J064744-441946), with an angular offset of $0.068 \degree$, less than one spatial bin width from the position of the binary. With a point source fitted to the position of 1WGA06, the blazar is only marginally detected, with a TS of 8.18 (slightly below $3 \sigma$). This is far less than the detection significance given in the 4FGL-DR2, which is $8.3 \sigma$. It is possible that the $\gamma$-ray excess observed at the position of 1WGA06 is due to source confusion with this blazar; temporal variability is used to test whether this is the case. 

A similar occurrence of source confusion between a $\gamma$-ray excess at the position of an X-ray binary (in this case V404 Cygni), and a blazar is dealt with in \cite{harvey_v404_2021}, where the apparent $\gamma$-ray excess was actually originating from a nearby flaring blazar. Given that we measure only two $\gamma$-ray flux points in the 6-month binned light-curve of the 1WGA06 excess, we can remove this excess from the model and generate a light-curve of the blazar to see whether these $\gamma$-ray flux points would otherwise be attributed to the blazar. If this is the case, then the excess at the position of 1WGA06 is not independent of the blazar, and source confusion is occurring. Figure \ref{fig:1WGA06_lc} shows both the light-curve of the 1WGA06 excess and the light-curve of the blazar. Given that we see only two $\gamma$-ray flux measurements amongst an otherwise complete set of upper limits for the excess, and we see corresponding flux points in the light-curve of the blazar, the $\gamma$-ray excess coincident with 1WGA06 is likely due to source confusion with the blazar.

\begin{figure}[h!]
    \centering
    \includegraphics[width=240pt]{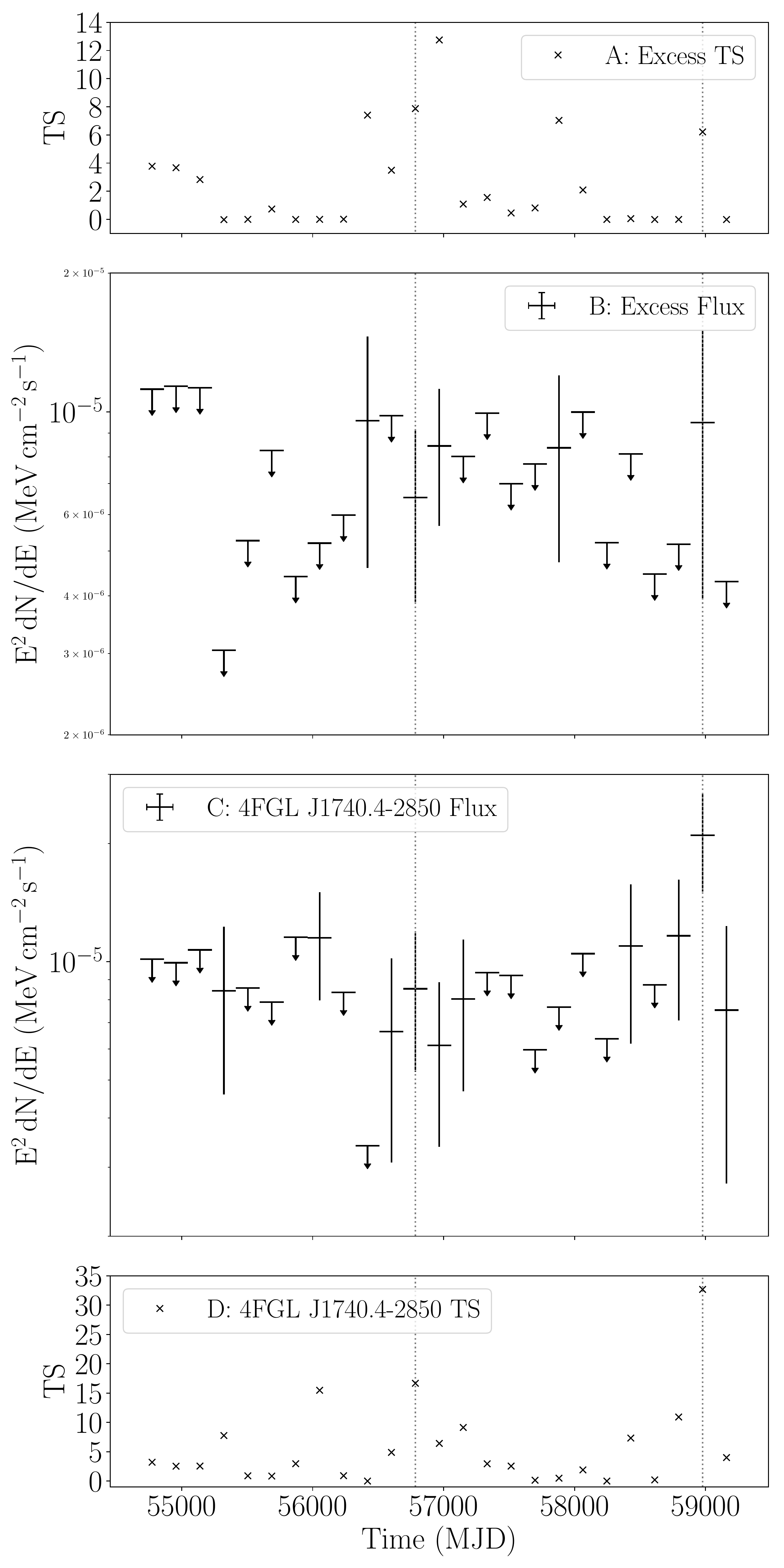}
    \caption{Panels B and A show the $\gamma$-ray flux and associated TS values of these flux points for the excess coincident with the optical position of AX\,J1740.1-2847. Panels C and D show the $\gamma$-ray flux and associated TS values of these flux values for the nearby $\gamma$-ray source 4FGL\,J1740.4-2850, without the excess in the model. We use approximately 6 month bins in each of these light-curves, and 95\% confidence upper limits on flux are used for any bin where the corresponding TS value is less than 4. The grey dotted lines indicate bins where source confusion between the 4FGL source and $\gamma$-ray excess are likely. } 
    \label{fig:AX17_lc}
\end{figure} 

\begin{figure}[h!]
    \centering
    \includegraphics[width=240pt]{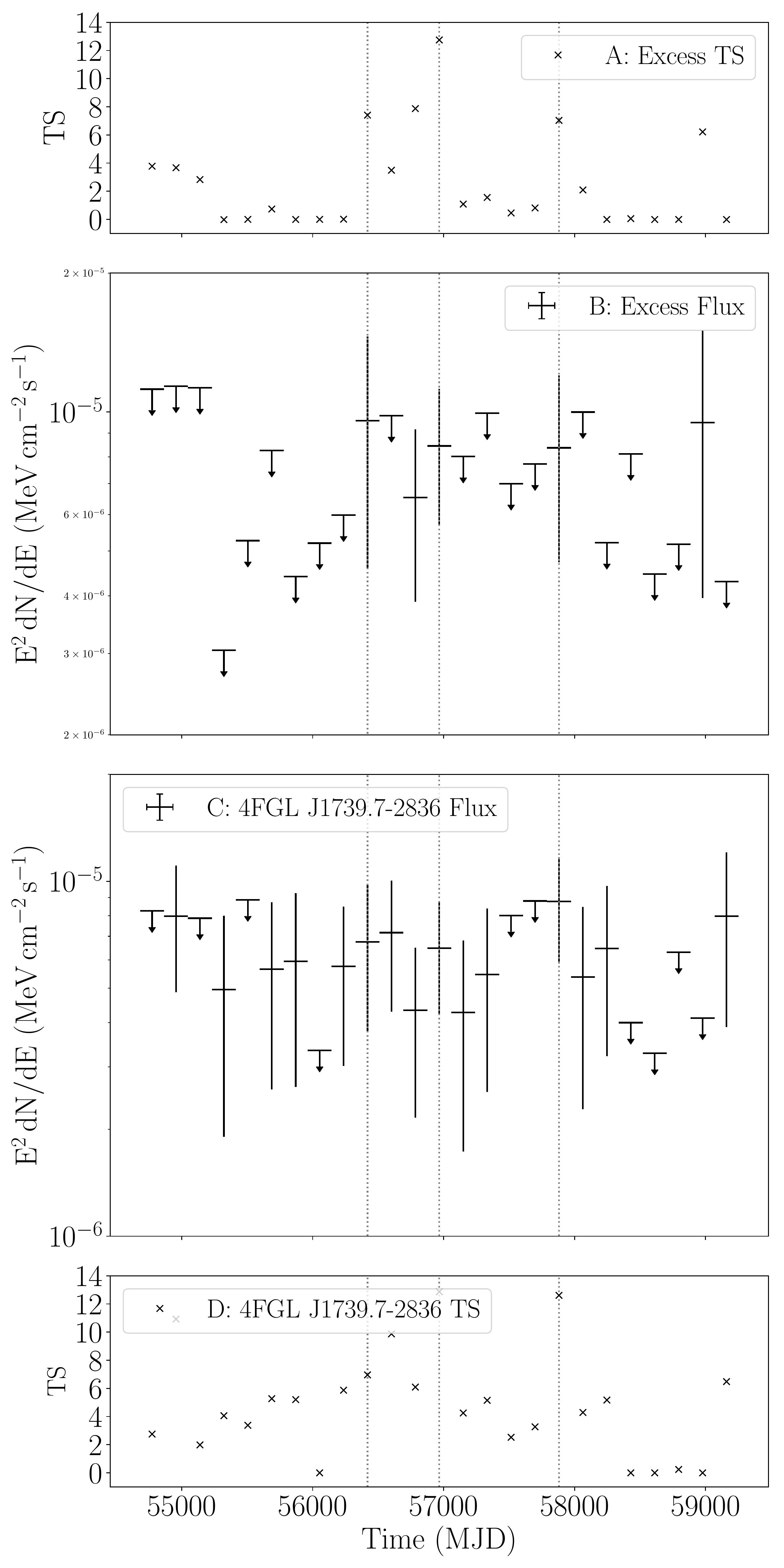}
    \caption{Panels B and A show the $\gamma$-ray flux and associated TS values of these flux points for the excess coincident with the optical position of AX\,J1740.1-2847. Panels C and D show the $\gamma$-ray flux and associated TS values of these flux values for the nearby $\gamma$-ray source 4FGL\,J1739.7-2836, without the excess in the model. We use approximately 6 month bins in each of these light-curves, and 95\% confidence upper limits on flux are used for any bin where the corresponding TS value is less than 4. The grey dotted lines indicate bins where source confusion between the 4FGL source and $\gamma$-ray excess are likely. } 
    \label{fig:AX17_lc_2}
\end{figure}

\subsection{AX\,J1740.1-2847}
\label{bin:AX17}
AX\,J1740.1-2847 (henceforth AX17) is a HMXB system with a long-period pulsar \citep{sakano_discovery_2000} in orbit around an unknown companion star \citep{kaur_near-infraredoptical_2010}. Over the entire observation window, we find a $\gamma$-ray excess coincident with the position of AX17 with a TS of 7.17, giving a $z$-score of $2.7 \sigma$. Across the 6 month binned light-curve we identify 5 bins with $\mathrm{TS} > 4$, with the peak TS being 12.8 (3.6 $\sigma$). This $\gamma$-ray excess is very likely due to source confusion given that the nearest sources are 4FGL\,J1740.4-2850 ($\mathrm{TS} = 100$ and an angular offset from AX17 of $0.082 \degree$) and 4FGL\,J1739.7-2836 ($\mathrm{TS} = 93.9$, offset $0.218 \degree$). Neither of these sources has any association with sources at other wavelengths. The primary source confusion counterpart for the $\gamma$-ray excess is likely to be 4FGL\,J1740.4-2850, with an angular offset that is less than one bin width. This presents a similar case to 1WGA06, dealt with in Section \ref{bin:1WGA06}, where we establish that the $\gamma$-ray excess is not independent of its nearest 4FGL neighbour. We calculate the 6 monthly binned light-curves of the AX17 excess and 4FGL\,J1740.4-2850, shown in Figure \ref{fig:AX17_lc}, and also 4FGL\,J1739.7-2836 (shown with the AX17 excess) in Figure \ref{fig:AX17_lc_2}. Unlike 1WGA06, source confusion in this case cannot be attributed to a single source, but rather to contamination from both nearby sources. 

Figures \ref{fig:AX17_lc} and \ref{fig:AX17_lc_2}, show that the bins from the $\gamma$-ray excess where a flux is measured (as opposed to an upper limit)  typically correlate with an enhancement in the TS values of one of the two nearby 4FGL sources. In particular, four of the five bins with measured $\gamma$-ray flux in the excess light-curve correlate in time with the two most significant bins from the 4FGL sources. These are indicated by the dotted grey lines. The evidence suggests that the $\gamma$-ray excess is not independent of either of its neighbouring sources, and therefore does not represent a legitimate detection of $\gamma$-rays from AX17.

\subsection{H\,1833-076}
\label{bin:H18}
H\,1833-076 (also known as Sct X-1 \citep{makino_scutum_1988} and henceforth referred to as H18) is a high mass X-ray binary system, thought to consist of an accreting pulsar and red supergiant donor star \citep{kaplan_lost_2007}. We identify a $\gamma$-ray excess coincident with the position of H18 with an angular offset from the X-ray position of H18 of $0.169 \degree$. This excess has a TS of 29.2 and a $95\%$ positional uncertainty of $0.175 \degree$. Given that the position of this excess lies right at the edge of the positional uncertainty bound, we localise its position to improve the uncertainty. Using the \texttt{localize} algorithm, we find a best fit position for the excess of $\mathrm{LII} = 24.5019  \degree \pm 0.0266 \degree$, $\mathrm{BII} = -0.0371 \degree \pm 0.0325 \degree$ with an overall $95\%$ positional uncertainty of $0.0717 \degree$. Given the shift in excess position and smaller, improved uncertainty, the $\gamma$-ray excess is no longer coincident with the position of H18 and we therefore believe it is unlikely to represent $\gamma$-ray emission from this X-ray binary. 

\subsection{GS\,1839-04}
\label{bin:GS18}
GS\,1839-04 (henceforth referred to as GS18) is an X-ray binary system with an unknown accretor and unknown companion star. We observe a $\gamma$-ray excess identified by the \texttt{gta.find\_sources} algorithm coincident with the position of GS18 which we designate PS J1842.0-0418. This has a TS value of 17.8 and an angular offset from the position of GS18 of $0.147 \degree$, although the 95\% positional uncertainty around PS J1842.0-0418 is unusually large at $1.01 \degree$. Contained within this uncertainty region are 7 4FGL sources, with the nearest neighbours to PS J1842.0-0418 being the unidentified source 4FGL J1842.5-0359c (TS = 318 and an angular offset from the position of GS18 of $0.498 \degree$) and 4FGL J1840.8-0453e, the young supernova remnant Kes\,73 \citep{gotthelf_kes_1997} (TS = 1050, offset $0.501 \degree$). Given that the positional uncertainty of PS J1842.0-0418 is so large, encompasses numerous, luminous $\gamma$-ray sources and that PS J1842.0-0418 is extended, we localise its position, even though the TS of PS J1842.0-0418 is below 25\footnote{Usually we lack photon statistics for more advanced analysis methods at such low source significance.}. 

\begin{figure}
    \centering
    \includegraphics[width=240pt]{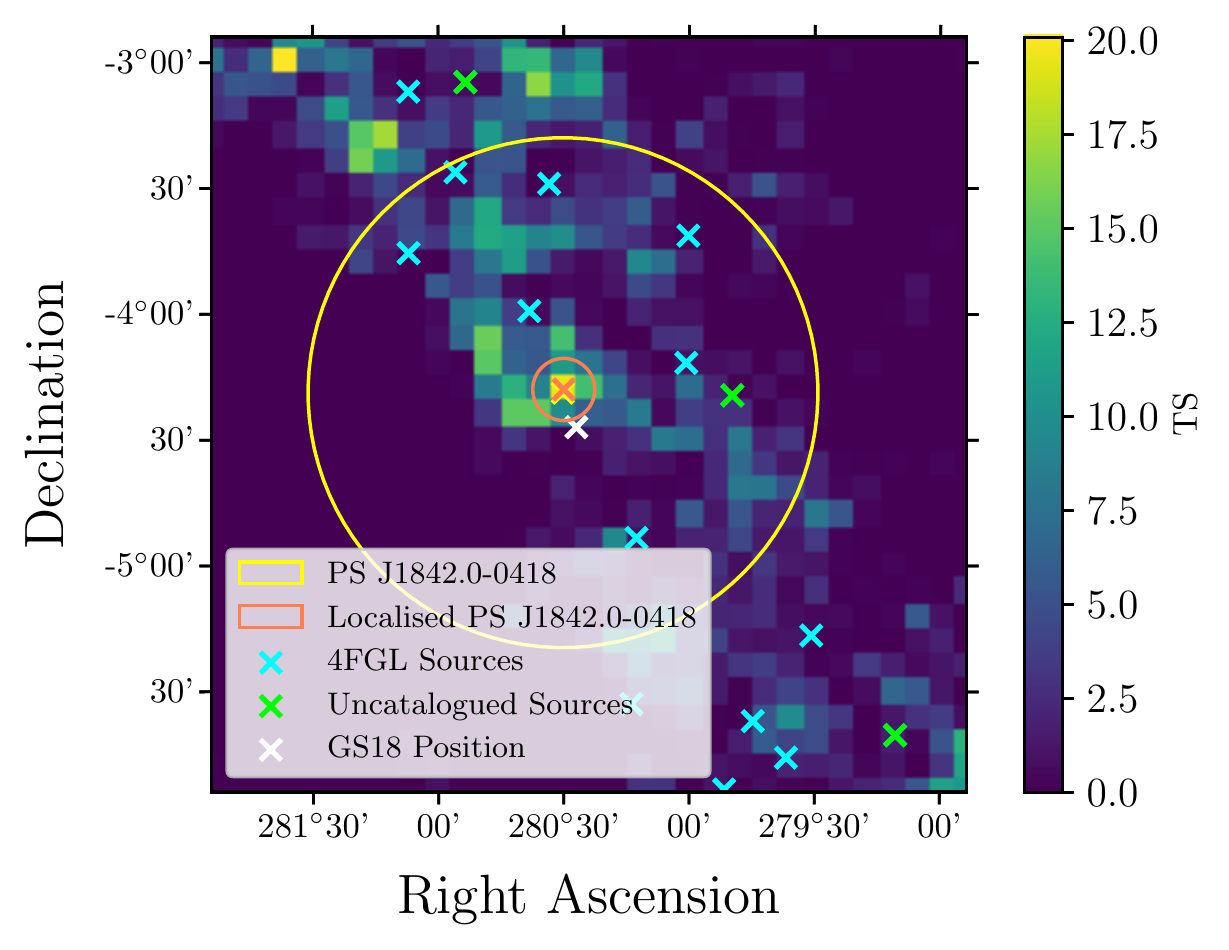}
    \caption{The TS map of the central $3\degree$ of the GS\,1839-04 ROI across the full 12.5 year observation time. Here, the positions of the closest 4FGL sources are indicated by blue crosses, whilst the positions of sources identified with the \texttt{gta.find\_sources} algorithm are indicated by green crosses. The centroid (shown with a cross) and 95\% positional uncertainty (shown with a circle) of PS J1842.0-0418 are given in yellow (before source localisation) and orange (after source localisation). The position of PS J1842.0-0418 barely shifts with localisation hence the orange and yellow markers overlap. The position of GS\,1839-04 itself is indicated by the white cross, and is no longer spatially coincident with PS J1842.0-0418 following localisation.  This TS map is generated after ROI optimization and fit, but before a point source for GS\,1839-04 is fitted to the model, to highlight the spatial coincidence between the excess and the position of GS\,1839-04. Bin widths are $0.1 \degree$ across. }
    \label{fig:GS18_TS}
\end{figure}

Figure \ref{fig:GS18_TS} shows a TS map centered on GS18, highlighting the extent of the PS J1842.0-0418 uncertainty and the sources within it. After localising the $\gamma$-ray emission from PS J1842.0-0418 we find that the 95\% positional uncertainty shrinks by approximately an order of magnitude to $0.1234 \degree$. As shown by Figure \ref{fig:GS18_TS}, this means that the X-ray position of GS18 is no longer within the 95\% positional uncertainty of PS J1842.0-0418 and therefore PS J1842.0-0418 is very unlikely to represent $\gamma$-ray emission from GS18. 

\begin{figure}
    \centering
    \includegraphics[width=240pt]{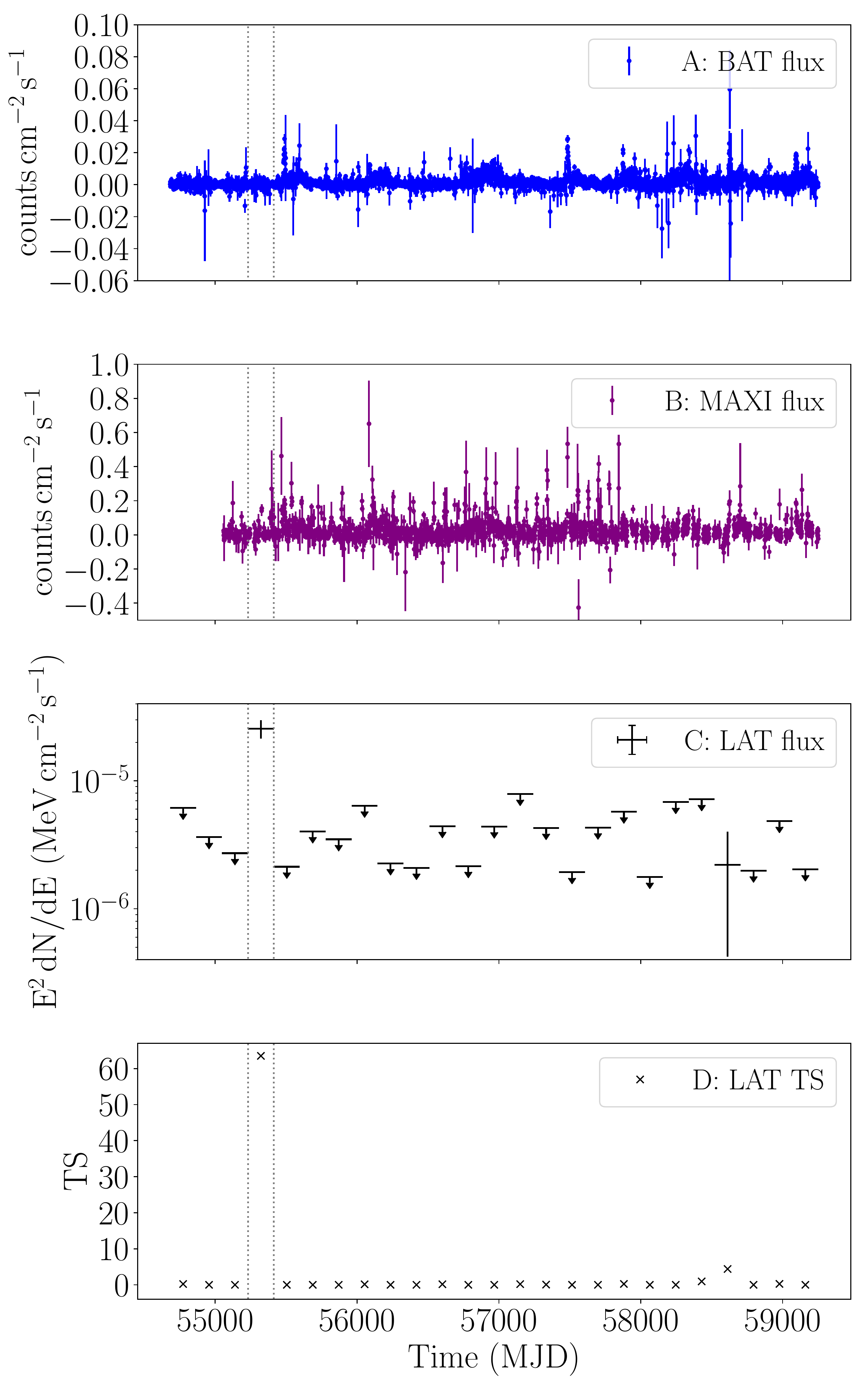}
    \caption{The \textit{Swift}-BAT and MAXI daily binned light-curves of SAX J2103.5+4545 are shown in Panels A and B respectively, with the 6-month energy flux measurements and TS values of the coincident $\gamma$-ray excess shown in Panels C and D respectively. We place 95\% confidence limits on any \textit{Fermi}-LAT energy flux bins with $\mathrm{TS} < 4$. The vertical dashed lines indicate the beginning and end of the 6-month period when there is a significant enhancement in the $\gamma$-ray flux. }
    \label{fig:SAX21_lc}
\end{figure}

\subsection{SAX J2103.5+4545}
\label{bin:SAX21}
SAX J2103.5+4545 (henceforth SAX21) is a pulsar \citep{hulleman_discovery_1998} accretor in orbit with a Be star companion \citep{reig_identification_2005} with an orbit of 12.7 days \citep{baykal_discovery_2000}. SAX21 is the closest known neutron star-Be star system to Earth \citep{blay_sax_2006}. SAX21 is a well-studied system, with outbursts well documented since its discovery, the most recent of which was in 2019 \citep{ducci_renewed_2019}, and persistent monitoring with \textit{Swift}-BAT meaning a wealth of multi-wavelength data is available for the source. Unlike the majority of X-ray binaries we report on here, we identify no persistent $\gamma$-ray excess coincident with the position of SAX21. However, we do identify significant transient emission in the 6-monthly binned light-curve. 

Figure \ref{fig:SAX21_lc} shows the X-ray light-curve of SAX21 together with the $\gamma$-ray light-curve generated at the position of SAX21. We measure 2 flux points, one with $\mathrm{TS} = 63.5$ ($z = 8.0 \sigma$) and the other with $\mathrm{TS} = 4.39$ ($z = 2.1 \sigma$). Given that we have 25 bins, we would expect one of these to be of $2 \sigma$ in a simply noise dominated distribution, however given the significance of the $8 \sigma$ bin, there is no statistical doubt that this represents a flaring $\gamma$-ray point source of some morphology. 

To test whether this transient point source represents $\gamma$-ray emission from SAX21, or another undiscovered source, we perform a full analysis using the same parameters as detailed in Table \ref{tbl:params}, with the exception that our time range now exclusively encompasses the bin containing the $8.0 \sigma$ $\gamma$-ray flare (MJD 55231 - MJD 55414). We then fit a $\gamma$-ray source to the position of SAX21 and localise its position. We then generate a TS map of the ROI, and plot the positions of all known sources, together with the position of SAX21 and the now localised position of the $\gamma$-ray excess, and its associated 95\% positional uncertainty.

\begin{figure}
    \centering
    \includegraphics[width=240pt]{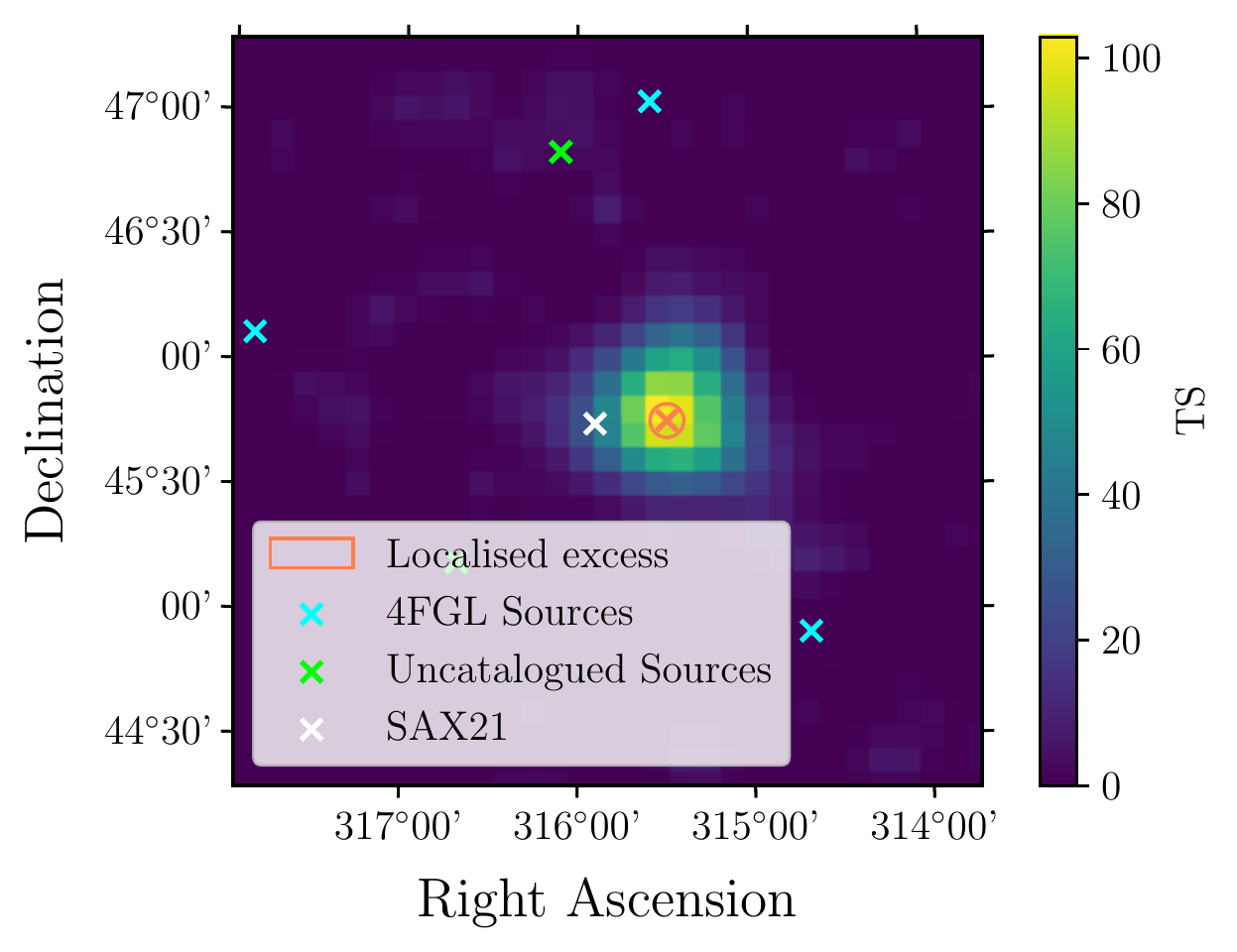}
    \caption{The TS map of the central $3\degree$ of the SAX J2103.5+4545 ROI, after our likelihood fit and the \texttt{gta.find\_sources} algorithm, over the MJD 55231 - MJD 55414 period. The blue crosses refer to the positions of 4FGL sources and the green crosses refer to the positions of uncatalogued sources. The white cross indicates the catalogued location of SAX21, whereas the orange cross and circle refer to the central position of the excess, and 95\% uncertainty after localisation. Our spatial bins have an angular width of $0.1 \degree$.}
    \label{fig:SAX21_TS}
\end{figure}

Figure \ref{fig:SAX21_TS} shows the TS map of the SAX21 ROI during the 6 month period of the $\gamma$-ray flare, which is the dominant feature of the plot. Whilst the white cross indicates the position of the binary itself, the orange cross indicates the localised position of the $\gamma$-ray flare, with the orange circle representing the bound of the 95\% positional uncertainty of the flare. As can be seen, SAX21 lies outside the positional uncertainty bound following localisation, so it is unlikely that SAX21 is the cause of this flare.

\bsp	
\label{lastpage}
\end{document}